\pdfoutput=1
\documentclass[onefignum,onetabnum]{siamart171218}

\usepackage{cite}

\usepackage{amsfonts}
\usepackage{graphicx}
\usepackage{epstopdf}
\usepackage{algorithmic}
\usepackage{hyperref}
\ifpdf
  \DeclareGraphicsExtensions{.eps,.pdf,.png,.jpg}
\else
  \DeclareGraphicsExtensions{.eps}
\fi
\usepackage{xcolor}
\colorlet{Barb}{magenta}

\newsiamremark{remark}{Remark}
\newsiamremark{hypothesis}{Hypothesis}
\crefname{hypothesis}{Hypothesis}{Hypotheses}
\newsiamthm{claim}{Claim}

\headers{Contagion maps on a class of networks embedded in a torus}{B. I. Mahler}

\title{Analysis of contagion maps on a class of networks that are spatially embedded in a torus\thanks{Submitted to the editors 28/12/2018.
}}

\author{Barbara I. Mahler\thanks{Mathematical Institute, University of Oxford, Oxford OX2 6GG, UK (\email{mahler@maths.ox.ac.uk}). The author acknowledges a studentship from the EPSRC under grant EP/G03706X/1.
 }}

\usepackage{amsopn}

\ifpdf
\hypersetup{
  pdftitle={Analysis of contagion maps on a class of networks that are spatially embedded in a torus},
  pdfauthor={B. I. Mahler}
}
\fi

\title{Analysis of contagion maps on a class of networks that are spatially embedded in a torus}
\begin{document}

\maketitle

\begin{abstract}
A spreading process on a network is influenced by the network's underlying spatial structure, and it is insightful to study the extent to which a spreading process follows such structure. We consider a threshold contagion model on a network whose nodes are embedded in a manifold and which has both `geometric edges', which respect the geometry of the underlying manifold, and `nongeometric edges' that are not constrained by that geometry. Building on ideas from Taylor et al. \cite{Taylor2015}, we examine when a contagion propagates as a wave along a network whose nodes are embedded in a torus and when it jumps via long nongeometric edges to remote areas of the network. We build a `contagion map' for a contagion spreading on such a `noisy geometric network' to produce a point cloud; and we study the dimensionality, geometry, and topology of this point cloud to examine qualitative properties of this spreading process. We identify a region in parameter space in which the contagion propagates predominantly via wavefront propagation. We consider different probability distributions for constructing nongeometric edges --- reflecting different decay rates with respect to the distance between nodes in the underlying manifold --- and examine the effect of such choices on the qualitative properties of the spreading dynamics. Our work generalizes the analysis in Taylor et al. and consolidates contagion maps both as a tool for investigating spreading behavior on spatial networks and as a technique for manifold learning.
\end{abstract}


\begin{AMS}
  {55N31, 05C82, 91D30, 82B43}
\end{AMS}



\section{Introduction}

Spreading dynamics are ubiquitous in many situations, including social settings and biological processes. The spreading of a contagious disease or of an idea between people are two obvious examples, and various other phenomena also give rise to spreading processes on networks \cite{LehmannSune2018,Newman2018,Pastor-Satorras2015}.  

The spreading of real-world contagions is often guided by the geometry of some underlying domain \cite{Hedstroem1994,Xu2007,Shannon1991,Corsi2014}. One example is the spread of contagious diseases. Historically, such diseases spread gradually along part of the earth's surface. Similarly, when means of transportation and communication are limited, information typically disseminates via entities that are physically close. In such cases, contagions often propagate as a wavefront, passing between geometrically close entities. However, with modern transportation and communication technology, there are now many scenarios where --- even in the presence of a well-defined underlying geometry, such as the spherical surface of the earth --- a contagion can also spread via connections that are not intrinsically geometric \cite{Balcan2009,Brockmann2013}. Examples of such scenarios include the spreading of an infectious disease via passengers traveling on a long-distance flight and the dissemination of information via social media. In these examples, a contagion jumps across space to distant locations, rather than following the geometry of an underlying domain. 

One way to study such phenomena is to consider contagion models on networks that are embedded in some underlying geometric space \cite{Rhodes1997}. In particular, one can consider networks that have both \emph{geometric edges} that respect the geometry of the underlying space, in the sense that they can only connect nodes that are close to each other according to the space's metric, and \emph{nongeometric} edges, which are not constrained by the underlying geometry and can connect nodes that are far from each other. Following terminology from \cite{Taylor2015}, we refer to such networks as \emph{noisy geometric networks}. It is interesting and important to ask \cite{LehmannSune2018,Pastor-Satorras2015,Rogers2010} what propagation pattern(s) a contagion follows and how much the structure of the underlying space influences such patterns. Two fundamental spreading mechanisms that can occur on a noisy geometric network are \emph{wavefront propagation} (WFP) and the \emph{appearance of new clusters} (ANC). Wavefront propagation is the spreading of a contagion along the structure of the underlying space via geometric edges. The appearance of new clusters occurs when long-range, nongeometric edges connect activated nodes with nodes in a region of the network that has been unaffected by the contagion and thereby lead to a new cluster of activated nodes in this previously unaffected region. We can view the nongeometric edges as `bridges' that can accelerate the spreading process considerably, especially if they are `long'. In a threshold model (a type of `complex contagion' \cite{Centola2007,LehmannSune2018}), in which a sufficient fraction or number of nodes in a focal node's neighborhood need to be active to activate that node, bridges also need to be sufficiently `wide' to encourage ANC.

Taylor et al. \cite{Taylor2015} used methods from topological data analysis and nonlinear dimensionality reduction to study spreading behavior of a threshold contagion model on noisy geometric networks. They explored the occurrence of WFP and ANC on a noisy ring lattice, and they examined the extent to which the spreading process follows the ring structure. To investigate the extent to which a complex contagion on a network adheres to the structure of the underlying space, they introduced the notion of a \emph{contagion map}, which maps each node of a network to a point in $\mathbb{R}^n$ based on its activation times in different realizations of a contagion process. It thereby produces a point cloud that one can view as a distortion of the network that reflects the contagion's spreading behavior. To see if they could identify the structure of the underlying space, Taylor et al. examined the geometry, dimensionality, and topology of such point clouds. They compared their results with a bifurcation analysis of the contagion on noisy ring lattices and found that the contagion map successfully recovers the geometry, dimensionality, and topology of the underlying space exactly when the contagion propagates predominantly by WFP. This illustrates that one can use contagion maps to illuminate propagation patterns of spreading processes on noisy geometric networks whose underlying space is known. Moreover, they found that on noisy ring lattices WFP occurs for a wide range of the network and contagion parameters, suggesting that contagion maps are a viable tool for inferring the structure of the underlying space of a noisy geometric network from contagion dynamics on it. That is, one may be able to use contagion maps as a technique for manifold learning. 

We follow the approach of \cite{Taylor2015}, and we build on their ideas through a study of a new example. We still use a threshold contagion model, but we consider a more complicated family of noisy geometric networks. Our networks, which can be interpreted as geometrically embedded in a flat torus, are similar to the Kleinberg small-world model \cite{Kleinberg2000small}. We use a contagion map to construct a point cloud that represents the dynamics of the contagion from a set of different initial conditions. We then examine the structure of this point cloud in three different ways: topologically (via the homology of a space that we build on the point cloud), geometrically (via distances between pairs of points), and with respect to dimensionality (via the approximate embedding dimension). We compare our findings to the topological and geometric structure, as well as the embedding dimension, of the torus. If the point cloud's structure resembles that of the torus geometrically, topologically, and in terms of its embedding dimension, this suggests that the contagion propagates via wavefront propagation along the structure of the underlying torus.

The motivation for our choice of network model is fourfold: (1) one can view it as a two-dimensional (2D) analogue of the noisy ring lattice that was studied in \cite{Taylor2015}; (2) the Kleinberg small-world model includes a nontrivial and adjustable spatial scaling in its probability distribution for constructing nongeometric edges; (3) the flat torus has locally Euclidean geometry and is entirely homogeneous (the local geometry at one point is the same as that at any other); and (4) the embedding of the network in the torus entails nontrivial topological features to take into account when comparing the topological structure of a contagion map to that of the underlying space. In our analysis, we find for a certain region of the parameter space that the structure of the contagion map resembles that of a torus in terms of topology, geometry and dimensionality. Further, this region corresponds to scenarios in which we can predict analytically that the contagion spreads predominantly via WFP, rather than via ANC. This consolidates the approach of \cite{Taylor2015} both as a way of determining spreading behavior of contagions on noisy geometric networks, and as a manifold-learning technique that is robust to noise. 

Our paper proceeds as follows. In sections~\ref{our_net} and \ref{contagion}, we define the network model and contagion model that we study, and we give some background on noisy geometric networks and contagions on networks in general. In section~\ref{methods}, we describe the employed methodology. We present a series of numerical experiments in section~\ref{results}, and we perform a bifurcation analysis in section~\ref{bifurcation}. In section~\ref{discussion}, we discuss our findings. We give background mathematical details on persistent homology in our supplementary material.


\section{Network model}\label{our_net}

A \emph{geometric network} is a network whose nodes are embedded in some metric space and whose edges, called \emph{geometric edges}, can occur only between pairs of nodes that are sufficiently close in this space \cite{Barthelemy2011,Barthelemy2018}. 

\begin{figure}[H]
    \centering
    \leftline{\hskip 0.3cm (a) \hskip 2.7cm (b) \hskip 2.7cm (c) \hskip 2.7cm (d)} 
    \begin{minipage}{0.25\textwidth}
        \centering
        \includegraphics[width=1\textwidth]{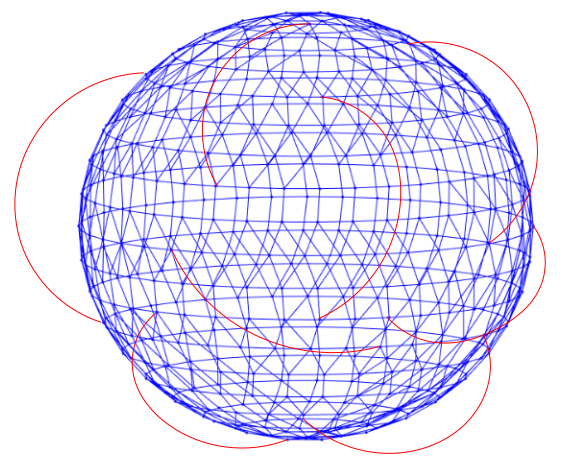} 
    \end{minipage}\hfill
    \begin{minipage}{0.25\textwidth}
        \centering
        \includegraphics[width=0.8\textwidth]{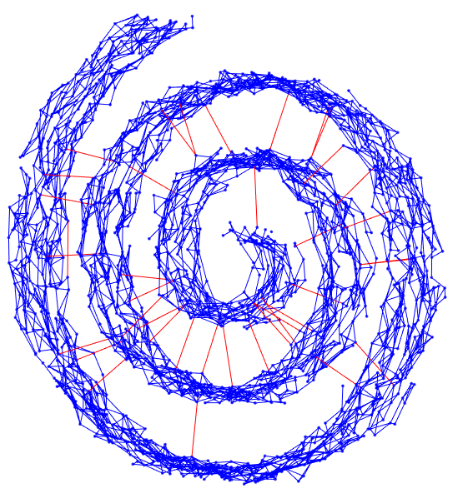} 
    \end{minipage}\hfill
    \begin{minipage}{0.25\textwidth}
        \centering
        \includegraphics[width=0.8\textwidth]{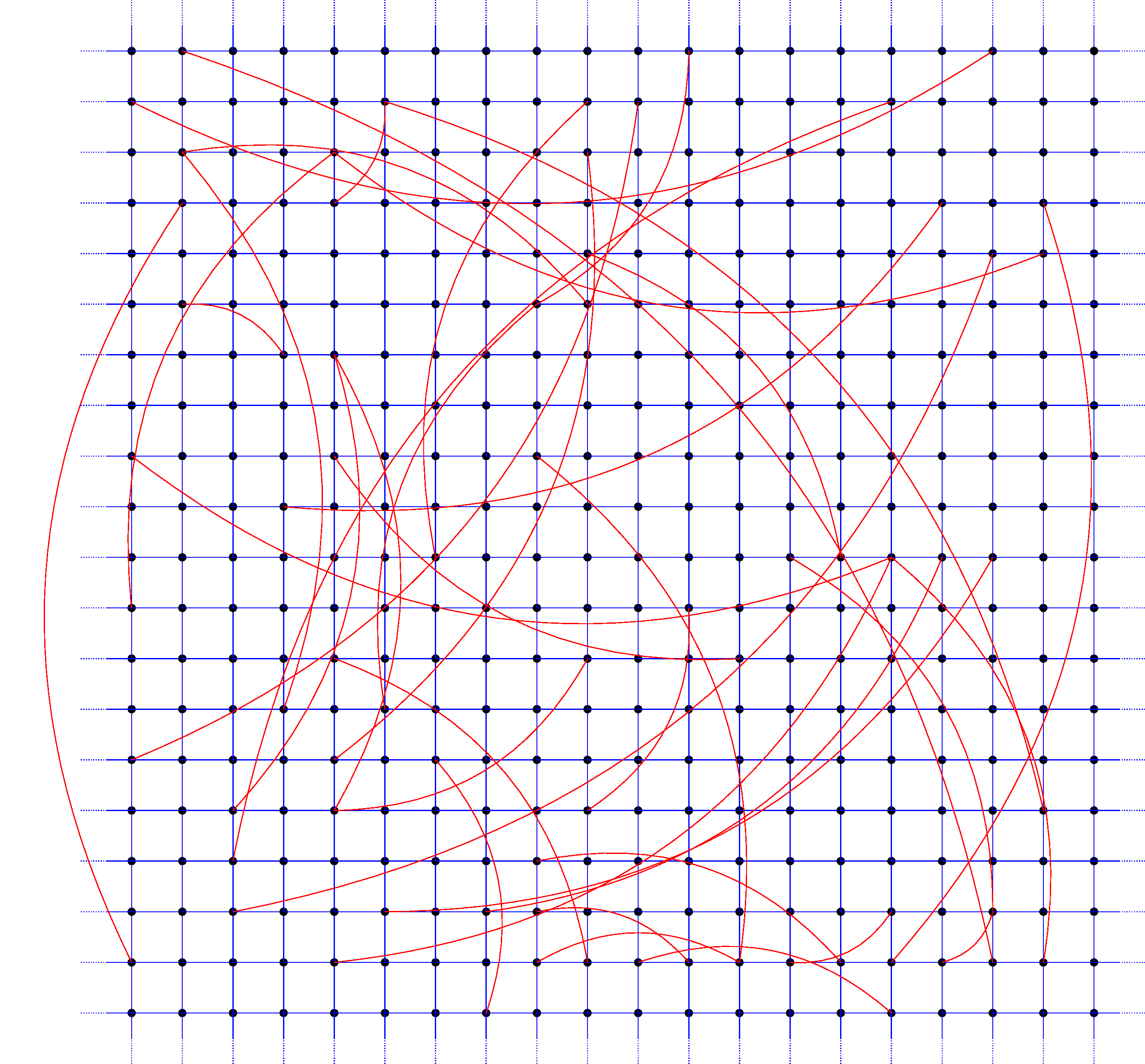} 
    \end{minipage}\hfill
    \begin{minipage}{0.25\textwidth}
        \centering
        \includegraphics[width=1\textwidth]{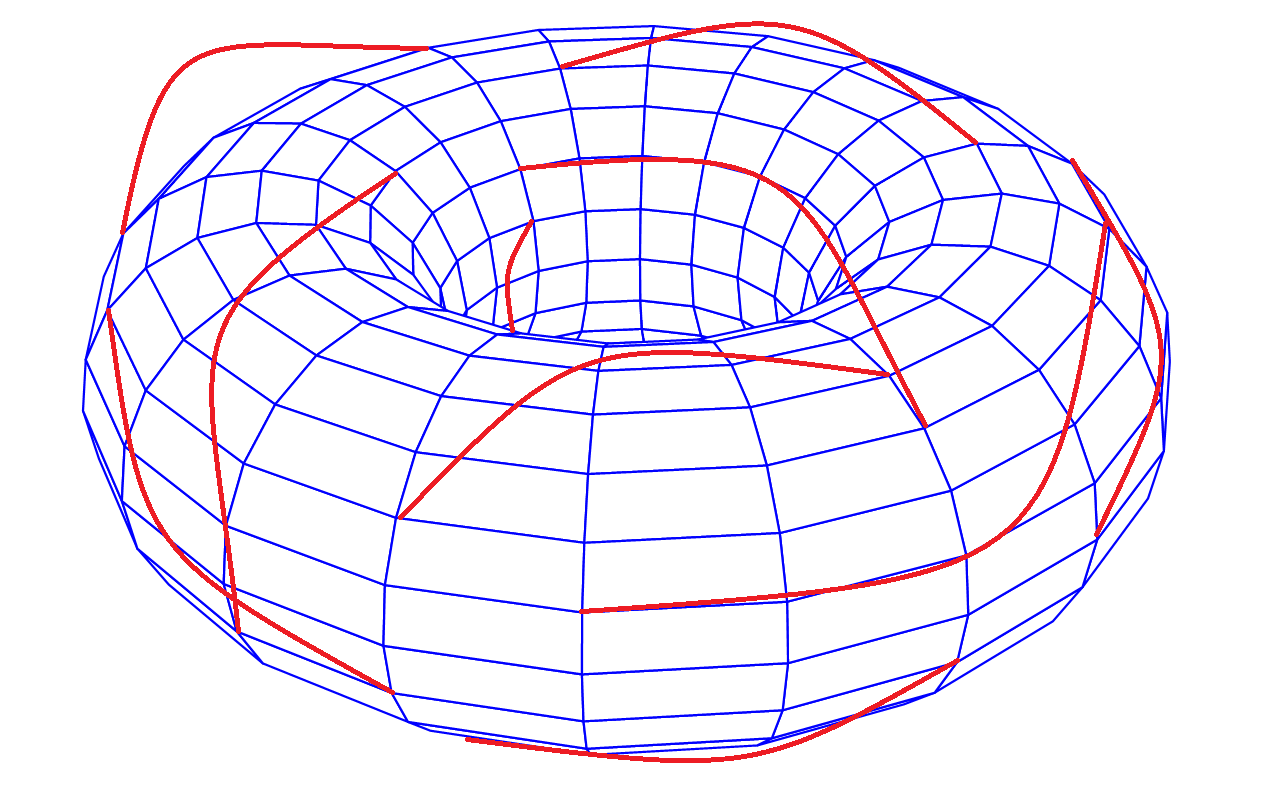} 
    \end{minipage}
    \caption{Examples of noisy geometric networks. (a) Nodes lie on a 2D sphere embedded in $\mathbb{R}^3$. There are geometric edges (blue) between nodes that are close to each other on the sphere, and we place nongeometric edges (red) uniformly at random. (b) The noisy Swiss roll. Nodes correspond to a noisy sample from a bounded 2D manifold that is embedded in $\mathbb{R}^3$ as a roll. There are geometric edges (blue) between nodes that are close to each other on the manifold, and there are nongeometric edges (red) between nodes that are close to each other in $\mathbb{R}^3$ but not close to each other on the manifold. (c,d) A Kleinberg-like small-world network. It consists of (c) a regular lattice of geometric edges (blue) that we (d) `wrap up' into a torus. We place nongeometric edges (red) according to some probability distribution. 
    }\label{fig:noisy}
\end{figure}

One can build a \emph{noisy geometric network} from a geometric network by adding so-called \emph{nongeometric} (i.e., `noisy') edges between pairs of nodes that can be distant from each other in the underlying metric space. For example, in synthetic noisy geometric networks, the nodes may be located on a manifold that is embedded in an ambient Euclidean space. One can place geometric edges between all or some of the node pairs that are at distances from each other that are below some fixed threshold, and one can then add nongeometric edges uniformly at random (see Figure~\ref{fig:noisy}(a)) or following some other probabilistic or deterministic rule. 

As another example (see Figure~\ref{fig:noisy}(b)), one can add noise to node locations in the ambient space and place a nongeometric edge between any two nodes that are close in the ambient space but not close with respect to geodesic distance along the manifold. Many nonlinear dimensionality-reduction techniques, such as diffusion maps and Isomap \cite{Belkin2002, Tenenbaum2000, Coifman2005, Gerber2007, Sorzano2014}, start by inferring a proximity network from point-cloud data, such as by connecting each point to its $k$ nearest neighbors, with the goal of finding underlying low-dimensional structure of the point cloud. Such proximity networks can be viewed as noisy geometric networks, as it is possible for nodes to be adjacent even when they are not close on the underlying manifold, and nonlinear dimensionality-reduction techniques seek to find purely geometric structures in such networks. 

We consider a family of noisy geometric networks that are embedded geometrically in a 2D manifold, with the nodes spread evenly on the surface of a flat torus. The torus has Betti numbers $\beta_0=1$, $\beta_1=2$, and $\beta_2 =1$ (we define Betti numbers in Definition~\ref{Betti} in the supplementary material), so there are multiple nontrivial topological features to take into account when comparing the structure of a contagion map to that of the underlying space. 

Our noisy geometric network is a variant of the Kleinberg small-world model \cite{Kleinberg2000small} (see Figure~\ref{fig:noisy}(c,d)). We start with a periodic square lattice of $N = n \times n$ nodes:
\begin{equation*}
	V = \frac{\mathbb{Z} \times \mathbb{Z} }{n\mathbb{Z} \times n\mathbb{Z}} \,,
\end{equation*}
so 
\begin{equation*}
	\overline{(i_x,i_y)} = \overline{(kn+i_x, ln+i_y)} \in V  \,\text{  for all  }\, k,l \in \mathbb{Z} \, .
\end{equation*}	
We define the \emph{periodic lattice distance} $\mu_{\rm per}$ between nodes $\overline{(i_x,i_y)}, \overline{(j_x,j_y)} \in V$ to be 
\begin{equation}\label{periodic_distance}
	\mu_{\rm per}(\overline{(i_x,i_y)},\overline{(j_x,j_y)})= |i_x-j_x|_{\mathrm{per}}+|i_y-j_y|_{\mathrm{per}}  \, ,
\end{equation}
where 
\begin{equation*}
	|a|_{\mathrm{per}}= \min \left\{ b \in \left\{0,1, \dots , n-2, n-1 \right\} \ | \ b=kn+a \,\text{  or   }\, b=kn-a \text{   for some   } k \in \mathbb{Z} \right\} \, ,
\end{equation*}
and the sum of the two residues in (\ref{periodic_distance}) is taken in $\mathbb{Z}$. The periodic lattice distance is the regular lattice distance with opposite sides of the lattice considered to be close to each other. We can thus think of the lattice as being `wrapped up' into a 2D torus, which has no boundary. In other words, we are using periodic boundary conditions. 

We fix $p \in \mathbb{R}_{>0}$ and place a geometric edge between any two nodes whose (periodic) Euclidean distance from each other is within $p$. That is, we place a geometric edge between nodes $i=\overline{(i_x, i_y)}$ and $j=\overline{(j_x,j_y)} \in V $ if and only if 
\begin{equation*}
	|i_x-j_x|_{\mathrm{per}}^2+|i_y-j_y|_{\mathrm{per}}^2 \leq p^2 \, .
\end{equation*}	
We call the number of geometric edges that are incident to a node $i$ its \emph{geometric degree}, which we denote by $d^{\rm{G}}(i)$. 

Each node also has $q \in \mathbb{N}$ `nongeometric stubs', and we connect pairs of stubs to build nongeometric edges as follows. We connect a nongeometric stub from node $i$ to a stub from node $j$ with a probability that is proportional to $\mu_{\rm per}(i,j)^{-\gamma}$, where $\gamma \in \mathbb{R}_{\geq 0}$ is a fixed parameter. We call the number of nongeometric edges that are incident to a node $i$ its \emph{nongeometric degree}, which we denote by $d^{\rm{NG}}(i)$ (where $d^{\rm{NG}}(i)=q$, by definition). The degree of a node $i$ is $d(i)=d^{\rm{G}}(i)+d^{\rm{NG}}(i)$, and the class of networks that we just defined consists of regular networks of uniform degree $d=d(i)$ for all $i \in V$. When $\gamma=0$, we match the nongeometric stubs uniformly at random. For $\gamma>0$, nongeometric edges tend to connect nodes that are close with respect to the periodic lattice distance, and this tendency becomes more pronounced for progressively larger $\gamma$.

\section{Contagion model}\label{contagion}
A \emph{contagion} on a network is a dynamical process in which nodes become successively `activated', starting from some initial condition. The most common type of initial condition is that  a set of `seed' nodes are active at time $t=0$ \cite{Porter2016, Kiss2017}. 
We examine the Watts threshold model (WTM) \cite{Watts2002} (see also \cite{Granovetter1978, Valente1995}), one of the simplest and best-studied models of a contagion on a network . 

Let $V$ (with $|V|=N$) denote the set of nodes of a network, and let $A=(A_{ij})_{i,j \in V}$ be the adjacency matrix of the network. In our contagion, each node can be either \emph{active} or \emph{inactive}, and we denote the state of node $i \in V$ at time $t$ by $\eta_i(t)$, which takes the value $1$ if it is active and the value $0$ if it is inactive. We call the set of nodes that are active at time $t=0$ a contagion \emph{seed}, and we denote the seed set by $S\subseteq V$. That is, $\eta_s(0) = 1$ for all $s \in S$ and $\eta_i(0) = 0$ for all $i \in V \backslash S$. If $S$ consists of a single node, the initial condition is called `node seeding'; if $S$ consists of a node together with its neighbors, the initial condition is called `cluster seeding'. For a given homogeneous threshold $T$, we update node states synchronously in discrete time steps according to the following rule. If $\eta_i(t)=1$, then $\eta_i(t+1)=1$. If $\eta_i(t)=0$, then 
\begin{equation*}
	\eta_i(t+1)=1 \, \text{ if and only if } \, f_i > T \,,\text{\hspace{5mm} where \hspace{3mm}} f_i = \frac{1}{d}\sum_{j \in V} A_{ij}\eta_j(t) \, .
\end{equation*}
In other words, a node activates at time $t+1$ if the fraction of its neighbors that are active is larger than $T$ at time step $t$. Once a node is active, it stays active forever. For a fixed homogeneous threshold $T$ and a given seed set $S$, this contagion is a deterministic and monotonic process, which eventually reaches a stable state in which either all of the nodes are active or some nodes are inactive and will never activate. (It is `monotonic' in the sense that a node that activates stays active forever.) For a given network and a given seed set $S$, this deterministic process is one `realization' of the contagion. Formally, a realization $R$ is the nested sequence of subsets of $V$ that are active at successive times: $R=\{S=S_0,S_1, \dots ,S_m,\dots\}$ such that $S_t=\{i \in V \mid\eta_i(t)=1\}$. Note that, due to the deterministic nature of the process, $S_0=S$ determines $S_m$ for all $m >0$. 


\section{Methods}\label{methods}

We construct a point cloud by mapping the nodes of a network to points in $\mathbb{R}^N$ based on their activation times in different realizations of the contagion dynamics. This so-called \emph{contagion map} was first studied in \cite{Taylor2015} and is inspired by approaches, such as diffusion maps and Isomap \cite{Belkin2002, Tenenbaum2000, Coifman2005, Gerber2007, Sorzano2014}, from nonlinear dimensionality reduction. A point cloud that is the image of a contagion map can be interpreted as a distortion of an underlying network structure that reflects the contagion dynamics. We analyze the structure of this point cloud from three different perspectives --- topologically, geometrically, and with respect to dimensionality --- and compare it to the structure of the underlying network. We expect the structure of the point cloud in $\mathbb{R}^N$ to resemble the structure of the underlying network when the contagion spreads predominantly via WFP. We perform a bifurcation analysis \footnote{The traditional use of the term `bifurcation' \cite{gucken1983} is to describe situations in dynamical-systems theory in which a system's qualitative behavior changes in a mathematically quantifiable way (e.g., as expressed using a normal form), such as the onset of a limit cycle for a critical value of a parameter, as a function of one or more parameters. The notion of bifurcation that we examine in the present paper is somewhat different in flavor from classical bifurcations, but we still examine qualitative changes in dynamics as we adjust parameters in a model.} to identify regions in parameter space for which WFP is the predominant mode of propagation, and we use the results of this analysis to validate that the structure of the underlying network is recovered in the point cloud whenever the contagion spreads predominantly via WFP. 

To compare the topology, geometry, and dimensionality of a point cloud to the structure of the network on which it is based, we need to specify this structure precisely. Specifically, we need to choose a metric space associated with the network and we need to specify the locations of the network's nodes in this metric space. 

We consider the torus as the Cartesian product of two circles: 
\begin{equation}\label{torus}
	\mathbb{T} = \frac{1}{2 \pi} \mathbb{S}^1 \times \frac{1}{2 \pi} \mathbb{S}^1 \subset \mathbb{C} \times \mathbb{C} \cong \mathbb{R}^4 \,.
\end{equation} 
We evenly distribute the nodes of our network on this torus $\mathbb{T}$. The $N=n^2$ nodes of our network are the points on $\mathbb{T}$ with coordinates
\begin{equation}\label{reg_points}
	\frac{1}{2 \pi} \left( \cos \frac{2 \pi x}{n}, \sin \frac{2 \pi x}{n}, \cos \frac{2 \pi y}{n}, \sin \frac{2 \pi y}{n} \right)\,, \qquad
	x,y \in \{0,1, \dots, n-1\} \, .
\end{equation}


\subsection{Contagion maps}\label{contagion_map}

Consider our WTM contagion, with homogeneous threshold $T$, on one instantiation of our Kleinberg-like network for some fixed parameter values $n$, $p$, $q$, and $\gamma$. For a given seed set, the contagion dynamics is a deterministic process, and we can record the activation times of the nodes. We consider several realizations of the contagion dynamics initialized with different seeds. We denote the set of realizations by $J=\{R_1,R_2, \dots, R_{|J|}\},$ and we denote the activation time of node $i \in V$ in realization $R_j \in J$ by $x_j^{(i)}$. If node $i$ is never activated in realization $R_j$, we set $x_j^{(i)}=2N$ (i.e., larger than any actual activation time). 

The \emph{regular contagion map} associated with the set $J$ of realizations is a function from the set $V$ of nodes to $\mathbb{R}^{|J|}$. It is defined by
\begin{equation*} 
	i \mapsto x^{(i)}=[x_1^{(i)}, x_2^{(i)}, \dots , x_{|J|}^{(i)} ]\,.
\end{equation*}
The regular contagion map associated with $J$ maps each node in $V$ to a vector in $\mathbb{R}^J$ that records its activation times in each of the realizations. 

We take $J$ to be the same size as $V$ and choose the seed sets to be the clusters around the different nodes, such that the seed that initializes realization $R_j \in J$ is $S^{(j)}=\{j\} \cup \{k \ | \ A_{jk} \neq 0\}$. In this case, the activation time of node $i$ in realization $R_j$ is a proxy for a distance between nodes $i$ and $j$. To see this, consider the realization of a contagion with homogeneous threshold $T=0$ initialized with a single seed node $\{j\}$ and observe that the activation time of node $i$ is exactly the length of a shortest path between $i$ and $j$. For cluster seeding of our contagion, the activation time of node $i$ in realization $R_j$ may not be precisely the shortest-path distance between $i$ and $j$; it depends on how the contagion spreads. Moreover, $x_j^{(i)} \neq x_i^{(j)}$ in general. With this in mind, we define the \emph{reflected contagion map}, which maps $i \mapsto y^{(i)}=[x_i^{(1)}, x_i^{(2)} \dots , x_i^{(|J|)}]$, and the \emph{symmetric contagion map}, which maps $i \mapsto [x_1^{(i)} + x_i^{(1)}, \dots , x_{|J|}^{(i)} + x_i^{(|J|)}]$. 


\subsection{Geometry}\label{geometry}

To quantify the similarity of the geometric structure of a contagion map to that of the network on which it is based, we calculate the Pearson correlation coefficient of pairwise distances between points of the point cloud and pairwise distances between corresponding nodes of the network. We use Euclidean distance in $\mathbb{R}^4$ for the nodes and Euclidean distance in $\mathbb{R}^{N}$ for points in the point cloud.

Recall that the nodes lie on the torus \eqref{torus} at points with coordinates \eqref{reg_points}. Let 
\begin{equation*}
	w^{(i)} = \frac{1}{2 \pi} \left(\cos \frac{2 \pi i_x}{n}, \sin \frac{2 \pi i_x}{n}, \cos \frac{2 \pi i_y}{n}, \sin \frac{2 \pi i_y}{n} \right)
\end{equation*} 
denote the point in $\mathbb{T}$ that is associated with node $i=\overline{(i_x,i_y)}$. 
The distance between two points, $w^{(i)}$ and $w^{(j)}$, in $\mathbb{T}$ is
\begin{equation*}
	d \left(w^{(i)}, w^{(j)} \right) =  \sqrt{\sum_{k=1}^4\left(w_k^{(i)} - w_k^{(j)} \right)^2}
= \frac{1}{\pi}\left(\sin^2  \frac{(i_x-j_x)\pi}{n} + \sin^2 \frac{(i_y-j_y) \pi}{n} \right)^{\frac{1}{2}}\,,
\end{equation*}
and the distance between the corresponding points, $x^{(i)}$ and $x^{(j)}$, in the point cloud is
\begin{equation*}
	d(x^{(i)},x^{(j)})=\sqrt{\sum_{k=1}^N\left(x_k^{(i)} - x_k^{(j)} \right)^2} \, .
\end{equation*}
Given ordered sets, $D^{\mathrm{net}}$ and $D^{\mathrm{map}}$, of pairwise distances between nodes of the network and points in the point cloud, respectively, we compute the Pearson correlation coefficient between these sets:
\begin{equation*}
	\rho= \frac{\sum\limits_{i=1}^N \sum\limits_{j=i+1}^N \left[ d(w^{(i)}, w^{(j)})-\overline{d(w^{(i)}, w^{(j)})} \right] \left[ d(x^{(i)},x^{(j)}) - \overline{d(x^{(i)},x^{(j)})}  \right]}{\sqrt{\sum\limits_{i=1}^N \sum\limits_{j=i+1}^N\left[ d(w^{(i)},w^{(j)})- \overline{d(w^{(i)},w^{(j)})} \right]^2} \sqrt{\sum\limits_{i=1}^N \sum\limits_{j=i+1}^N\left[ d(x^{(i)},x^{(j)})- \overline{d(x^{(i)},x^{(j)})} \right]^2}} \, ,
\end{equation*}
where 
\begin{equation*}
	\overline{d(w^{(i)}, w^{(j)})}= \frac{\sum_{i=1}^N \sum_{j=i+1}^N  d(w^{(i)}, w^{(j)})}{(N^2-N)/2}
\end{equation*}
denotes the mean pairwise distance between nodes and $\overline{d(x^{(i)}, x^{(j)})}$ denotes the mean pairwise distance between points in the contagion map. Progressively larger Pearson correlation coefficients $\rho$ indicate progressively more similar geometric structures between a contagion map and its associated network. 


\subsection{Topology}\label{topology}

We examine the topology of a contagion map by studying the persistent homology (PH) of the Vietoris--Rips (VR) filtration (see Definition~\ref{Rips_definition} in the supplementary material) on its associated point cloud. 
We calculate PH using the software package {\sc Ripser}\footnote{\textsc{Ripser} is publicly available at \url{https://github.com/Ripser/ripser}.}. We seek to quantify the extent to which topological features of a torus appear in the barcode that represents PH in a given dimension. To do this, we calculate the Wasserstein distance $W_2[d]$ (see Definition~\ref{Wasserstein} in the supplementary material) between this barcode and a `model barcode' that represents topological features of a torus in the given dimension. As a `model  barcode', we choose the one that corresponds to the PH of the VR filtration on the regular point cloud on the torus in formula~\eqref{reg_points}. 
Smaller Wasserstein distances correspond to more `torus-like' point clouds, recovering the topology of the manifold in which the network's nodes are embedded. Roughly speaking, a barcode exhibits the topological features of a torus when it has two dominant bars in dimension $1$ and one dominant bar in dimension $2$ (as well as one bar that never dies in dimension $0$). We work with networks of $N=50 \times 50$ nodes (see section~\ref{our_net}). Due to the computational complexity of computing 2D persistent homology of a VR filtration on $2500$ points (it involves building up to $1.6263 \times 10^{12}$ simplices), we compute PH only up to dimension $1$, which requires building only up to $2.6042 \times 10^9$ simplices for a given point cloud.   

\begin{figure}[H]
    \centering
    \leftline{\hskip 0.00cm (a) \hskip 3.9cm (b) \hskip 3.9cm (c)} 
    \begin{minipage}{.31\textwidth}
        \centering
        \includegraphics[width=1\textwidth]{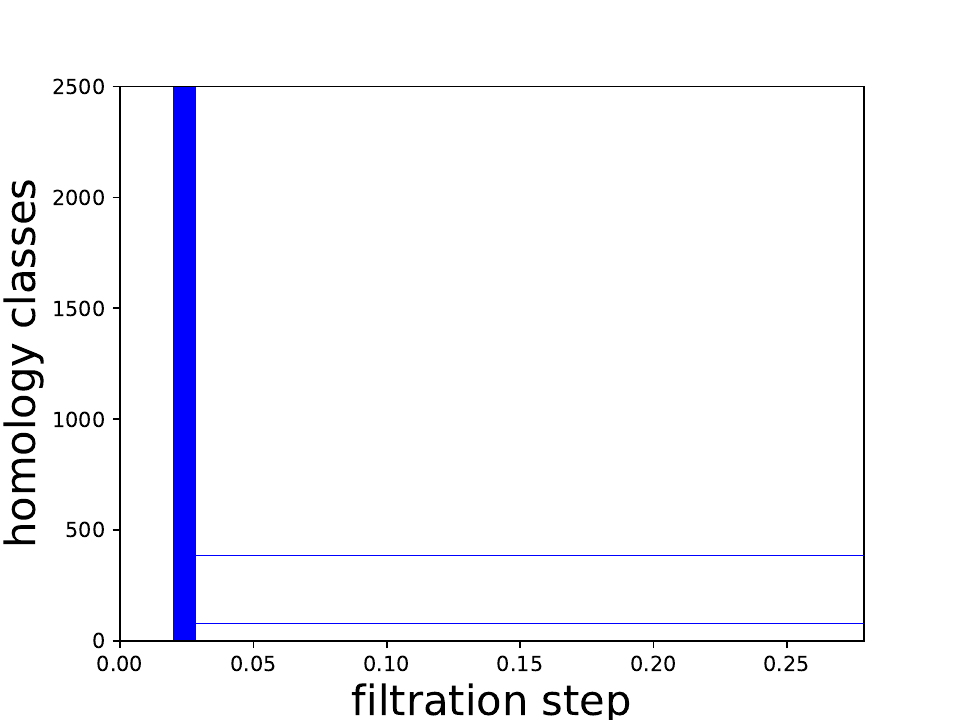} 
    \end{minipage}\hfill
    \begin{minipage}{0.31\textwidth}
        \centering
        \includegraphics[width=1\textwidth]{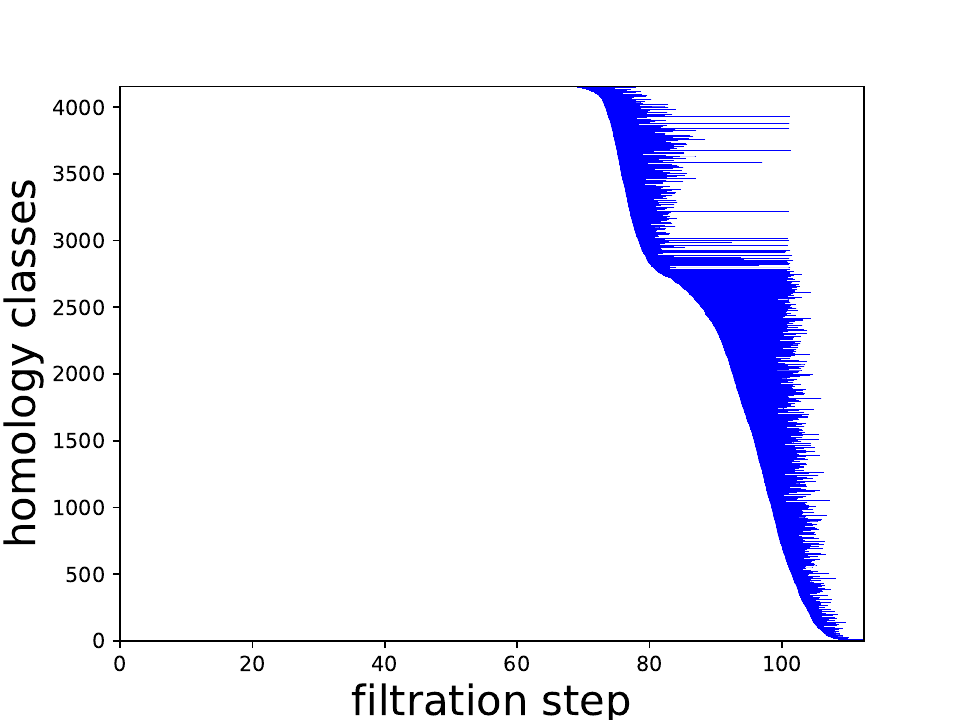} 
    \end{minipage}\hfill
    \begin{minipage}{0.31\textwidth}
        \centering
        \includegraphics[width=1\textwidth]{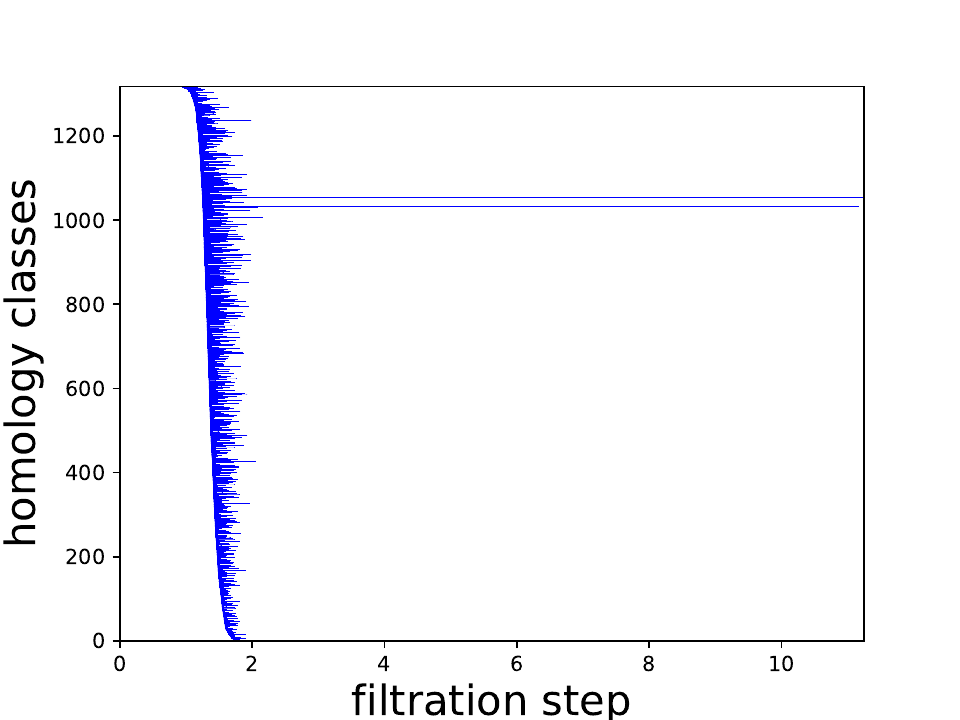} 
    \end{minipage}\hfill
    
    \leftline{\hskip 0.00cm (d) \hskip 3.9cm (e) \hskip 3.9cm (f)} 
    \begin{minipage}{.31\textwidth}
    \centering
        \includegraphics[width=1\textwidth]{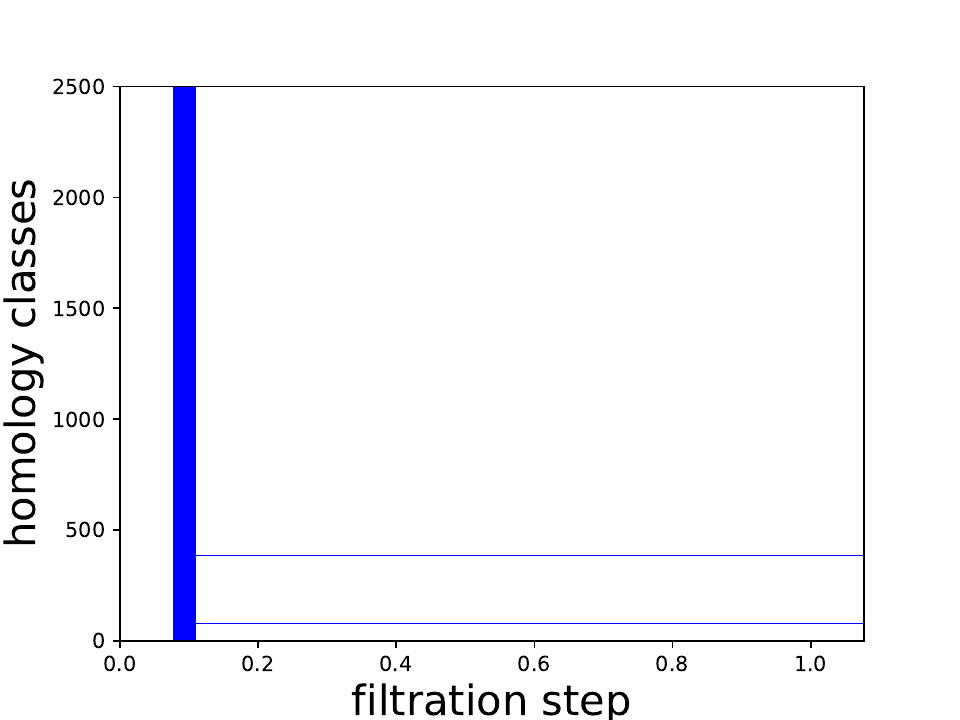}
    \end{minipage}\hfill
    \begin{minipage}{.31\textwidth}
    \centering
        \includegraphics[width=1\textwidth]{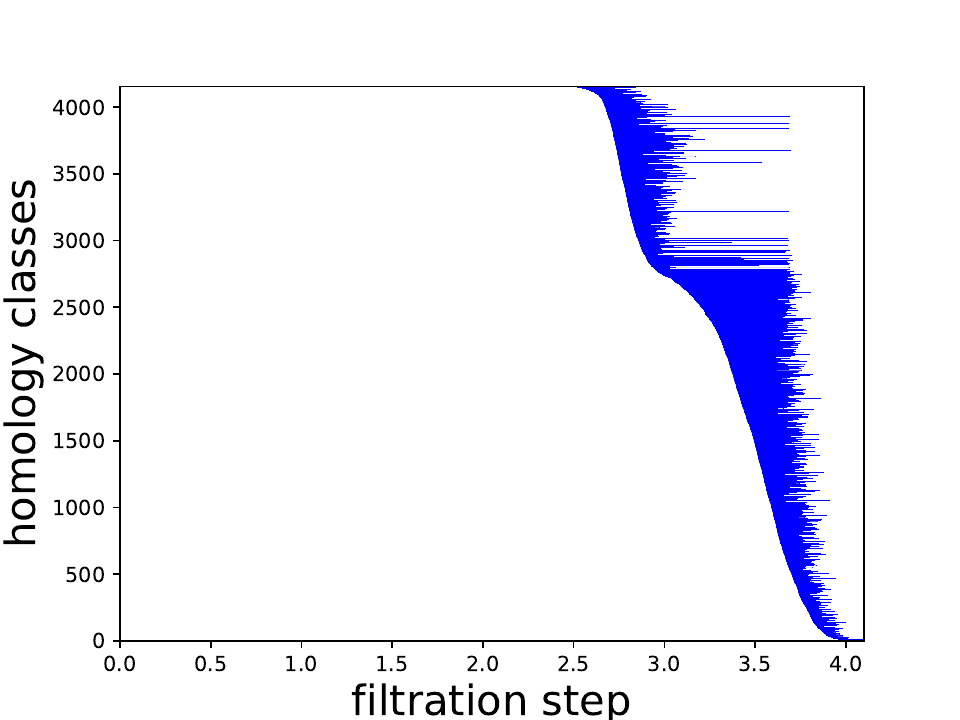}
    \end{minipage}\hfill
    \begin{minipage}{0.31\textwidth}
        \centering
        \includegraphics[width=1\textwidth]{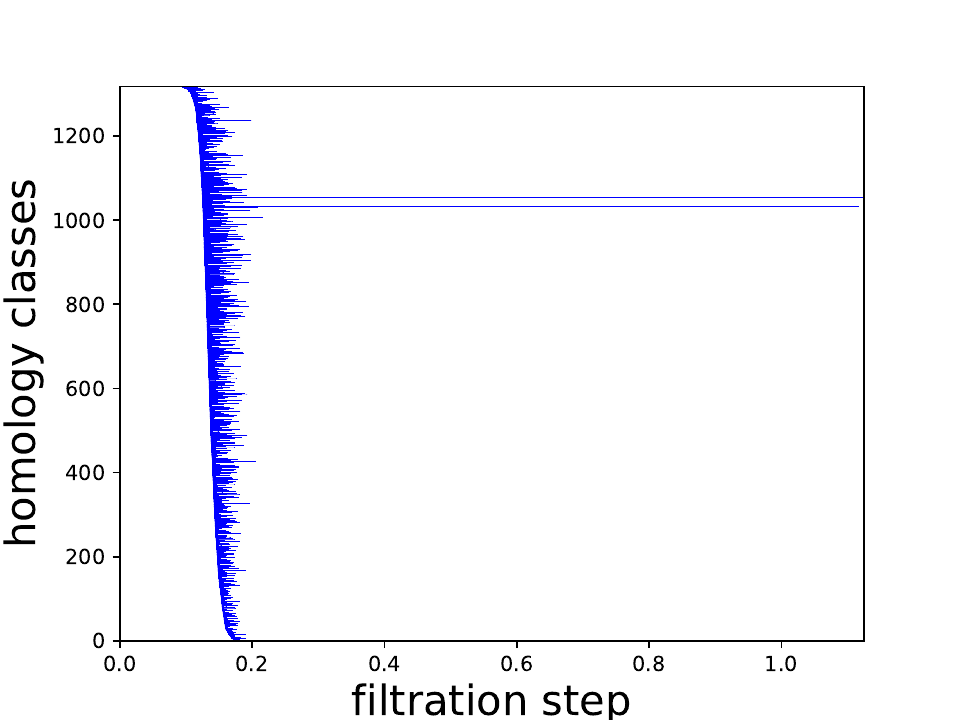} 
    \end{minipage}\hfill
    \caption{Illustrative examples to demonstrate the sensitivity of the Wasserstein distance $W_2[d]$ to barcode `scales'. Barcodes (d), (e), and (f) are the calibrated versions of barcodes (a), (b), and (c), respectively. For example, the Wasserstein distance $W_2[d]$ between the barcodes in (a) and (d) is about $51.55$, although they are identical aside from their scales. Barcodes (a) and (c) (and consequently (d) and (f)) represent 1D topological features of a torus, whereas barcode (b) (and consequently barcode (e)) does not. However, in the uncalibrated versions of the barcodes, the Wasserstein distance $W_2[d]$ between barcodes (a) and (b) is about $12.01$ and the Wasserstein distance $W_2[d]$ between barcodes (a) and (c) is about $138.86$, suggesting that (a) and (b) --- rather than (a) and (c) --- have similar topological features. By contrast, the Wasserstein distances between the calibrated versions of the barcodes illustrate the true topological proximities. The distance is about $350.41$ between barcodes (d) and (e) and about $51.85$ between barcodes (d) and (f).
}
    \label{scalability}
\end{figure}

The Wasserstein distance between barcodes is sensitive to scaling. Consider, for instance, two barcodes that have the same number of bars, such that the relative lengths of the bars within each barcode are the same. Although these two barcodes represent the exact same topological features --- albeit of different sizes --- the Wasserstein distance between them is nonzero. Similarly, two barcodes that represent very similar topological features, but are at very different scales, may have a larger Wasserstein distance between each other than two barcodes that represent different features but are close in `scale'\footnote{Our use of the term `scale' differs from existing uses in topological data analysis. Two example uses of `scale' in TDA are for the persistence of a topological feature in a filtration and for the point in a filtration at which a feature appears.}. 
See Figure~\ref{scalability} for an illustration of this phenomenon. 

In the present application, this sensitivity to scaling can manifest as follows. The model-torus barcode corresponding to regularly spaced points on a torus constructed as the Cartesian product of two circles of circumference $1$ has relatively short bars. 
For our contagion, larger values of $T$ entail slower spreading. 
Therefore, if we have two contagion maps which both arise from spreading via WFP without ANC, but one with large $T$ (entailing slow propagation) and the other with small $T$ (entailing fast propagation), then both contagion maps have the same (torus-like) shape, but the former is much `larger' than the latter. This implies, in turn, that the former's corresponding barcode is farther away than the latter from the model-torus barcode. 
Similarly, when there are many nongeometric edges, there is fast spreading via ANC. Therefore, although the shape of the contagion map should not look torus-like in this case, but instead should look like a cluster of points, the Wasserstein distance from the corresponding barcode to the model torus barcode may still be small (simply by virtue of the size of the point cloud, rather than because of its shape). 

\begin{figure}[H]
\leftline{\hskip -0.3cm (a) \hskip 3.9cm (b) \hskip 3.9cm (c)} 
\hskip -5mm\begin{minipage}{0.31\textwidth}
\includegraphics[width=1\textwidth]{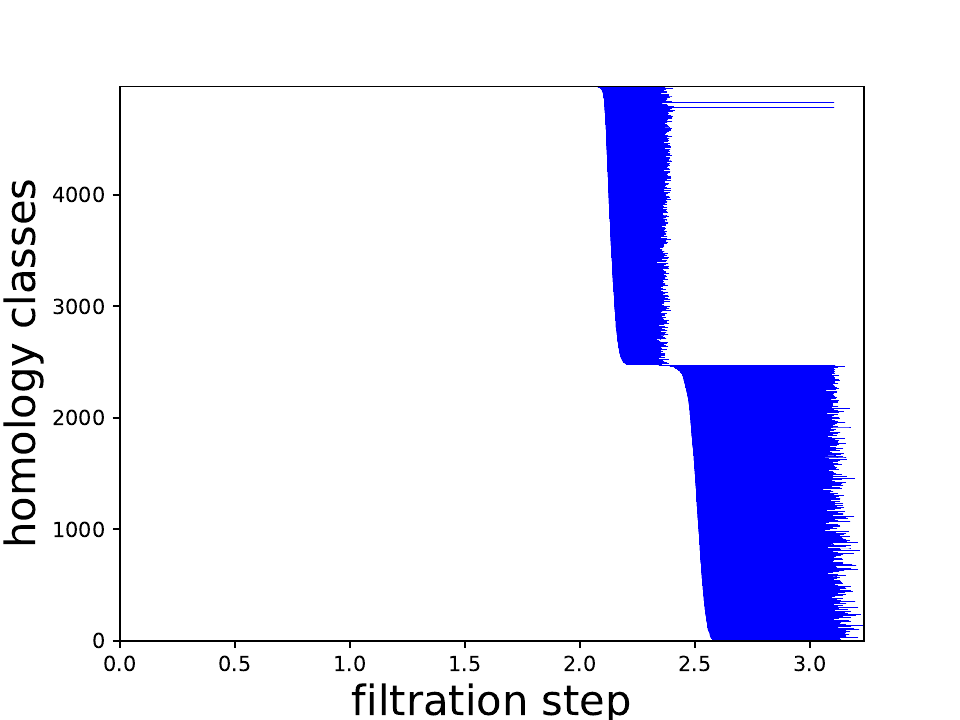}
\end{minipage}
\hskip 4mm\begin{minipage}{0.31\textwidth}
\includegraphics[width=1\textwidth]{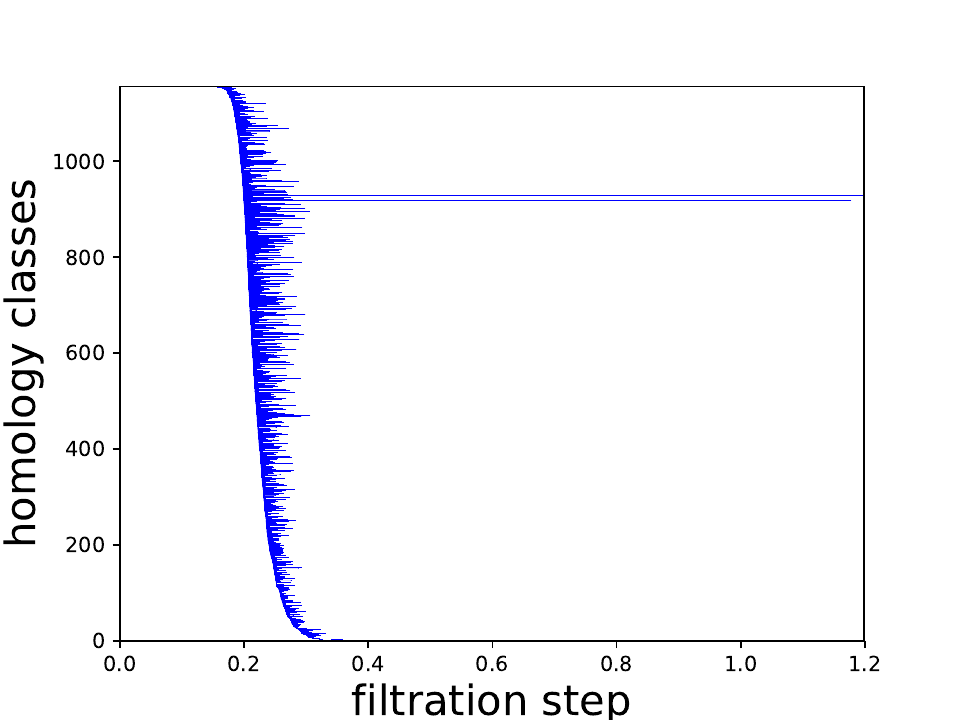}
\end{minipage}
\hskip 4mm\begin{minipage}{0.31\textwidth}
\includegraphics[width=1\textwidth]{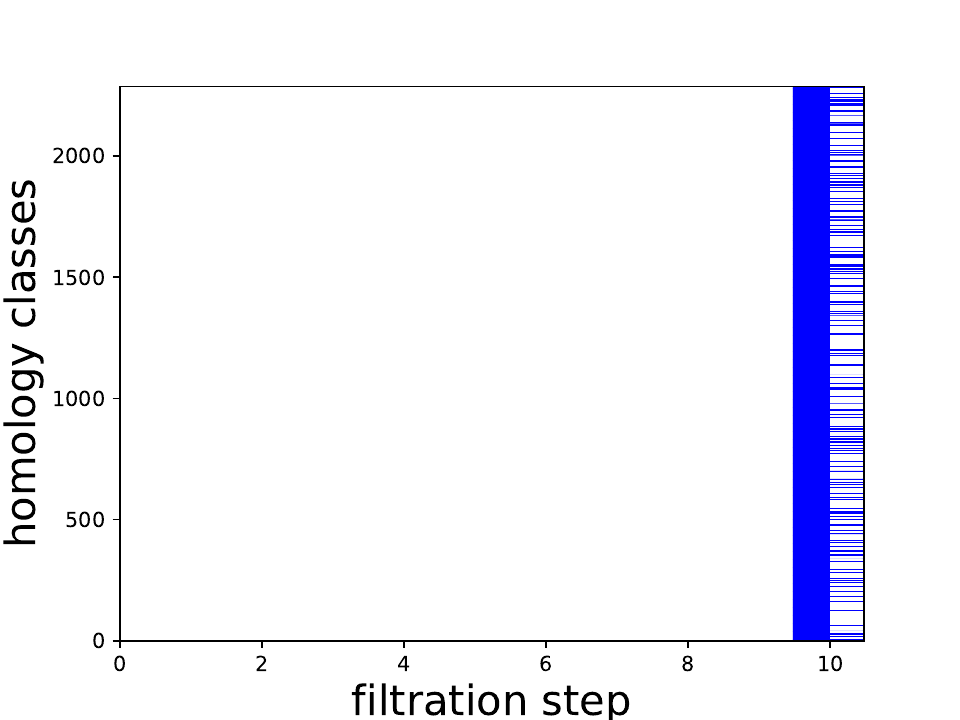}
\end{minipage}
\caption{Calibrated barcodes of the PH of the VR filtrations on contagion maps arising from a Kleinberg-like small-world network with parameter values $N=2500$, $d^{\rm{G}}=8$, $d^{\rm{NG}}=2$ and $\gamma=0$ for different values for the contagion threshold $T$. In (a), we use a small threshold ($T=0.05$), and we observe fast spreading via both WFP and ANC.  In (b), we use a threshold of $T=0.25$, for which ANC is unlikely and we expect spreading to follow WFP. In (c), $T=0.4$, which is a large threshold, for which we expect little spreading of a contagion. (See Figure~\ref{full_bifurcation} to locate these parameter values in the $(d^{\rm{NG}},T)$-parameter space and identify their associated spreading regimes.) In panel (b), we observe two relatively long bars, which represent the 1D topological features of the torus. In panel (a), we still observe some torus-like features (in the form of two slightly dominant bars), despite the fast spreading via ANC. This illustrates that WFP can still affect contagion maps noticeably, even in the presence of ANC. In panel (c), there are many bars that are born and die at the same time. This reflects the fact that little spreading occurs, as most nodes do not activate in most realizations of the contagion. The Wasserstein distances from the calibrated barcode for the PH of the VR filtration on the regularly spaced point cloud on the torus (see Figure~\ref{scalability}(d)) are about $943.25$ in panel (a), about $52.01$ in panel (b), and about $669.68$ in panel (c).
}\label{calibrated}
\end{figure}

 To counteract the above scaling issue, we `calibrate' all barcodes before calculating the Wasserstein distance (see Figure~\ref{calibrated}). We find the longest bar in each barcode and divide the birth and death times of all bars by that length. This yields barcodes whose longest bar is exactly $1$, so they can be considered to be at the same `scale'. Consequently, the Wasserstein distance between these calibrated barcodes can serve as a measure for comparing the topological features of the corresponding point clouds without taking absolute distances (i.e.~geometry) into account (see Figure~\ref{scalability} (d)--(f)).


To compute Wasserstein distance, we use the software package \textsc{Hera}\footnote{\textsc{Hera} is publicly available at \url{https://bitbucket.org/grey_narn/hera}.}. \textsc{Hera} currently provides the fastest algorithm for computing Wasserstein distances. 


\subsection{Dimensionality}\label{dimensionality}

We determine the approximate embedding dimension $P$ of a point cloud by finding the smallest dimension such that we lose less than $5\%$ of the variance when projecting to that dimension using principal component analysis (PCA) \cite{Sorzano2014}. That is, for each $p \in \{1, 2, \dots \}$, we project the point cloud $\{x^{(i)} \in \mathbb{R}^N\}_{i \in V}$ to $\mathbb{R}^p$ using PCA, resulting in a point cloud $\{\hat{x}^{(i)}_p  \in \mathbb{R}^p\}_{i \in V}$. 

We estimate the extent to which this projection preserves the original point cloud by calculating the residual variance \cite{Tenenbaum2000,Cox2010}
\begin{equation*}
	R_p = 1 - \left(\rho^{(p)}\right)^2 \, ,
\end{equation*}
where $\rho^{(p)}$ is the Pearson correlation coefficient between the pairwise Euclidean distances of points in $\{x^{(i)}  \in \mathbb{R}^N\}_{i \in V}$ and corresponding pairwise Euclidean distances between points in $\{\hat{x}^{(i)}_p  \in \mathbb{R}^p\}_{i \in V}$ (see section~\ref{geometry}). 
The approximate embedding dimension $P$ is the smallest dimension for which the residual variance is less than 5\%; that is, $P=\min \{p \ | \ R_p < 0.05 \}$.

In practice, we put a cap of $100$ on $P$, so if the approximate embedding dimension is $100$ or larger, we record it to be $100$.  
Because we consider the torus to be embedded in $\mathbb{R}^4$, an approximate embedding dimension of $P = 4$ indicates that the contagion map recovers the dimensionality of the torus. 

\pagebreak
\section{Numerical Experiments}\label{results}

\subsection{Experiments in $(d^{\rm{NG}},T)$ parameter space}\label{num}

We construct Kleinberg-like small-world networks (as detailed in section~\ref{our_net}) for the following parameter values: 
$N=2500$ nodes (i.e., $n=50$), geometric degrees of $d^{\rm{G}}=4,8,12$ (corresponding to $p=1,\sqrt{2},2$), nongeometric degrees of $d^{\rm{NG}}=0,1,2,\dots, 25$, and distance decay parameter $\gamma=0,0.1,0.2,0.3,\dots,3$. For each of these $3 \times 26 \times 31=2418$ networks and for each threshold value $T=0,0.01,0.02,0.03,\dots,1$, we examine a WTM contagion (see section~\ref{contagion}) with cluster seeding around each of its $2500$ nodes and record the activation times of each node. Using the activation times of each node in each of these realizations as coordinates, we map the nodes of each network to a point cloud in $\mathbb{R}^{2500}$ via the symmetric contagion map (see section~\ref{contagion_map}). 

\begin{figure}[H]

     \leftline{\hskip 0.0cm (a) \hskip 3.8cm (b)} 
    \begin{minipage}{.33\textwidth}
        \centering
        \includegraphics[width=1.00\textwidth]{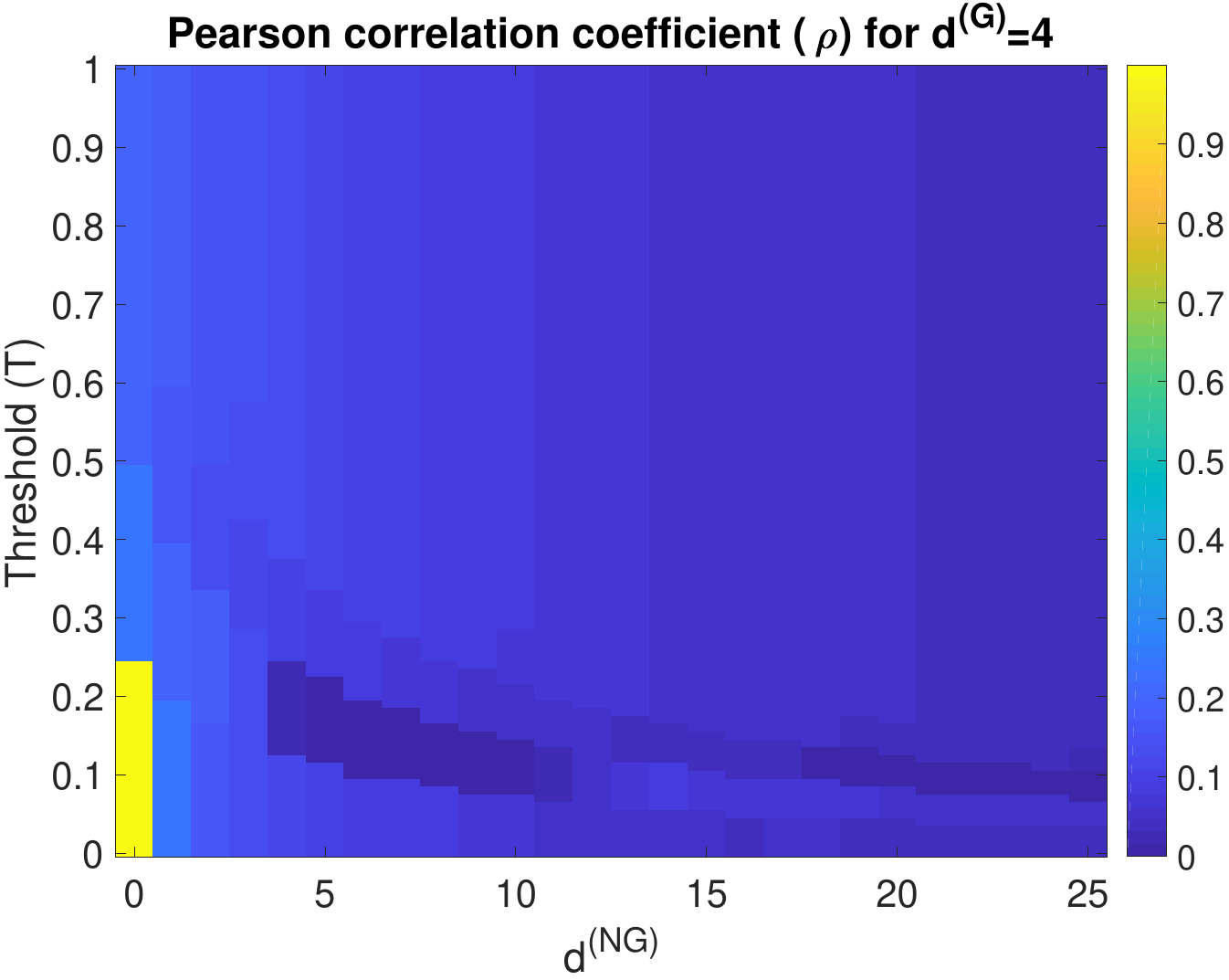} 
    \end{minipage}
    \begin{minipage}{0.33\textwidth}
        \centering
        \includegraphics[width=1.00\textwidth]{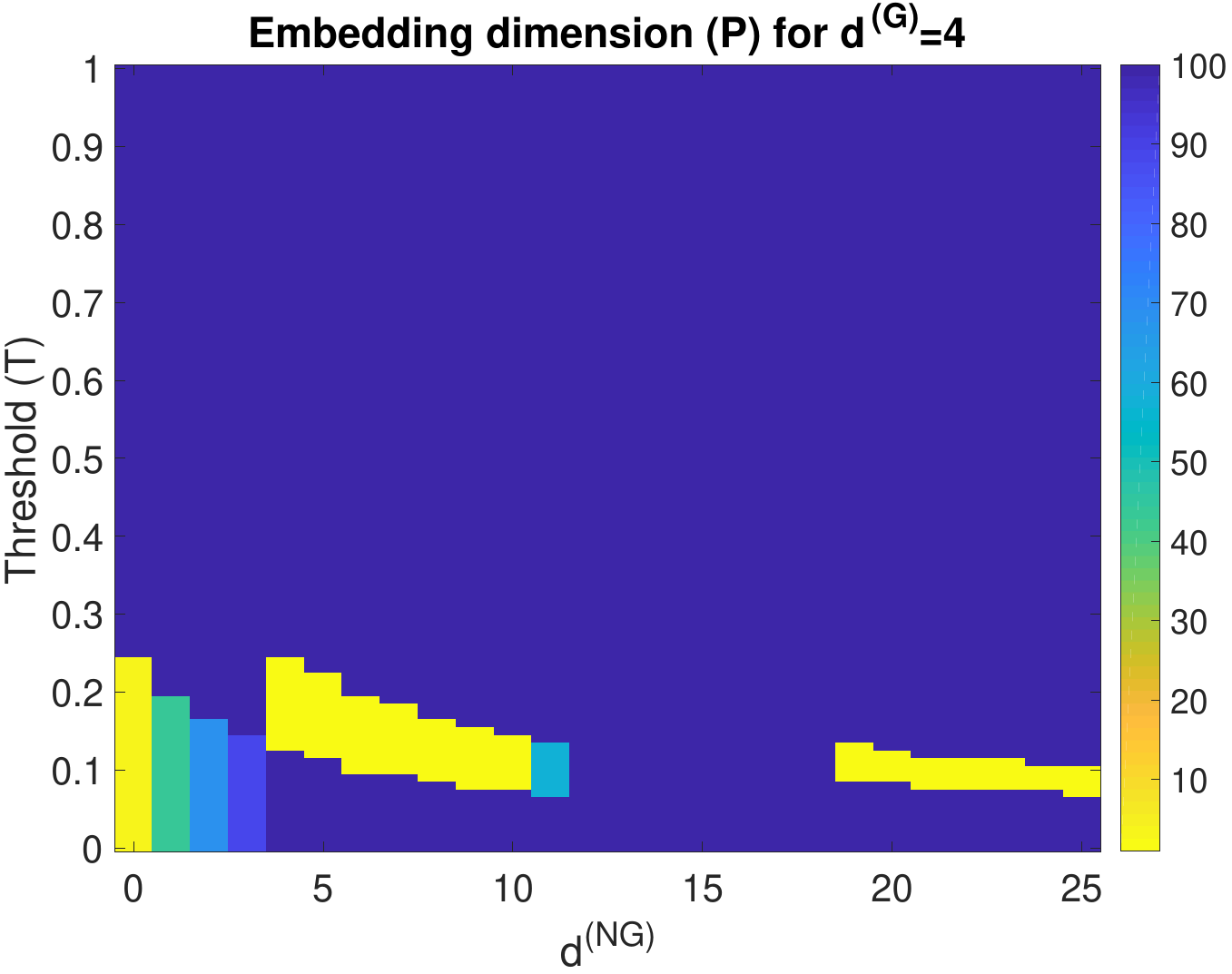} 
    \end{minipage}\hfill
    
 \centering
     \leftline{\hskip 0.1cm (c) \hskip 3.8cm (d) \hskip 3.8cm (e)} 
    \begin{minipage}{0.33\textwidth}
        \centering
        \includegraphics[width=1.00\textwidth]{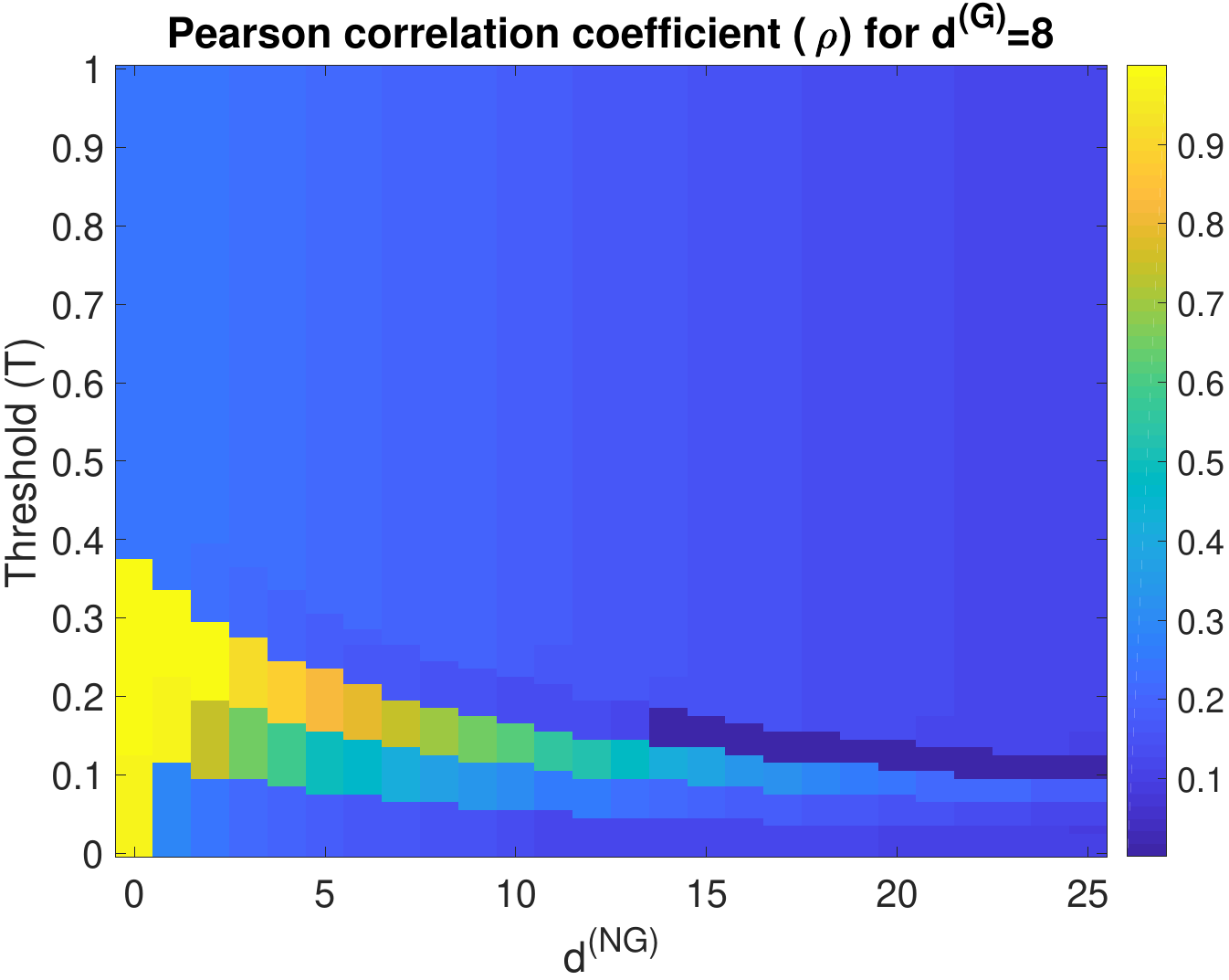}
    \end{minipage}\hfill
    \begin{minipage}{0.33\textwidth}
        \centering
        \includegraphics[width=1.00\textwidth]{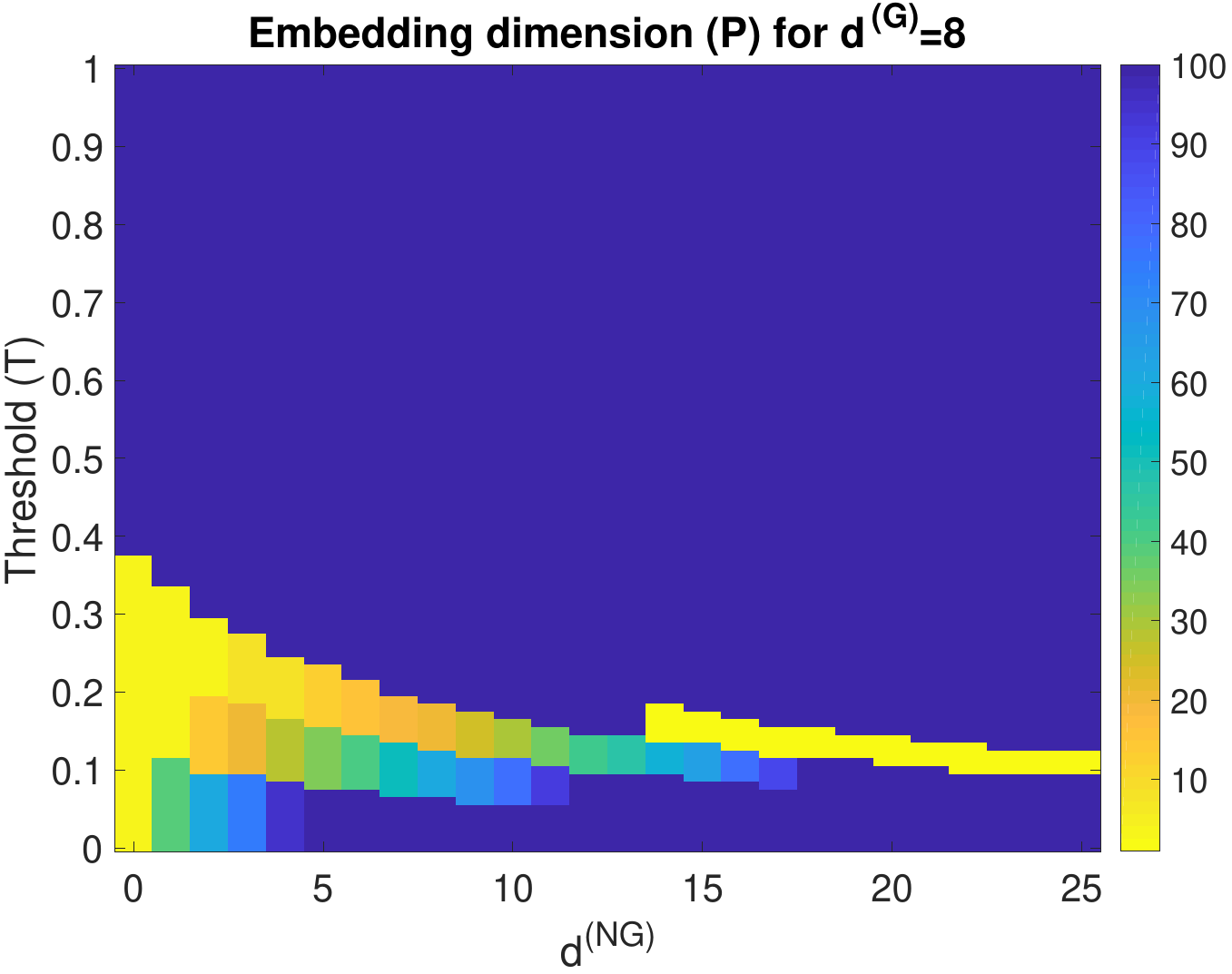} 
    \end{minipage}\hfill    
    \begin{minipage}{0.33\textwidth}
        \centering
    \includegraphics[width=1.00\textwidth]{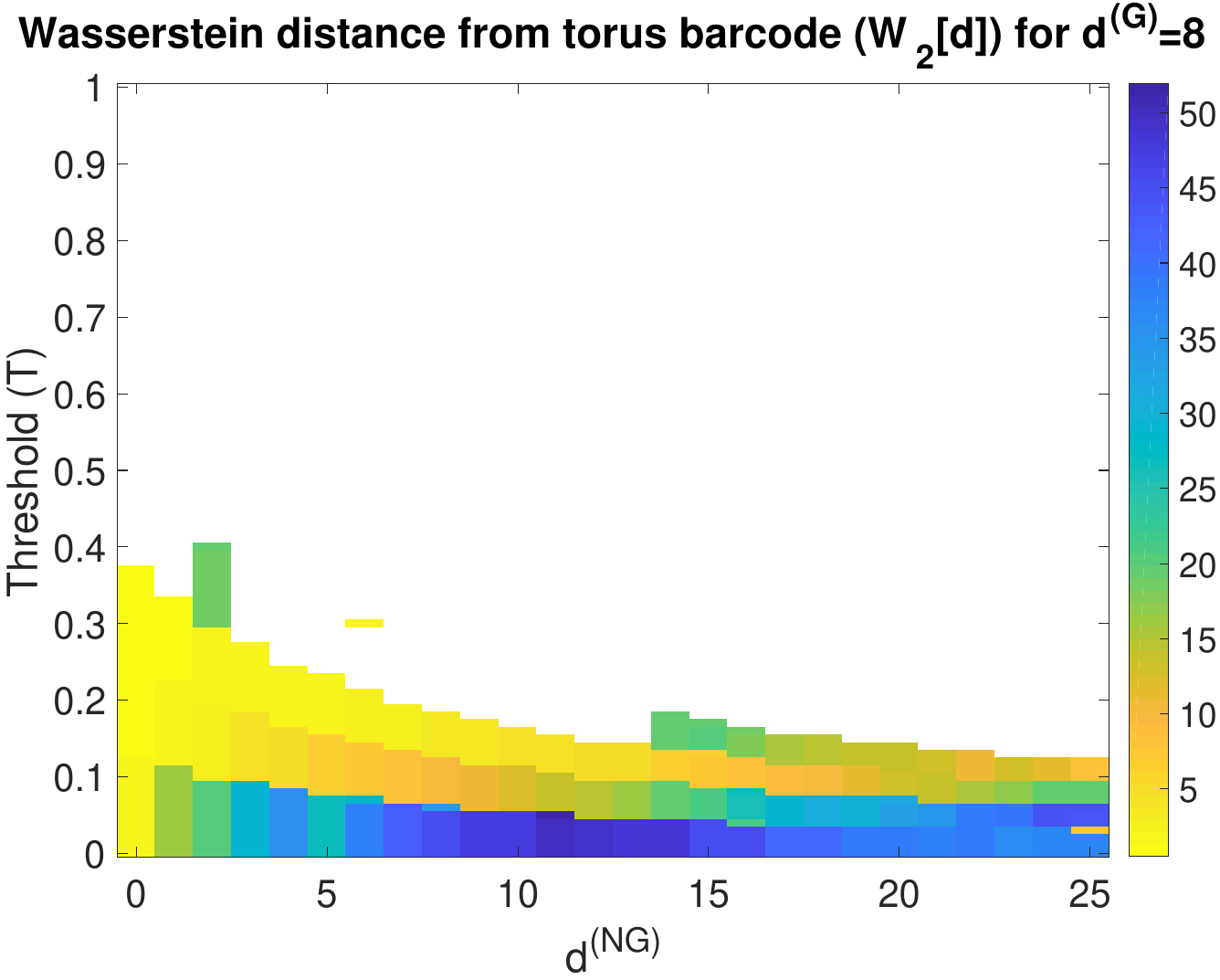}
    \end{minipage}\hfill
   
    \centering
     \leftline{\hskip 0.0cm (f) \hskip 3.8cm (g)} 
    \begin{minipage}{.33\textwidth}
        \centering
        \includegraphics[width=1.00\textwidth]{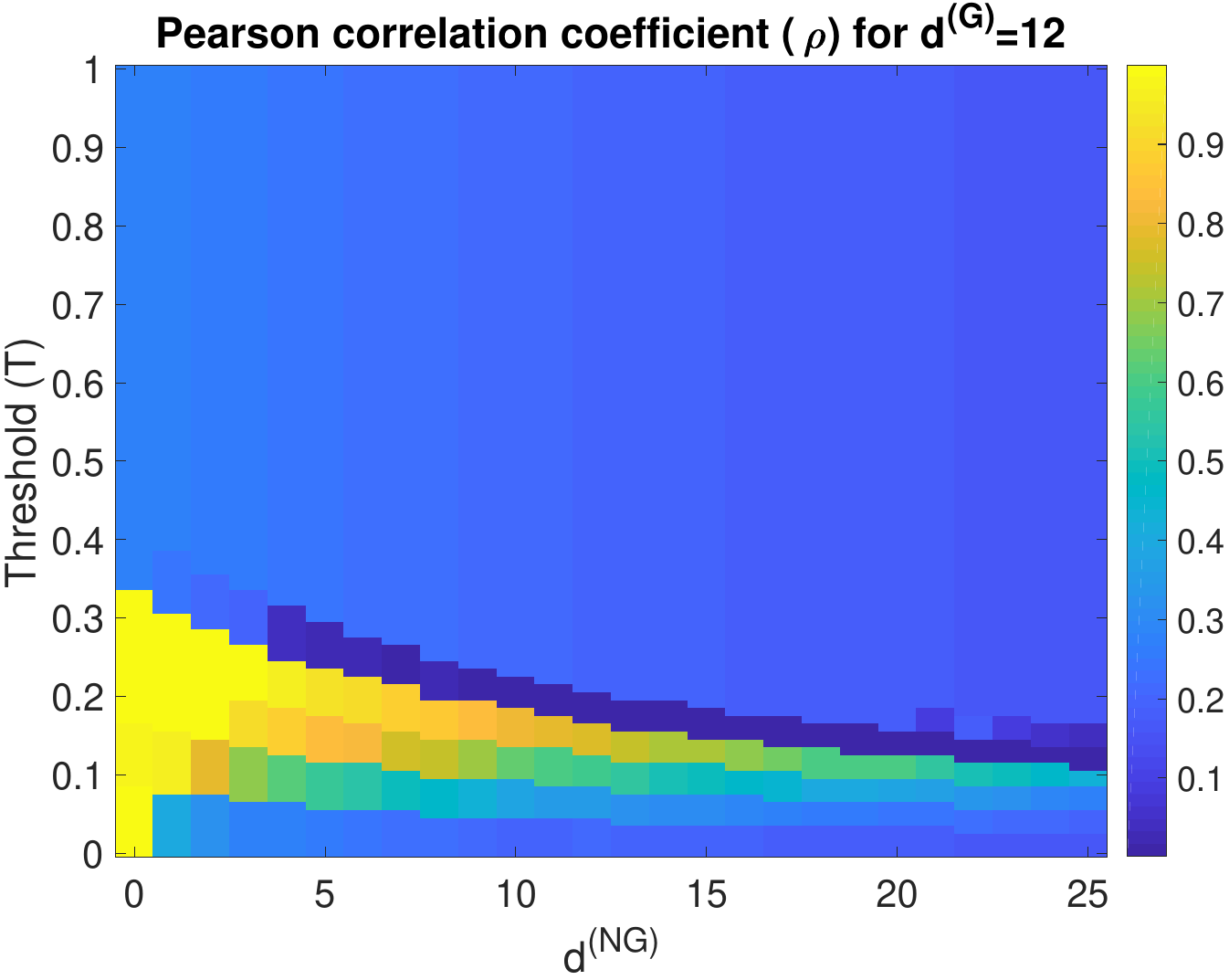} 
    \end{minipage}
    \begin{minipage}{0.33\textwidth}
        \centering
        \includegraphics[width=1.00\textwidth]{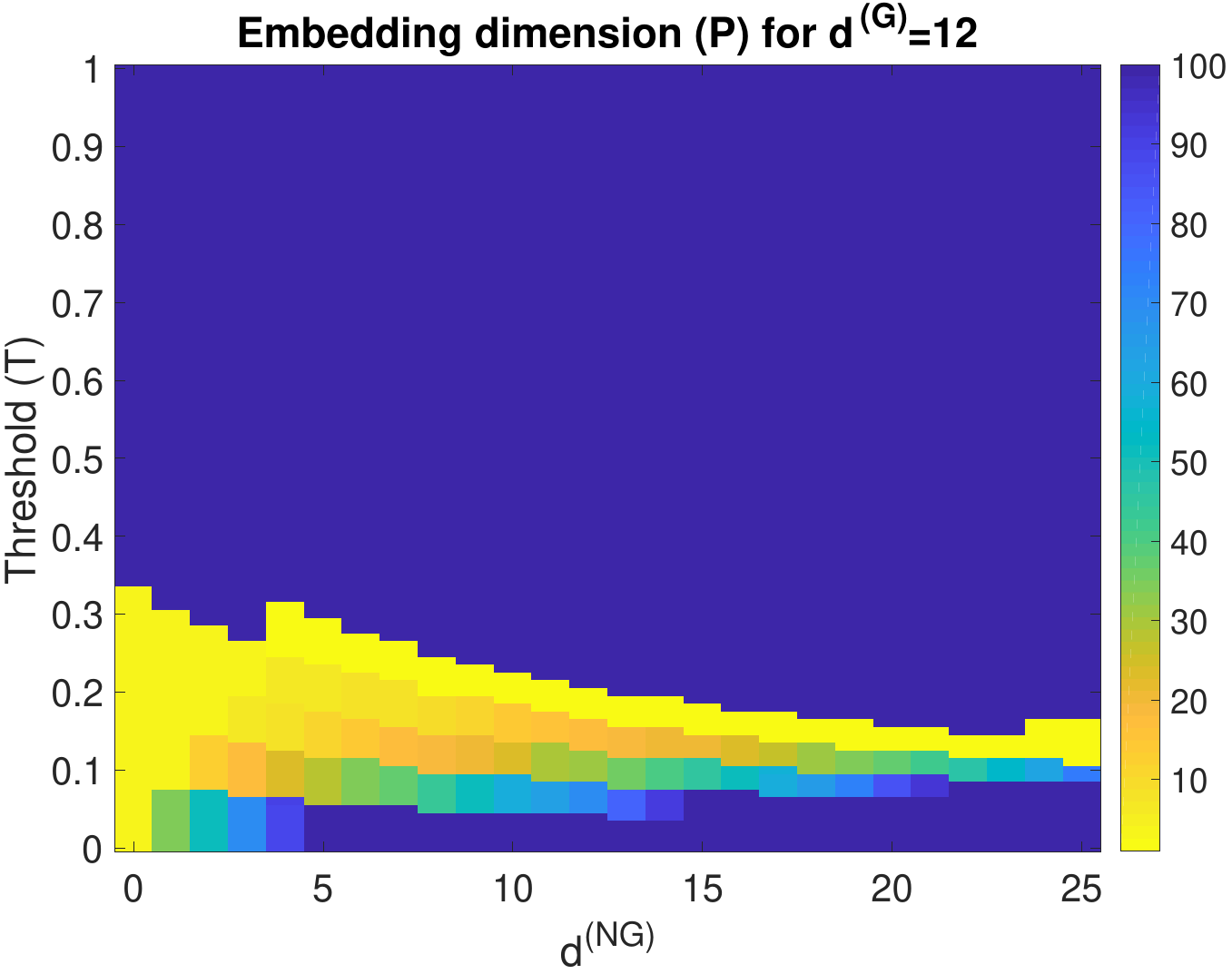} 
    \end{minipage}\hfill
    \caption{(a) Geometry (as quantified by the Pearson correlation coefficient) and (b) dimensionality (as determined by the approximate embedding dimension, which we cap at $100$) of contagion maps from WTM contagion dynamics on Kleinberg-like small-world networks with parameter values $N=2500$, $d^{\rm{G}}=4$, $\gamma=0$, and $d^{\rm{NG}}=0,1,2,\dots,25$ and contagion thresholds of $T=0,0.01,0.02,0.03,\dots,1$.
    (c) 
    Pearson correlation coefficient and (d) approximate embedding dimension (which we cap at $100$)
    of contagion maps from WTM contagion dynamics on Kleinberg-like small-world networks with parameter values $N=2500$, $d^{\rm{G}}=8$, $\gamma=0$, and $d^{\rm{NG}}=0,1,2,\dots,25$ and contagion thresholds of $T=0,0.01,0.02,0.03,\dots,1$. (e) Wasserstein distances between the scaled barcode of the PH of the VR filtrations on the regularly spaced point cloud on the torus (see formula~\eqref{reg_points}) and the scaled barcodes of the PH of the VR filtration on the contagion maps from WTM contagion dynamics on Kleinberg-like small-world networks with parameter values $N=2500$, $d^{\rm{G}}=12$, $\gamma=0$, and $d^{\rm{NG}}=0,1,2,\dots,25$ and contagion thresholds of $T=0,0.01,0.02,0.03,\dots,1$. The white regions correspond to parameter combinations for which there are nodes that do not activate (i.e., they have `infinite' activation times) in some realizations of the contagion.
    (f) Pearson correlation coefficient and (g) approximate embedding dimension (which we cap at $100$) of contagion maps from WTM contagion dynamics on Kleinberg-like small-world networks with parameter values $N=2500$, $d^{\rm{G}}=8$, $\gamma=0$, and $d^{\rm{NG}}=0,1,2,\dots,25$ and contagion thresholds of $T=0,0.01,0.02,0.03,\dots,1$.
    }
    \label{numerics_combined}
\end{figure}

We compute quantitative measures of the similarity of these point clouds to the underlying torus in terms of geometry (see section~\ref{geometry}), topology (see section~\ref{topology}), and dimensionality (see section~\ref{dimensionality}) when we place nongeometric edges uniformly at random (i.e., when $\gamma=0$). We illustrate our results by separately displaying the values of the Pearson correlation coefficient $\rho$, the Wasserstein distance $W_2[d]$, and the embedding dimension $P$ in $(d^{\rm{NG}},T)$ parameter space for each value of $d^{\rm{G}}$ (see Figure~\ref{numerics_combined}).
 When examining topological similarity, we only cover the case $d^{\rm{G}}=8$, as computing PH of the VR filtration on a point cloud of 2500 points is extremely time-consuming because of the large number of simplices involved. Brighter regions in our plots signify larger Pearson correlation coefficients $\rho$ (in the geometry computation), smaller Wasserstein distances $W_2[d]$ (in the topology computation), and lower approximate embedding dimensions $P$. In each plot, we can identify a region in the parameter space for which $\rho$ is large, and $W_2[d]$ and $P$ are small, indicating that WFP dominates for these parameter values. 

The first column of each plot in Figure~\ref{numerics_combined} (e.g., see the yellow bar in panel (a)) shows our results for $d^{\rm{NG}}=0$, which corresponds to a purely geometric network. In this case, network formation is deterministic and we can analytically determine the WTM dynamics (in particular, the presence versus absence of WFP). (See section~\ref{bifurcation} for details.) We see in the first column of each plot that $\rho$, $W_2[d]$, and $P$ take only extreme values for $d^{\rm{NG}}=0$ and that the transition between extreme values occurs at the same threshold $T$ for all three quantities. Below this threshold, the Pearson correlation coefficients are large and the Wasserstein distances and approximate embedding dimensions are small ($P = 4$, to be precise). Above this threshold, the Pearson correlation coefficients are small and $P$ is large (at the cap of $100$), and these values of $T$ yield `infinite activation times' of nodes in the plot for the Wasserstein distance. The observations described in this paragraph are consistent with our analytical considerations, which demonstrate that spreading (by WFP) can occur only below this threshold.

There is a band along the transition between the region in which we expect WFP (see Figures~\ref{WFP_4}--\ref{WFP_12}) and the region in which we do not. This band is dark in the plots of the geometric and the topological structure, and it is bright in the plot of dimensionality. This implies that the point cloud is low-dimensional for the corresponding parameter combinations, but that it does not exhibit torus-like structure in terms of geometry or topology. Although this was not discussed in \cite{Taylor2015}, one can also observe such a band for WTM contagions on the noisy ring lattices in that study. 

Some irregularities and outliers in our figures are likely due to the probabilistic nature of nongeometric edges in our network construction. One example is that of the nonwhite spots in the white region of Figure~\ref{numerics_combined}(e). These correspond to parameter combinations for which our bifurcation analysis (see section~\ref{bifurcation}) suggests that we should expect infinite activation times, but all nodes have finite activation times in practice in all realizations.

\subsection{Effect of the distance-decay parameter $\gamma$ on contagion maps}

In our Kleinberg-like small-world networks (see section~\ref{our_net}), recall that we regulate the range of nongeometric edges using the decay parameter $\gamma \in \mathbb{R}_{\geq 0}$. Each node has a fixed number of nongeometric stubs, and we connect two stubs that emanate from nodes $i$ and $j$ to form a nongeometric edge with a probability that is proportional to $\mu_{\rm per}(i,j)^{-\gamma}$. For $\gamma=0$, we match the nongeometric stubs uniformly at random, regardless of the distance between the corresponding nodes, so the length of the nongeometric edges can take any value with equal probability. For $\gamma>0$, nongeometric edges have a bias to connect nodes that are close to each other with respect to the periodic lattice distance. This bias becomes more pronounced for progressively larger $\gamma$, so larger values of $\gamma$ tend to yield shorter nongeometric edges.  

\begin{figure}[H]
    \centering
    \leftline{\hskip 1mm (a) \hskip 2.7cm (b) \hskip 2.7cm (c) \hskip 2.5cm (d)} 
    \begin{minipage}{.25\textwidth}
        \centering
        \includegraphics[width=1.00\textwidth]{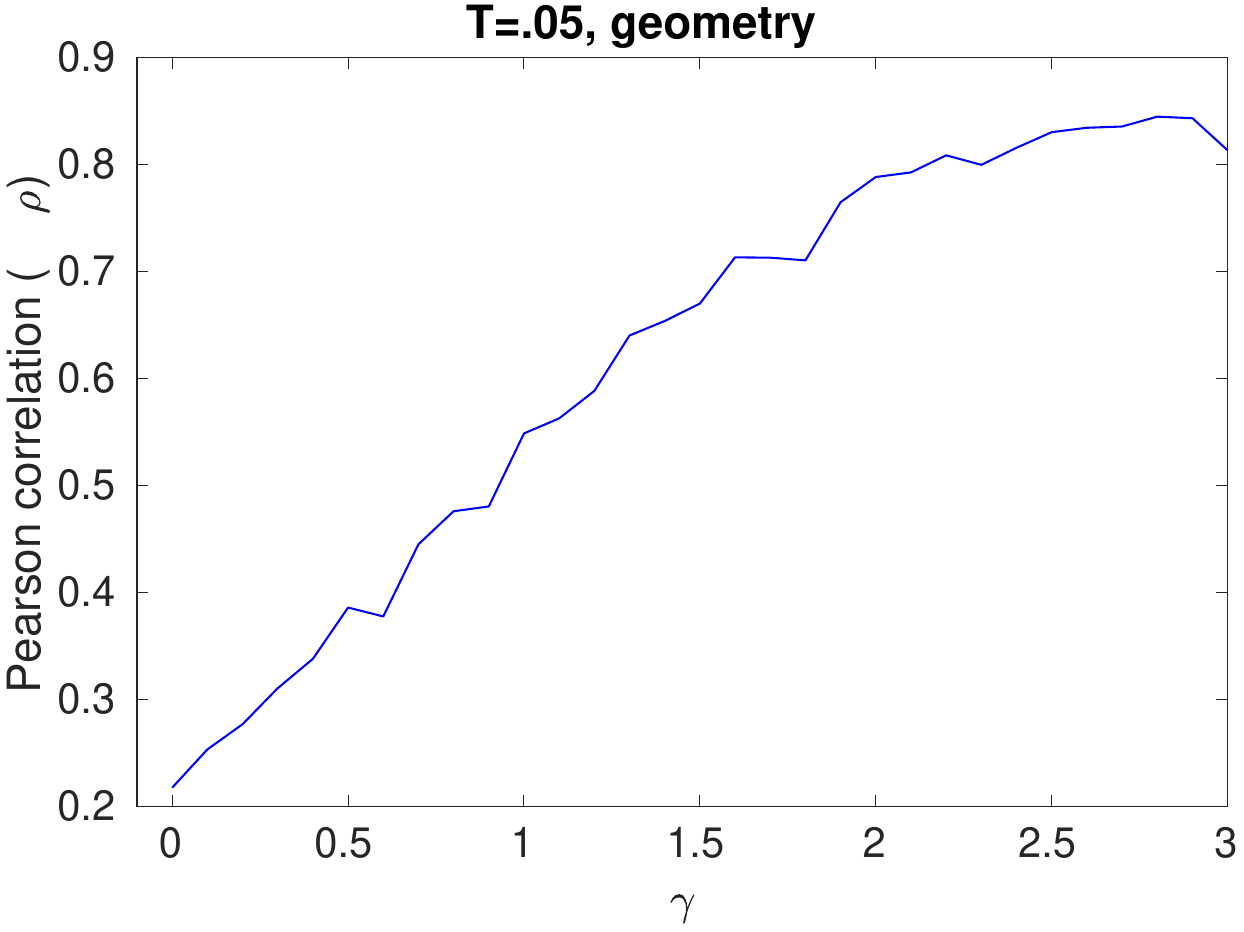} 
    \end{minipage}\hfill
    \begin{minipage}{0.25\textwidth}
        \centering
        \includegraphics[width=1.00\textwidth]{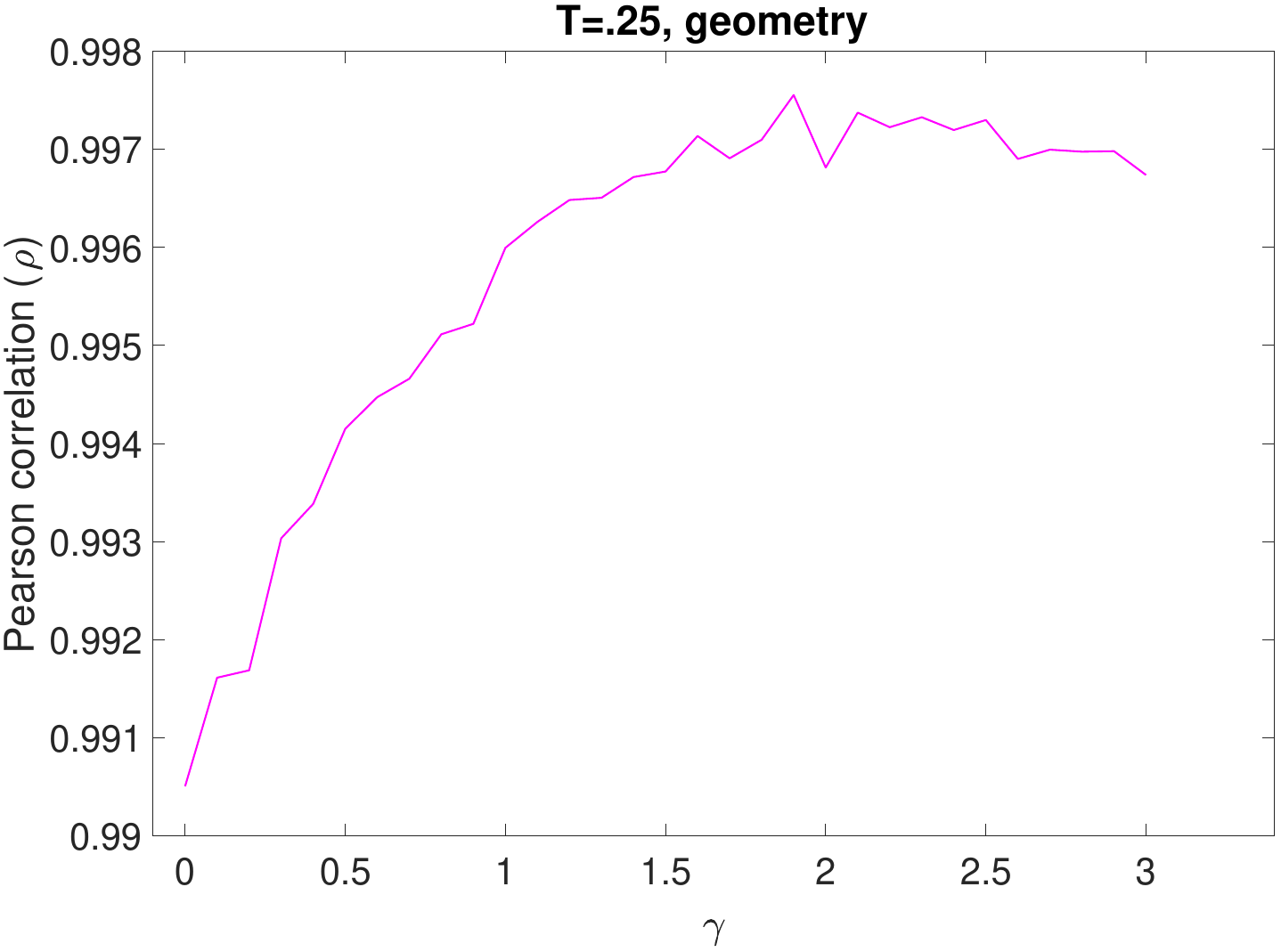} 
    \end{minipage}\hfill   
    \begin{minipage}{.25\textwidth}
    \centering
        \includegraphics[width=1.00\textwidth]{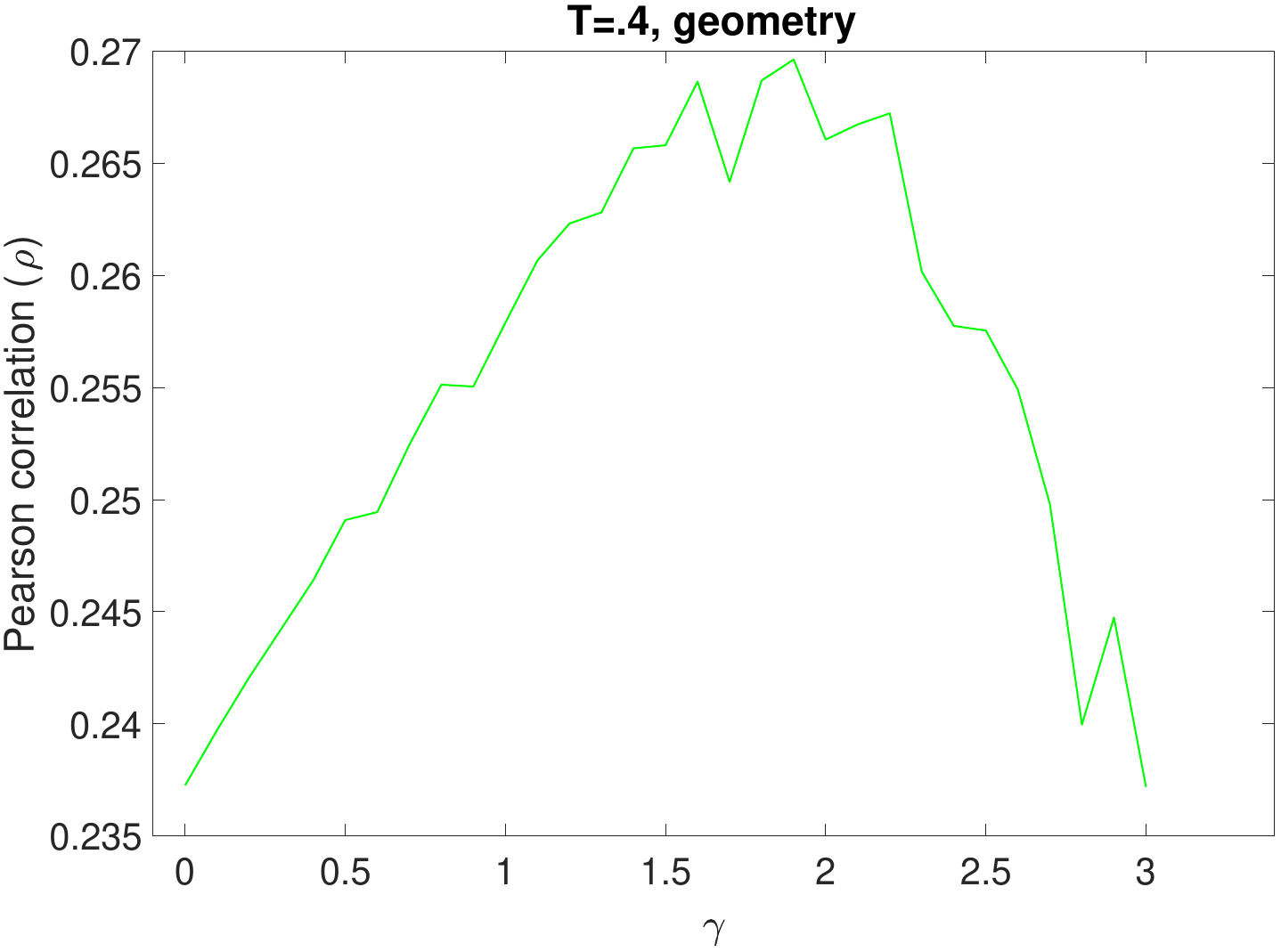}
    \end{minipage}\hfill
    \begin{minipage}{.25\textwidth}
    \centering
        \includegraphics[width=1.00\textwidth]{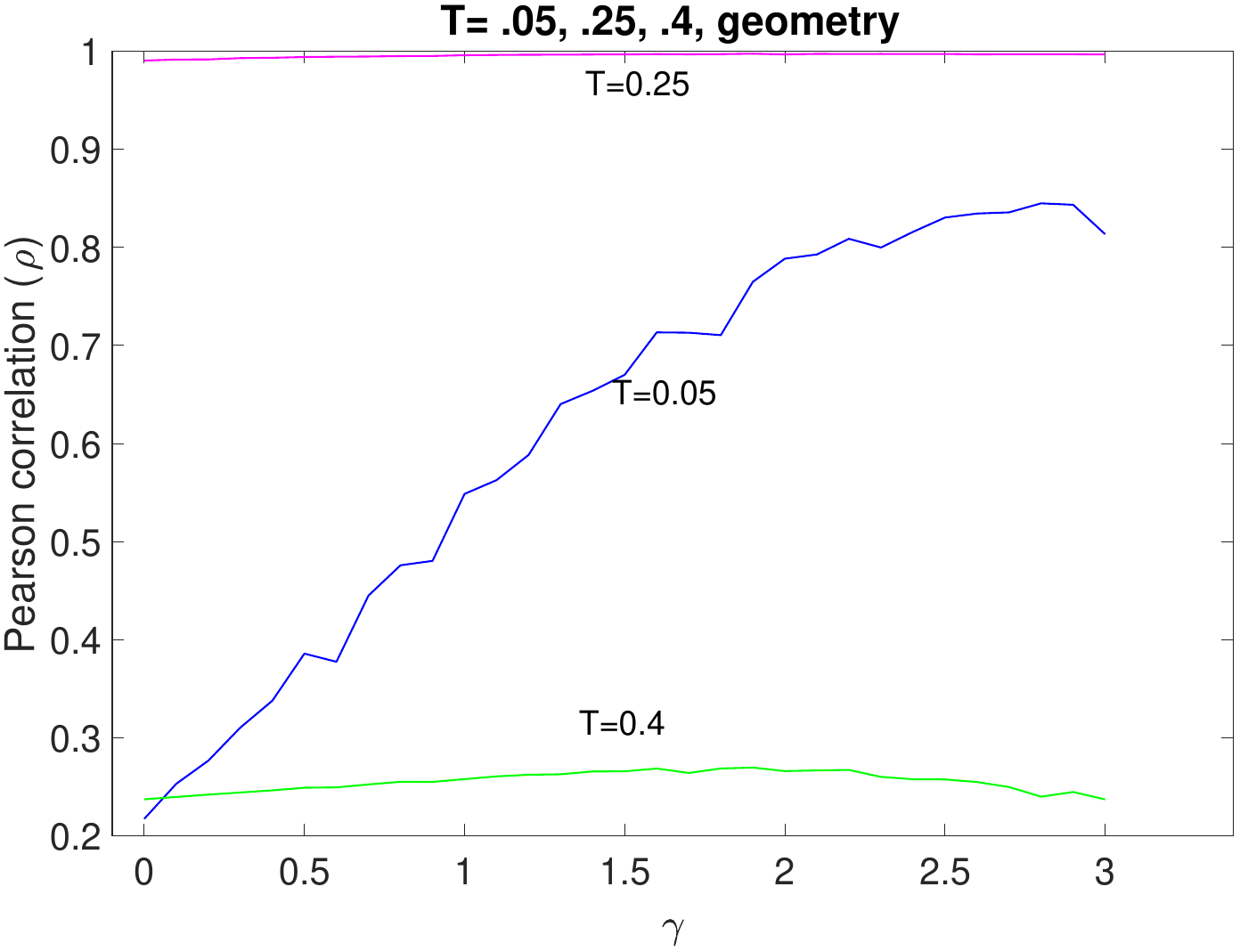}
    \end{minipage}\hfill
    \caption{Pearson correlation coefficient between point-cloud distances and node--node distances as we increase $\gamma$ from $0$ to $3$ in increments of $0.1$ for Kleinberg-like small-world networks with parameters $N=2500$, $d^{\rm{G}}=8$, and $d^{\rm{NG}}=2$ and contagion thresholds of (a) $T=0.05$, (b) $T=0.25$, and (c) $T=0.4$. In panel (d), we show the plots for all three values of $T$.
    }   
    \label{gamma_geometry}
\end{figure}

\begin{figure}[H]
    \centering
   \leftline{\hskip 1mm (a) \hskip 2.7cm (b) \hskip 2.7cm (c) \hskip 2.7cm (d)} 
    \begin{minipage}{.25\textwidth}
        \centering
        \includegraphics[width=1.00\textwidth]{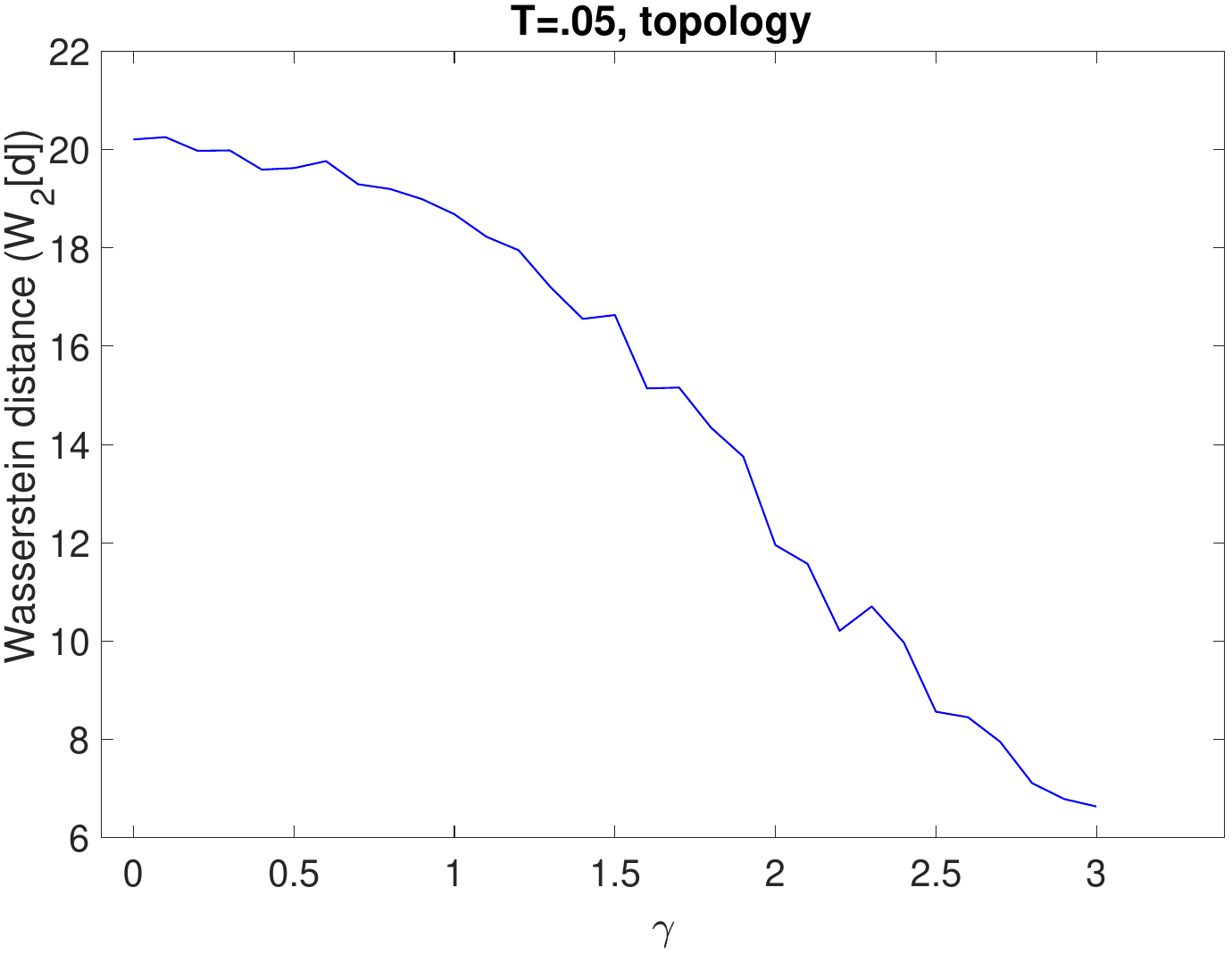}
    \end{minipage}\hfill
    \begin{minipage}{0.25\textwidth}
        \centering
        \includegraphics[width=1.00\textwidth]{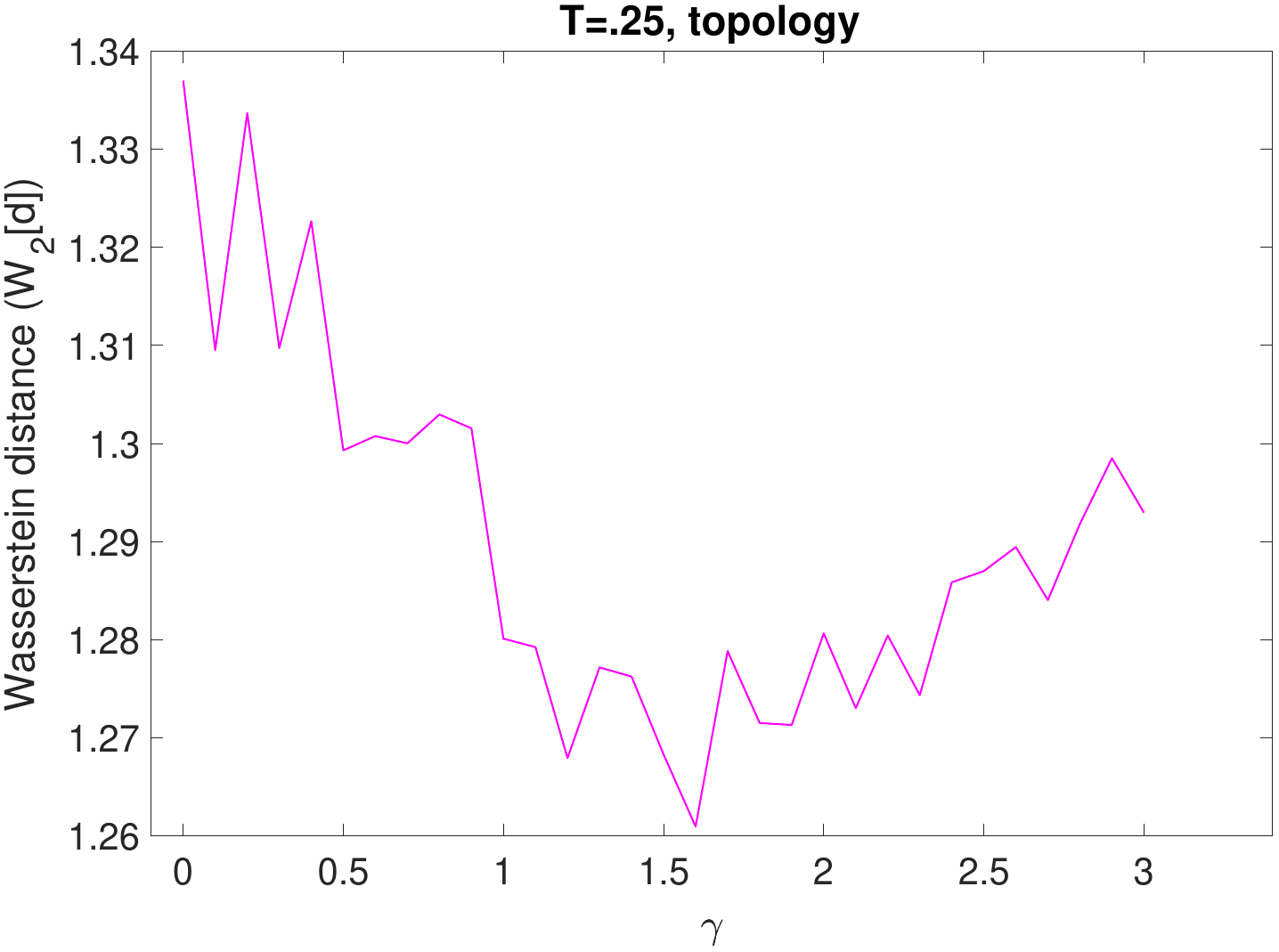}
    \end{minipage}\hfill
    \begin{minipage}{.25\textwidth}
    \centering
        \includegraphics[width=1.00\textwidth]{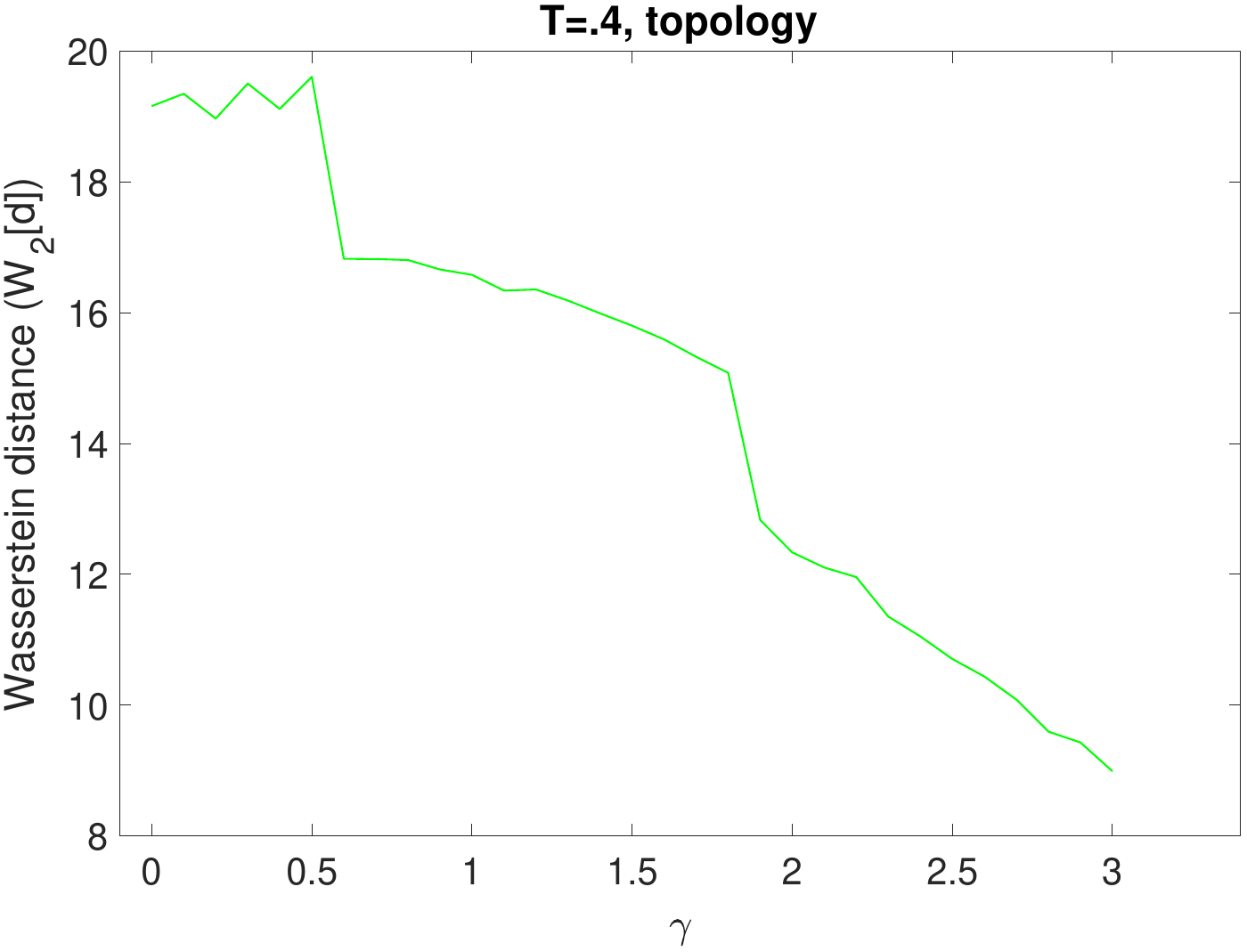}
    \end{minipage}\hfill
    \begin{minipage}{.25\textwidth}
    \centering
        \includegraphics[width=1.00\textwidth]{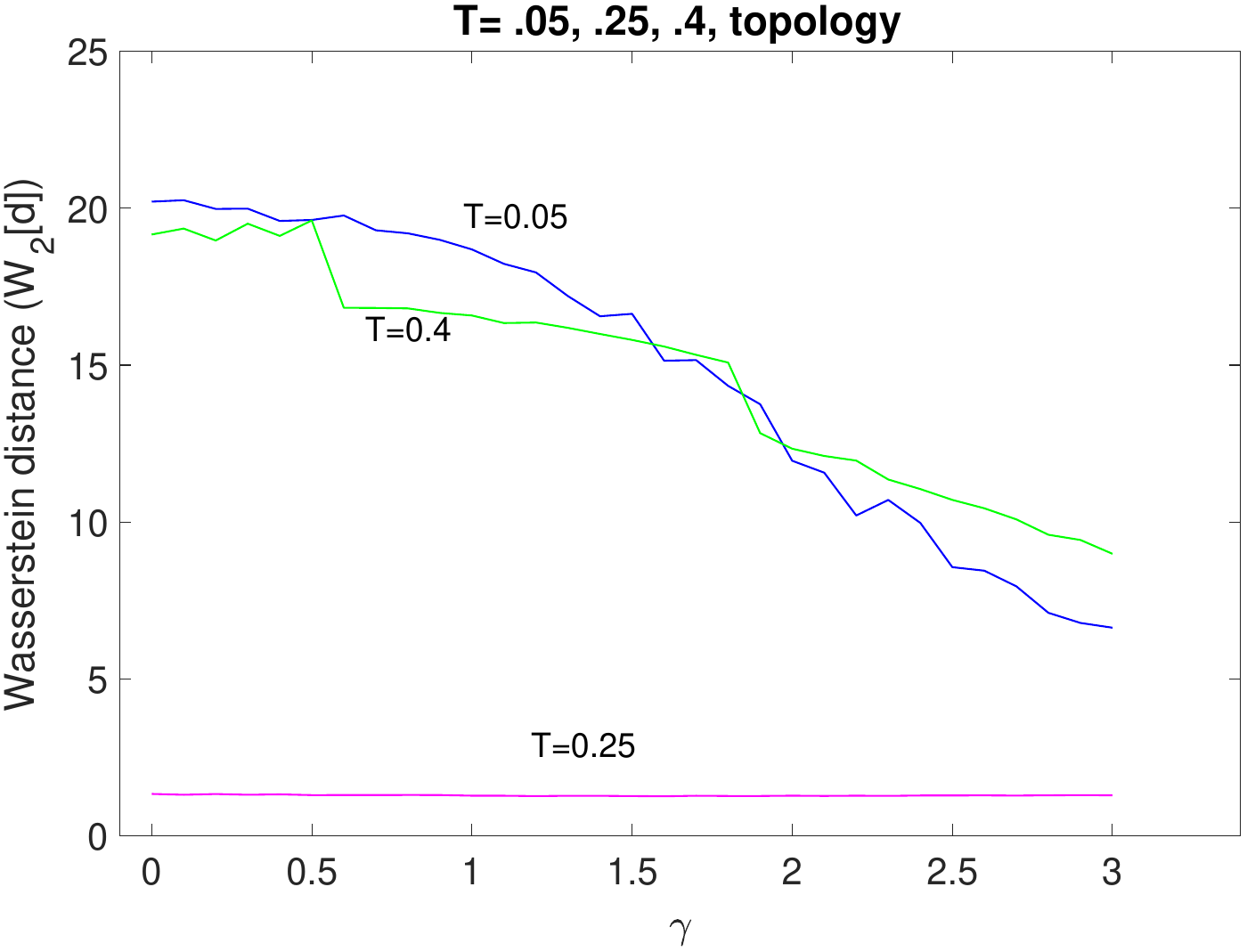}
    \end{minipage}\hfill
    \caption{Wasserstein distance between scaled barcodes as we increase $\gamma$ from $0$ to $3$ in increments of $0.1$ for Kleinberg-like small-world networks with parameters $N=2500$, $d^{\rm{G}}=8$, and $d^{\rm{NG}}=2$ and contagion thresholds of (a) $T=0.05$, (b) $T=0.25$, and (c) $T=0.4$. In panel (d), we show the plots for all three values of $T$.
    }
    \label{gamma_topology}    
\end{figure}

\begin{figure}[H]
    \centering
   \leftline{\hskip 1mm (a) \hskip 2.7cm (b) \hskip 2.7cm (c) \hskip 2.7cm (d)} 
    \begin{minipage}{.25\textwidth}
        \centering
        \includegraphics[width=1.00\textwidth]{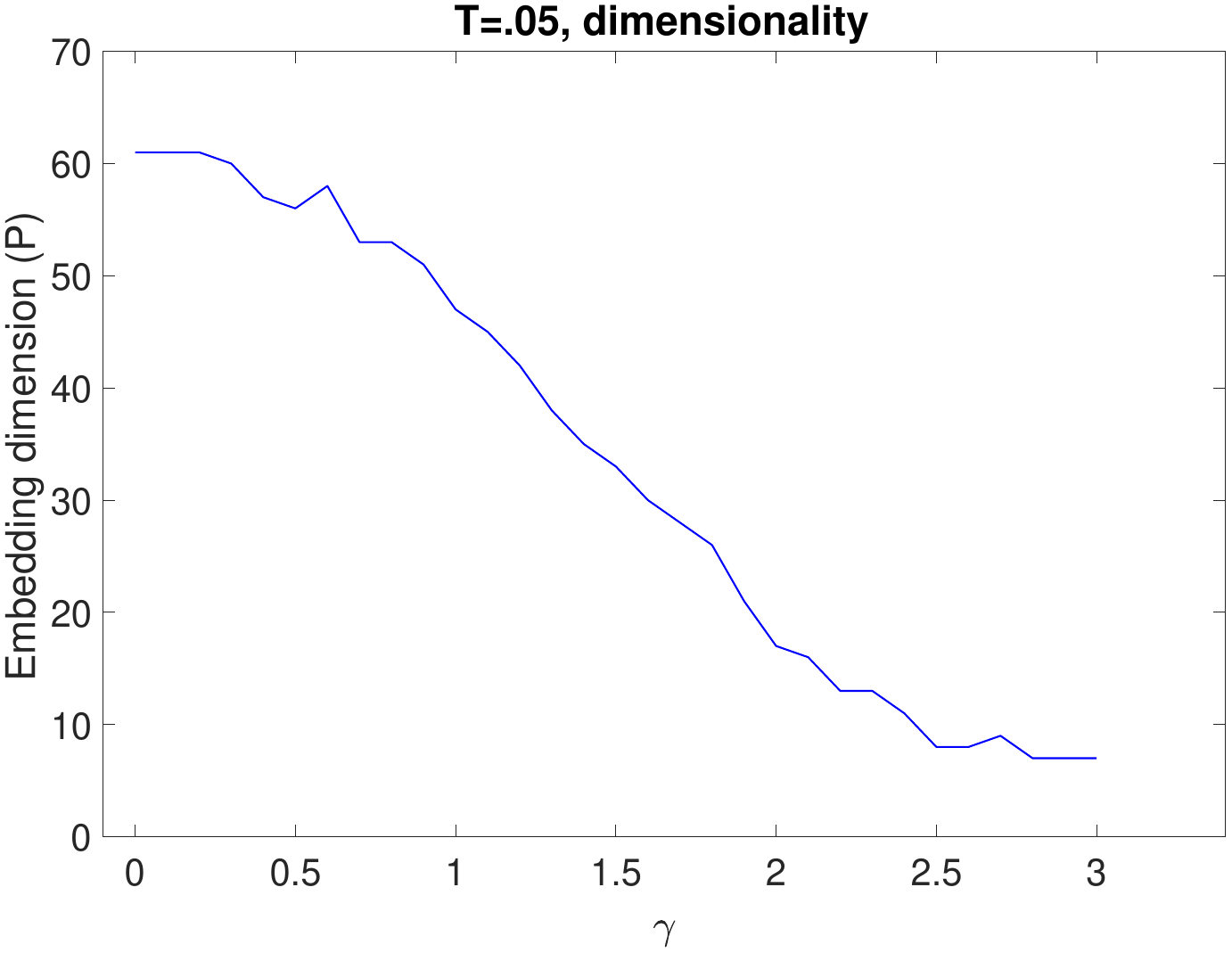}
    \end{minipage}\hfill
    \begin{minipage}{0.25\textwidth}
        \centering
        \includegraphics[width=1.00\textwidth]{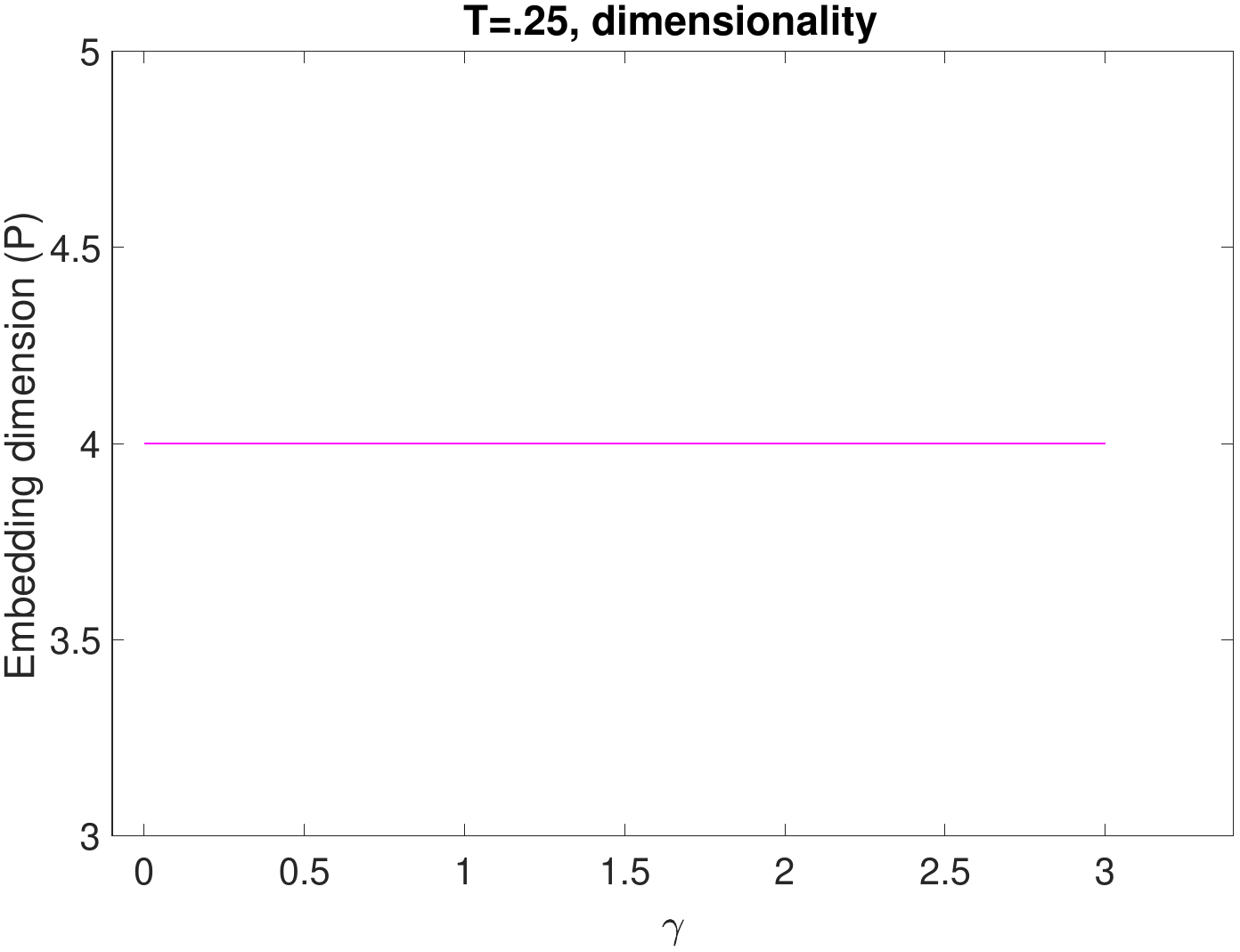}
    \end{minipage}\hfill
    \begin{minipage}{.25\textwidth}
    \centering
        \includegraphics[width=1.00\textwidth]{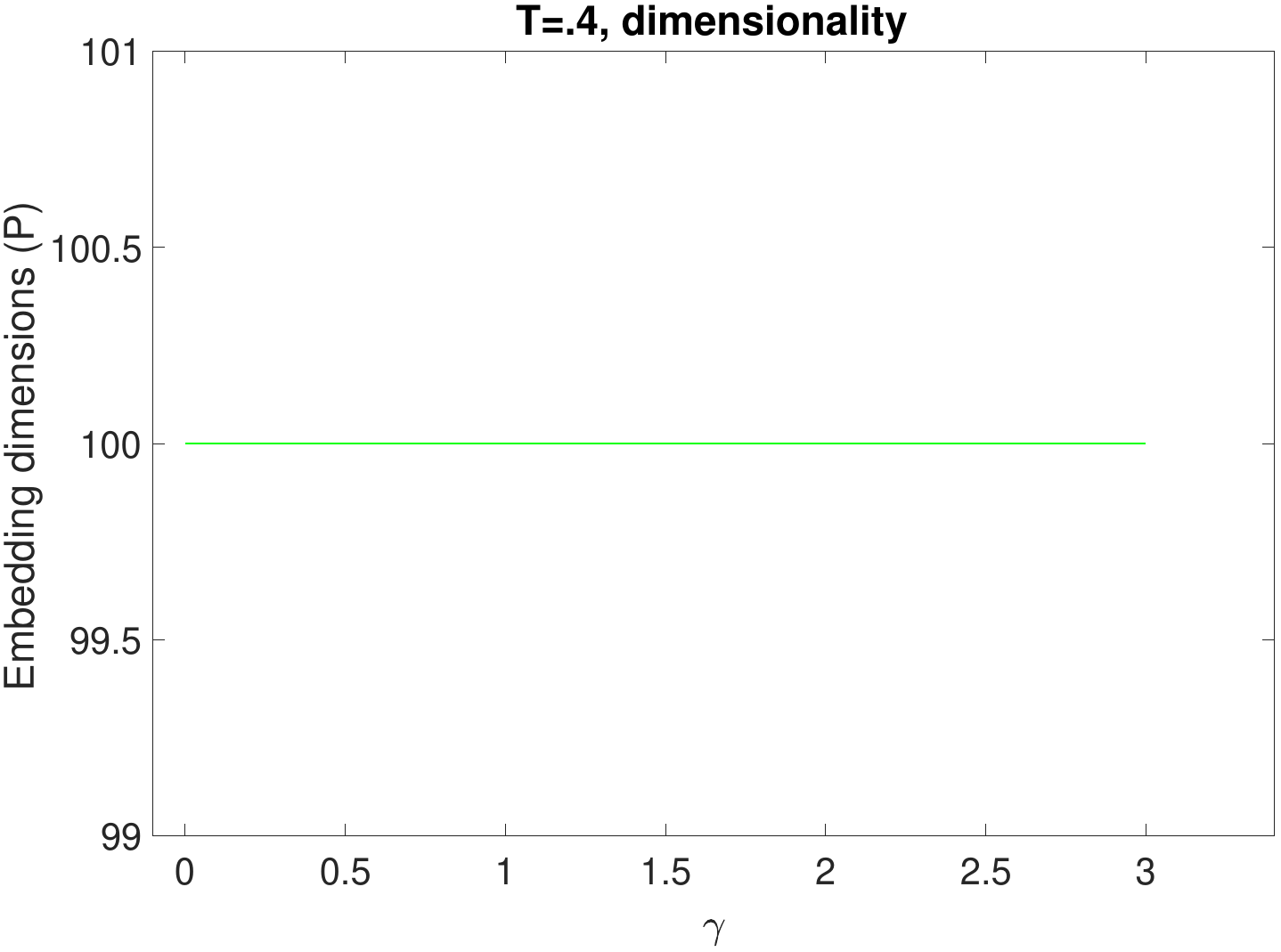}
    \end{minipage}\hfill
    \begin{minipage}{.25\textwidth}
    \centering
        \includegraphics[width=1.00\textwidth]{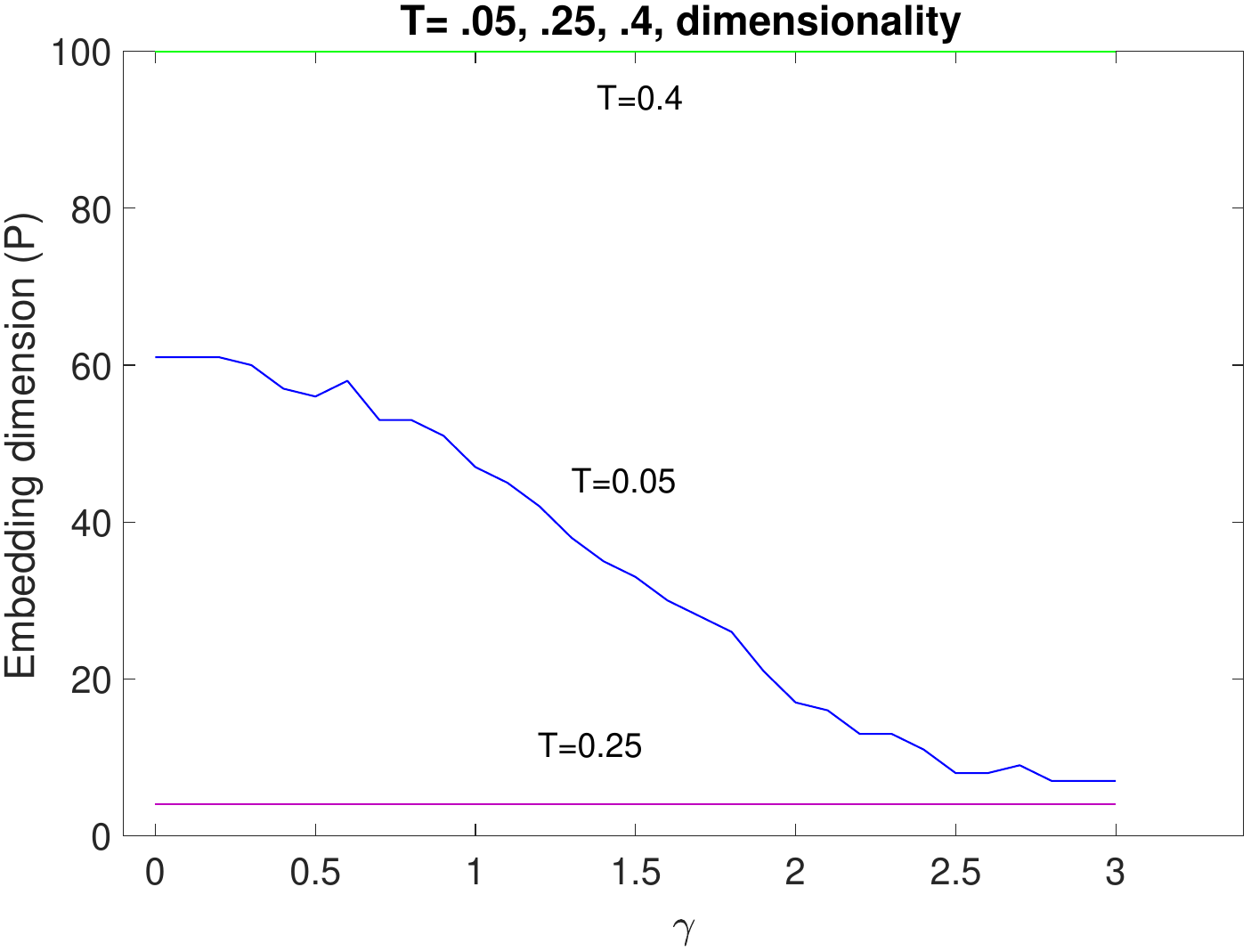}
    \end{minipage}\hfill
    \caption{Approximate embedding dimension as we increase $\gamma$ from $0$ to $3$ in increments of $0.1$ for Kleinberg-like small-world networks with parameters $N=2500$, $d^{\rm{G}}=8$, and $d^{\rm{NG}}=2$ and contagion thresholds of (a) $T=0.05$, (b) $T=0.25$, and (c) $T=0.4$. In panel (d), we show the plots for all three values of $T$.
    }
    \label{gamma_dimensionality}
\end{figure}

The speed of a WTM contagion on a Kleinberg-like network depends significantly on the parameter $\gamma$ \cite{Ghasemiesfeh2013,Ebrahimi2015}. We examine the effect of $\gamma$ on the shape of a contagion map. We show our results for geometry in Figure~\ref{gamma_geometry}, for topology in Figure~\ref{gamma_topology}, and for dimensionality in Figure~\ref{gamma_dimensionality}. 
For fixed values of the geometric degree $d^{\rm{G}}$ and nongeometric degree $d^{\rm{NG}}$ of our networks and threshold $T$ of our contagion, we vary the value of $\gamma$. Specifically, we choose a geometric degree of $d^{\rm{G}}=8$, a nongeometric degree of $d^{\rm{NG}}=2$, and one value for the contagion threshold $T$ for each predicted spreading regime when $\gamma=0$. We use the value $T= 0.05$ for the regime in which we expect both WFP and ANC when $\gamma=0$, the value $T=0.25$ for the regime in which we expect WFP but no ANC when $\gamma=0$, and the value $T=0.4$ for the regime in which we expect neither WFP nor ANC when $\gamma=0$. See Figure~\ref{full_bifurcation} to locate these values in $(d^{\rm{NG},T})$ parameter space and thereby identify their associated spreading regimes. For each of these three values for $T$, we let $\gamma$ vary from $0$ to $3$ in increments of $0.1$. For each value for $\gamma$, we map the nodes of the associated network via the contagion map using the given value of $T$ and analyze the resulting point cloud as described in section~\ref{methods}. 

For $T=0.05$, the Pearson correlation coefficient increases significantly in an almost linear fashion as we increase $\gamma$, while the Wasserstein distance and the approximate embedding dimension both decrease. This arises from the fact that nongeometric edges change in function from drivers of ANC to contributors to WFP. For $\gamma=0$, we expect fast spreading of the contagion that is dominated by ANC. This leads to a contagion map whose image is a cluster of tightly bunched points that are distributed fairly evenly in the region that they occupy. In particular, the pairwise distances between the points are not influenced much by the pairwise distances between their corresponding nodes. Such a cluster of points has a high approximate embedding dimension $P$, because its points are distributed with roughly constant density across the region that they occupy, so the point cloud does not have a lower intrinsic dimension than its ambient space. 
For progressively larger $\gamma$, the nongeometric edges tend to become shorter and contribute increasingly to WFP, instead of facilitating spreading across large distances in a network. They thereby produce a point cloud that is still contained in a small volume (because the spreading is still fast with such a low threshold), but with pairwise distances between points that become increasingly faithful to the pairwise distances between their corresponding nodes. 
 
 For $T=0.25$, the Pearson correlation coefficient starts out large and increases further for progressively larger $\gamma$. By contrast, the Wasserstein distance is small throughout the range of $\gamma$, with a slight decrease at the lower end of the range. The approximate embedding dimension is $P = 4$ for all values of $\gamma$ that we considered. This relative stability of all three measures stems from the fact that, for this value of $T$, WFP dominates over ANC even when $\gamma=0$, as the nongeometric edges are not sufficiently numerous to drive ANC. For progressively larger $\gamma$, the gradually shortening nongeometric edges only contribute increasingly to WFP. 

For $T=0.4$, the Pearson correlation coefficient and the approximate embedding dimension remain fairly small and fairly large, respectively, for all values of $\gamma$. The Wasserstein distance decreases steadily as we increase $\gamma$.

\section{Bifurcation analysis}\label{bifurcation}

We conduct a bifurcation analysis of the spreading behavior of the WTM contagion (see section~\ref{contagion} for its definition), which we initialize with cluster seeding on the family of Kleinberg-like small-world networks that we described in section~\ref{our_net}. The results of this bifurcation analysis give a guideline for interpreting our prior numerical computations. We want to determine analytically which combinations of network parameter values, $d^{\rm{G}}$ and $d^{\rm{NG}}$ (see section~\ref{our_net}), and threshold parameter value $T$ (see section~\ref{contagion}) allow the contagion to spread by WFP and which allow it to spread by ANC. That is, we want to identify regions in parameter space for which the spreading dynamics follow specific regimes that are characterized by the presence and absence of WFP and ANC. We are especially interested in the region of parameter space for which there is WFP but no ANC, as this region should comprise the parameter combinations for which the contagion map exhibits structural features of a torus. 

We consider $N=2500$ exclusively, as --- at least locally and sufficiently early in the contagion process --- the total size of a network should not affect contagion behavior. At later stages, a contagion that saturates a network will speed up earlier for smaller networks, as the active region of the network is now proportionately larger with respect to the total network size. Additionally, we restrict our analysis to the case $\gamma=0$ (i.e., when we place nongeometric edges uniformly at random). 

We fix the geometric degree $d^{\rm{G}}$ and examine the spreading behavior as we vary the nongeometric degree $d^{\rm{NG}}$ and the threshold $T$. The possible values for $d^{\rm{G}}$ are constrained by the number of nodes that are within a distance $p\in \mathbb{R}_{>0}$ of a given node (see section~\ref{our_net}). For a given $p$, the corresponding $d^{\rm{G}}$ is $1$ less than the number of integer lattice points that lie inside a circle of radius $p$ that is centered at the origin. This number is approximately equal to the area of the circle, and the problem of determining it is known as ``Gauss' Circle Problem'' \cite{Hardy1915}. The three smallest values of nongeometric degree $d^{\rm{G}}$ are $4$, $8$, and $12$, which correspond to $1 \leq p < \sqrt{2}$, $\sqrt{2} \leq p < 2$, and $2 \leq p < \sqrt{5}$, respectively, in the definition of our Kleinberg-like small-world network.


\subsection{Wavefront Propagation (WFP)}

We consider $d^{\rm{G}}=4,8,12$ individually, and we work out the maximum threshold for which a Kleinberg-like network with $\gamma = 0$ can support sustained spreading via only geometric edges. 

\begin{figure}[H]
    \centering
    \leftline{\hskip 0.00cm (a) \hskip 3.5cm (b) \hskip 2.9cm (c)} 
    \begin{minipage}{0.3\textwidth}
        \centering
\includegraphics[width=.8\textwidth]{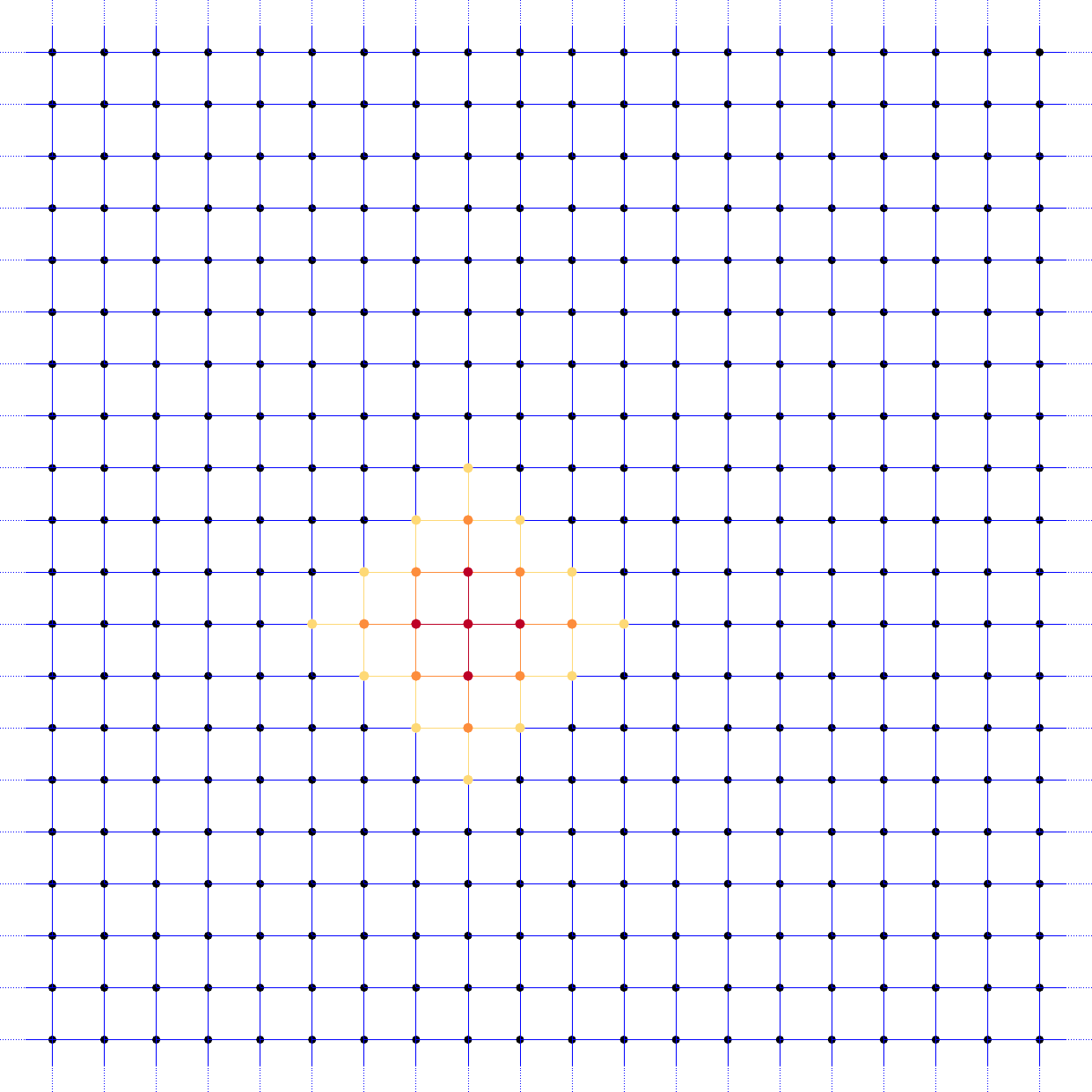}
\end{minipage}\hfill
    \begin{minipage}{0.2\textwidth}
        \centering
\includegraphics[width=.7\textwidth]{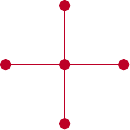}        
\end{minipage}\hfill
\begin{minipage}{0.45\textwidth}
\centering
\includegraphics[width=.7\textwidth]{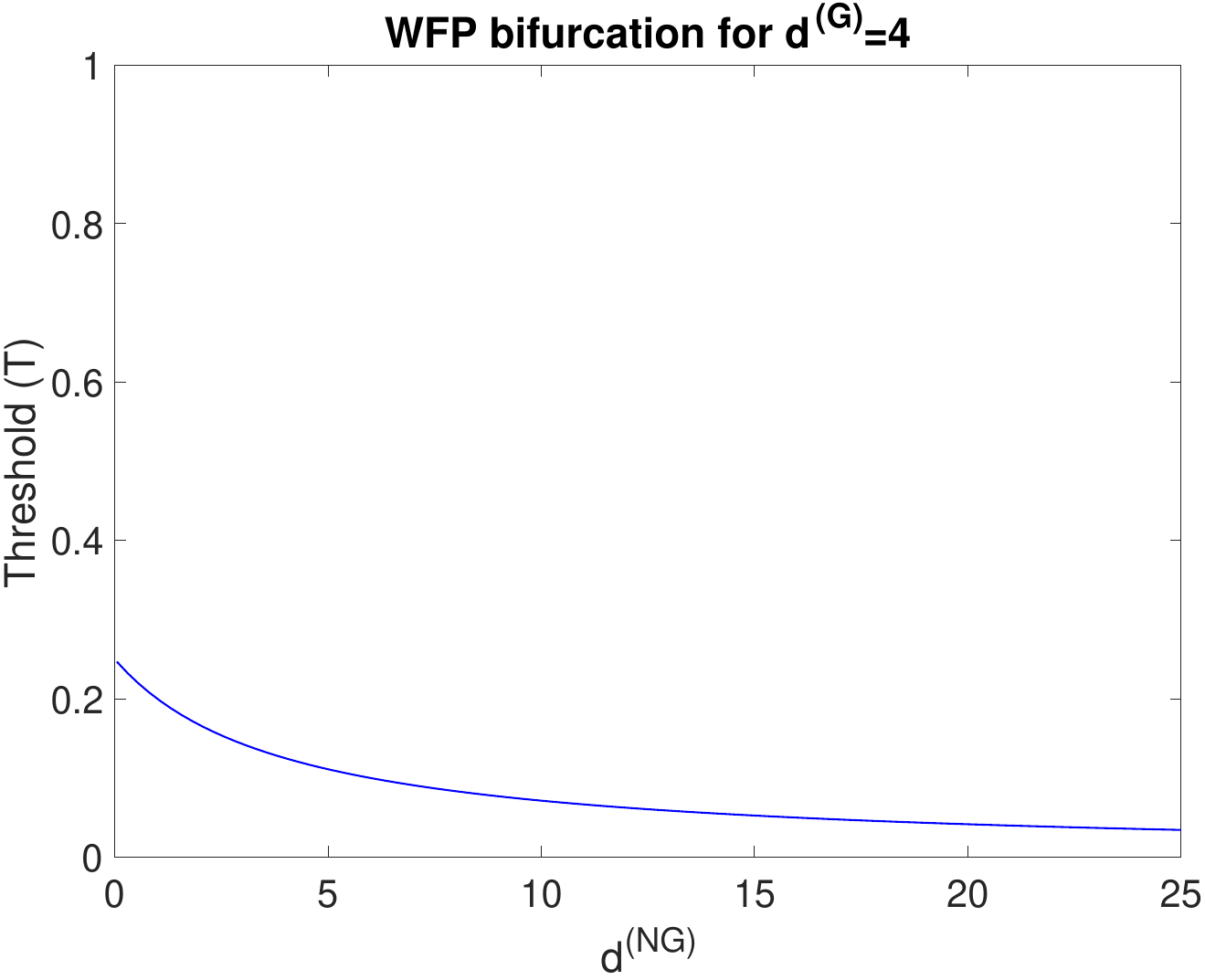}
\end{minipage}\hfill
\caption{(a) Purely geometric network with geometric neighbors up to radius $r=1$ around a node. (This corresponds to $p=1$ and $q=0$ in our Kleinberg-like small-world networks.) The geometric degree is $d^{\rm{G}}=4$, and the nongeometric degree is $d^{\rm{NG}}=0$. We color the seed nodes of the contagion in dark red, the nodes that activate in the first time step in a moderately dark color, and the nodes that activate in the second time step in a light color. (b) A node with its four direct neighbors, which are the nodes that are within Euclidean distance $p=1$ from it. (c) Bifurcation diagram for the occurrence of WFP in a WTM contagion on a Kleinberg-like network with $d^{\rm{G}}=4$. We vary the nongeometric degree $d^{\rm{NG}}$ from $0$ to $25$, and we vary the contagion threshold $T$ from $0$ to $1$. WFP occurs only in the region below the curve.
}\label{WFP_4}
\end{figure}

If $d^{\rm{G}}=4$ (see Figure~\ref{WFP_4}), then for WFP to occur, the threshold $T$ needs to be small enough to allow spreading via a single edge. Therefore, for variable $d^{\rm{NG}}$, for WFP to occur, the threshold $T$ needs to be smaller than
\begin{equation*}
	T^{\rm{WFP}}=\frac{1}{4+d^{\rm{NG}}} \, .
\end{equation*}

\begin{figure}[H]
    \centering
    \leftline{\hskip 0.00cm (a) \hskip 3.5cm (b) \hskip 2.9cm (c)} 
    \begin{minipage}{0.3\textwidth}
        \centering
\includegraphics[width=.8\textwidth]{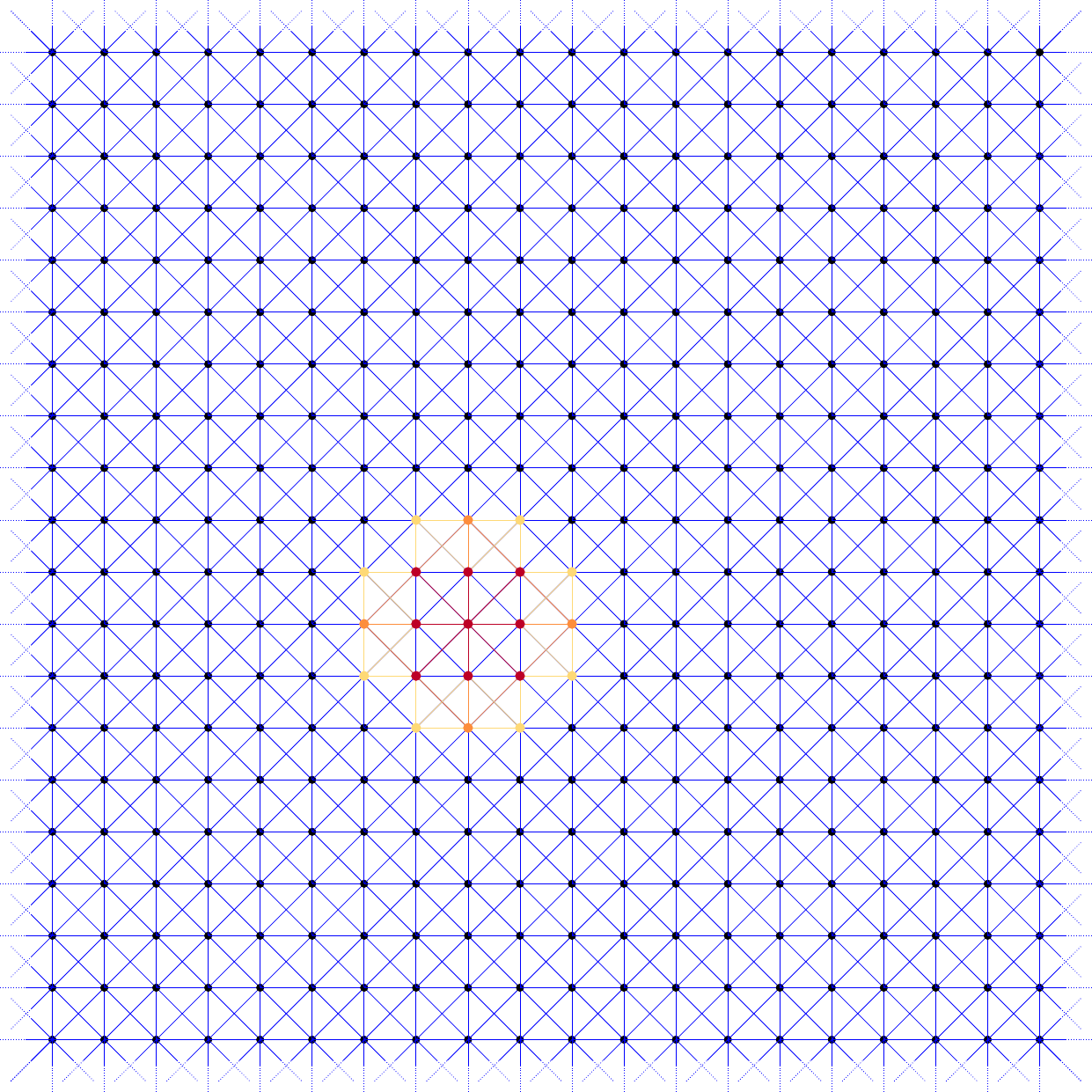}
\end{minipage}\hfill
    \begin{minipage}{0.2\textwidth}
        \centering
\includegraphics[width=.7\textwidth]{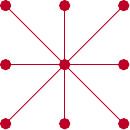}        
\end{minipage}\hfill
\begin{minipage}{0.45\textwidth}
\centering
\includegraphics[width=.7\textwidth]{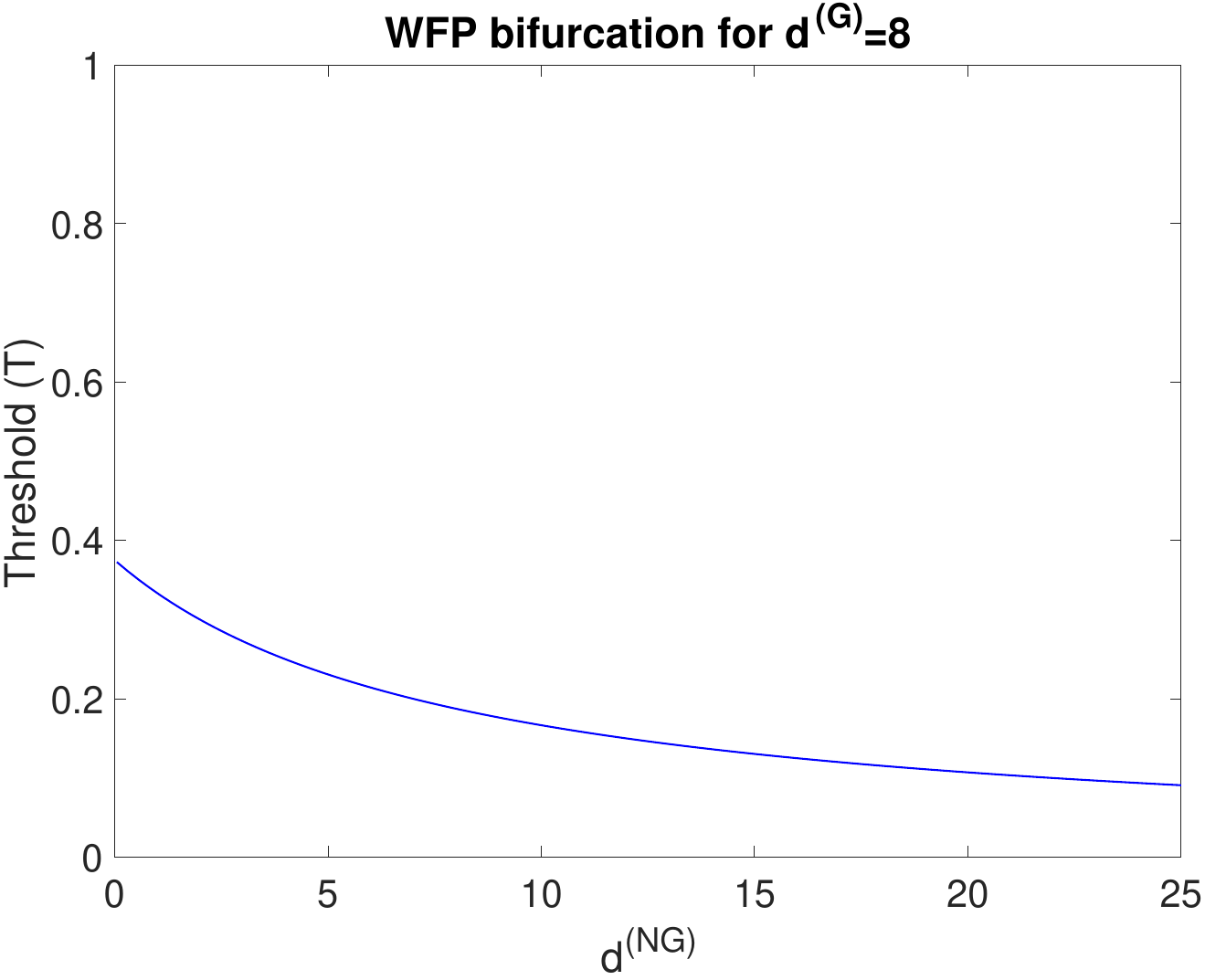}
\end{minipage}\hfill
\caption{(a) Purely geometric network with geometric neighbors up to radius $r=\sqrt{2}$ around a node. (This corresponds to $p=\sqrt{2}$ and $q=0$ in our Kleinberg-like small-world networks.) The geometric degree is $d^{\rm{G}}=8$, and the nongeometric degree is $d^{\rm{NG}}=0$. We color the seed nodes of the contagion in dark red, the nodes that activate in the first time step in a moderately dark color, and the nodes that activate in the second time step in a light color. (b) A node with its eight direct neighbors, which are the nodes that are within Euclidean distance $p=\sqrt{2}$ from it. (c) Bifurcation diagram for the occurrence of WFP in a WTM contagion on a Kleinberg-like network with $d^{\rm{G}}=8$. We vary the nongeometric degree $d^{\rm{NG}}$ from $0$ to $25$, and we vary the contagion threshold $T$ from $0$ to $1$. WFP occurs only in the region below the curve.
}\label{WFP_8}
\end{figure}

If $d^{\rm{G}}=8$ (see Figure~\ref{WFP_8}), then for WFP to occur, the threshold $T$ needs to be small enough to allow spreading via three edges. Therefore, for variable $d^{\rm{NG}}$, for WFP to occur, the threshold $T$ needs to be smaller than
\begin{equation*}
	T^{\rm{WFP}}=\frac{3}{8+d^{\rm{NG}}} \, .
\end{equation*}

\begin{figure}[H]
    \centering
    \leftline{\hskip 0.00cm (a) \hskip 3.5cm (b) \hskip 2.9cm (c)} 
    \begin{minipage}{0.3\textwidth}
        \centering
\includegraphics[width=.8\textwidth]{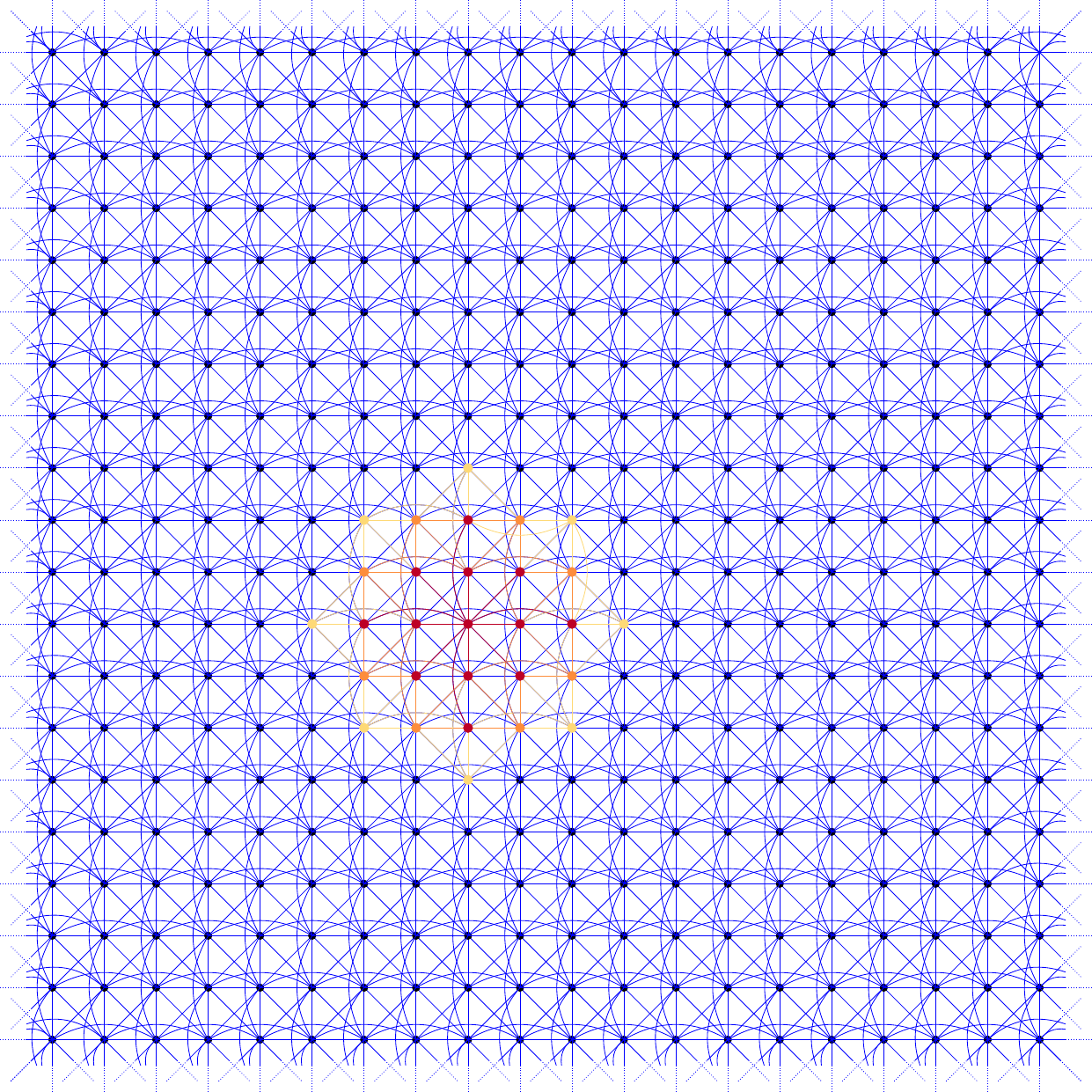}
\end{minipage}\hfill
    \begin{minipage}{0.2\textwidth}
        \centering
\includegraphics[width=.8\textwidth]{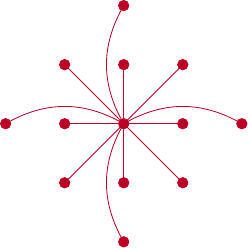}        
\end{minipage}\hfill
\begin{minipage}{0.45\textwidth}
\centering
\includegraphics[width=.7\textwidth]{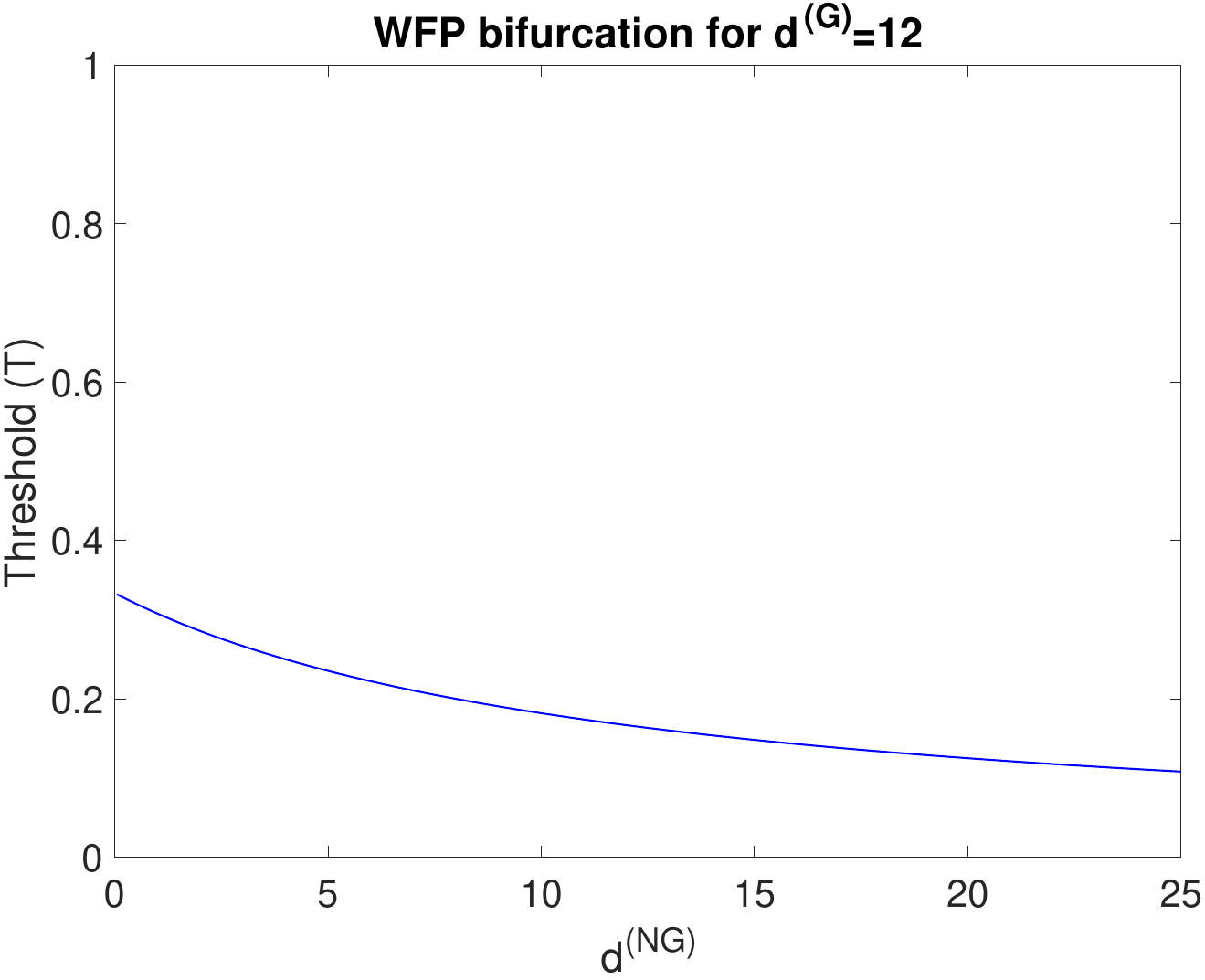}
\end{minipage}\hfill
\caption{(a) Purely geometric network with geometric neighbors up to radius $r=2$ around a node. (This corresponds to $p=2$ and $q=0$ in our Kleinberg-like small-world networks.) The geometric degree is $d^{\rm{G}}=12$, and the nongeometric degree is $d^{\rm{NG}}=0$. We color the seed nodes of the contagion in dark red, the nodes that activate in the first time step in a moderately dark color, and the nodes that activate in the second time step in a light color. (b) A node with its twelve direct neighbors, which are the nodes that are within Euclidean distance $p=2$ from it. (c) Bifurcation diagram for the occurrence of WFP in a WTM contagion on a Kleinberg-like network with $d^{\rm{G}}=12$. We vary the nongeometric degree $d^{\rm{NG}}$ from $0$ to $25$, and we vary the contagion threshold $T$ from $0$ to $1$. WFP occurs only in the region below the curve.
}\label{WFP_12}
\end{figure}

If $d^{\rm{G}}=12$ (see Figure~\ref{WFP_12}), then for WFP to occur, the threshold $T$ needs to be small enough to allow spreading via four edges. Therefore, for variable $d^{\rm{NG}}$, for WFP to occur, the threshold $T$ needs to be smaller than
\begin{equation*}
	T^{\rm{WFP}}=\frac{4}{12+d^{\rm{NG}}} \, .
\end{equation*}

There does not seem to be a closed form for $T^{\rm{WFP}}$ that holds for general values of $d^{\rm{G}}$. One needs to find the maximum threshold that allows spreading by WFP for each value of $d^{\rm{G}}$ individually by finding the edges that can support spreading from the contagion seed. 


\subsection{Appearance of new clusters (ANC)}

The activation of an inactive node by ANC occurs, by definition, exclusively via nongeometric edges. That is, a node is activated via ANC if it is adjacent to at least $T \times (d^{\rm{G}}+d^{\rm{NG}})$ active nodes by nongeometric edges and all of its geometric neighbors are inactive. Consequently, if the threshold $T$ is larger than or equal to the ratio of the nongeometric degree to the total degree (i.e., $T \geq \frac{d^{\rm{NG}}}{d^{\rm{G}}+d^{\rm{NG}}} $), then ANC is impossible. If $T < \frac{d^{\rm{NG}}}{d^{\rm{G}}+d^{\rm{NG}}} $, then ANC is possible. When $\frac{d^{\rm{NG}}-1}{d^{\rm{G}}+d^{\rm{NG}}} \leq T < \frac{d^{\rm{NG}}}{d^{\rm{G}}+d^{\rm{NG}}} $, ANC can occur in principle, but only if all of the nongeometric edges of an inactive node that has no active geometric neighbors `reach into' contagion clusters. This is very unlikely to occur in practice, so a threshold $T$ for which ANC is possible in principle is not a good indicator in practice for the presence of ANC. We will explore this issue.  

We define the \emph{horizon} 
\begin{equation*}
	H^{\rm{ANC}} = \frac{d^{\rm{NG}}}{d^{\rm{G}}+d^{\rm{NG}}} 
\end{equation*}
of ANC to be the boundary between thresholds for which ANC is possible in theory and thresholds for which ANC is impossible. 
Using the horizon as a boundary curve for ANC generates an `idealized' bifurcation diagram that tends to overestimate the size of the region of the parameter space for which ANC occurs.

In practice, one needs to think about the probability that a number $k$ among all nongeometric edges of a given node reach into clusters of active nodes. If nongeometric edges are placed uniformly at random (which occurs when $\gamma=0$ in the construction of our Kleinberg-like networks), the expected probability for a nongeometric edge of an inactive node to be incident to an active node at time $t$ is $\frac{q(t)}{N-1}$, where $q(t)$ is the number of active nodes at time $t$. Consequently, if the nongeometric degree is $d^{\rm{NG}}$, the expected number of nongeometric edges of an inactive node that are incident to active nodes is $\frac{q(t)}{N-1}d^{\rm{NG}}$. It follows that the maximum threshold for which one can expect every node that is inactive before time $t$ to activate via ANC at time $t$ is 
\begin{equation}\label{approximate}
	T = \frac{\frac{q(t)}{N-1} d^{\rm{NG}}}{d^{\rm{G}}+d^{\rm{NG}}}\, .
\end{equation} 
However, for ANC to occur at a certain time, it is not necessary for every inactive node to activate via ANC at that time. It suffices for any inactive node that is sufficiently far away from the contagion to activate via ANC, and the requirements for that to occur are generally lower than \eqref{approximate}. Consequently, we expect \eqref{approximate} to be a lower bound for $T^{\rm{ANC}}$, the critical threshold for ANC to occur. 

The numerator in \eqref{approximate} depends linearly on the (time-dependent) number $q(t)$ of active nodes. This raises the question of what may be a sensible choice for $t$ (and $q(t)$). Intuitively, if ANC occurs towards the end of a spreading process, when large parts of the network are already active, then its contribution to the spreading of the contagion is a minor one and it has only a negligible distortive effect on the contagion map. The activation times of nodes that are infected via ANC late in a contagion process are only mildly shorter than what occurs for spreading purely via WFP, so the points in the image of the contagion map are perturbed only slightly. To make \eqref{approximate} a meaningful bound for $T^{\rm{ANC}}$, we thus seek to work out the latest point in the spreading process up to which the occurrence of ANC plays a significant role in the overall spreading behavior and accordingly has a noticeable effect on the contagion map. The later this point occurs, the larger $\frac{q(t)}{N-1}d^{\rm{NG}}$ will be and the larger we expect the critical threshold (i.e., bifurcation point) $T^{\rm{ANC}}$ to be. 
We have
\begin{equation}\label{inequal}
	\delta \, H^{\rm{ANC}}=\frac{\delta \, d^{\rm{NG}}}{d^{\rm{G}}+d^{\rm{NG}}} < T^{\rm{ANC}} < \frac{d^{\rm{NG}}}{d^{\rm{G}}+d^{\rm{NG}}}=H^{\rm{ANC}}
\end{equation}
for some $\delta \in (0,1)$, where the parameter $\delta$ is determined by how late in the spreading process the occurrence of ANC plays a significant role. 
If, for instance, the occurrence of ANC plays a significant role in overall spreading behavior and thus has a noticeable distortive effect on a contagion map only if it takes place by the time that three fifths of the nodes in the network are active, then the bifurcation curve for ANC is bounded below as follows: 
\begin{equation*}
	T^{\rm{ANC}} > \frac{\frac{1500}{2499} d^{\rm{NG}}}{d^{\rm{G}}+d^{\rm{NG}}} \, .
\end{equation*}

\begin{figure}[H]
        \leftline{\hskip 1.7mm (a) \hskip 6cm (b)} 
    \begin{minipage}{.49\textwidth}
    \centering
        \includegraphics[width=.9\textwidth]{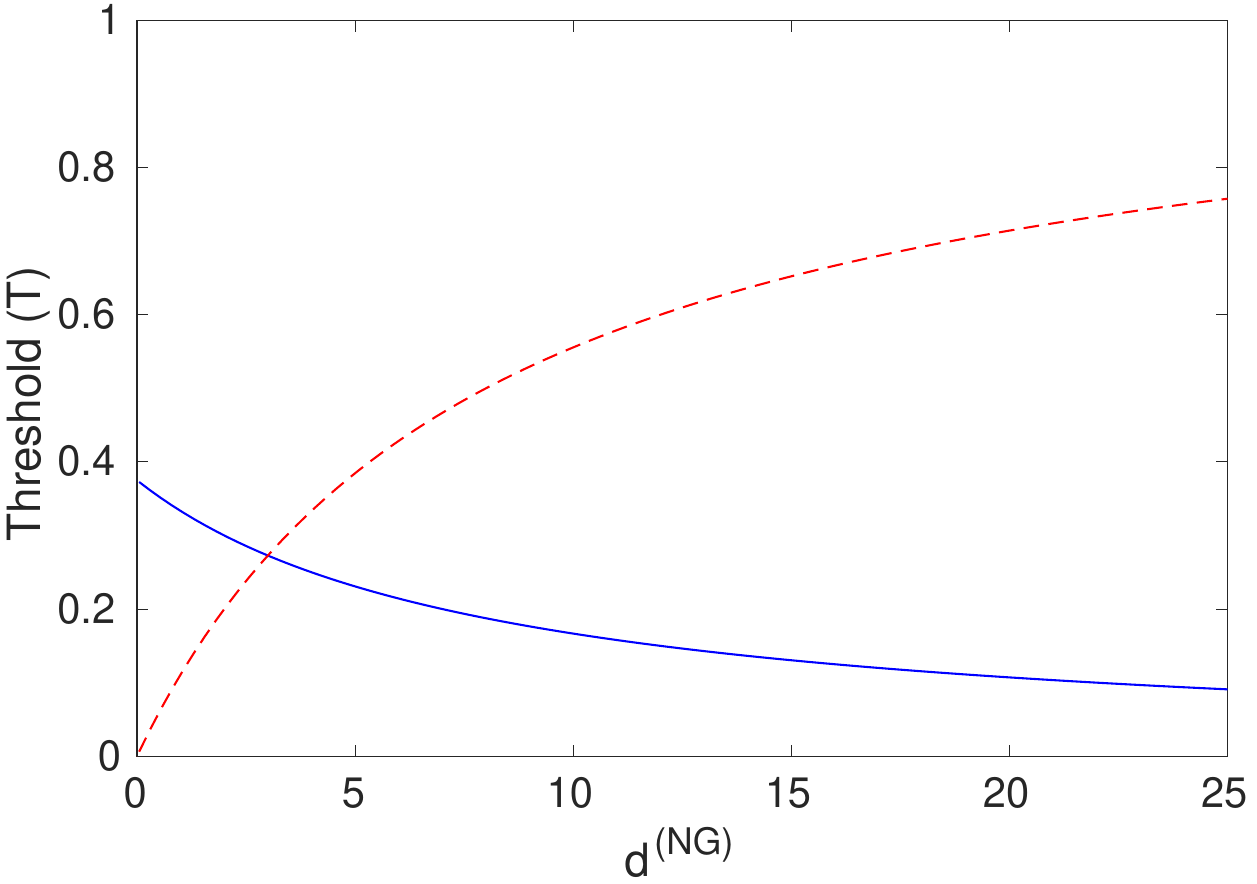}
    \end{minipage}  \hfill 
    \begin{minipage}{.49\textwidth}
    \centering
        \includegraphics[width=.9\textwidth]{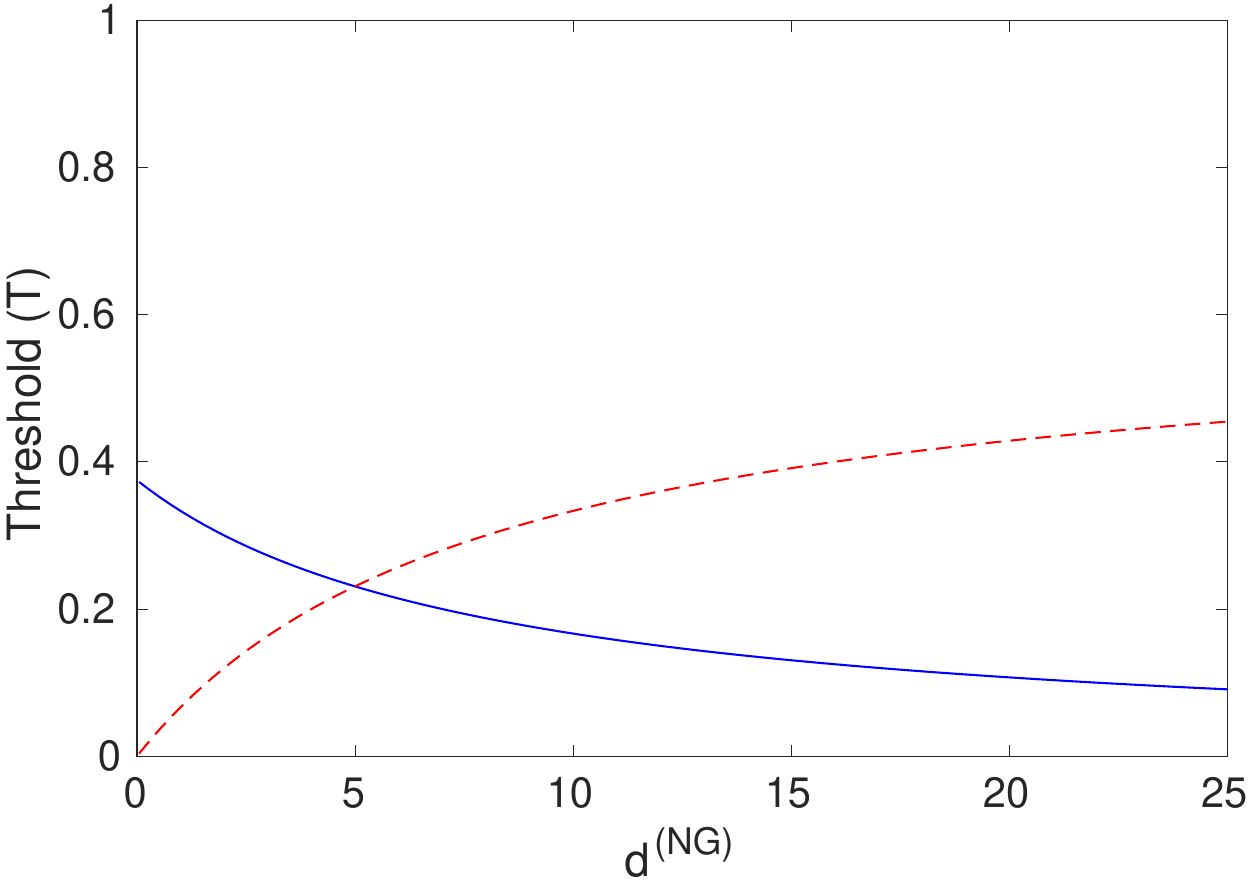}
    \end{minipage}\hfill
    \caption{(a) Idealized bifurcation curve and (b) approximate bifurcation curve for WFP and ANC of a WTM contagion on a Kleinberg-like small-world network with geometric neighbors up to distance $r=\sqrt{2}$ around a node (i.e., with geometric degree $d^{\rm{G}}=8$). The blue curve shows $T^{\rm{WFP}}$; the red curve shows the idealized and approximate $T^{\rm{ANC}}$ in panels (a) and (b), respectively. 
    }\label{compare_upper_correct}
\end{figure}

We compare the idealized bifurcation diagram (using the horizon of ANC as its bifurcation curve) and the diagram that we obtain from \eqref{approximate} with $q(t)=(3/5)N$, where $N$ is the number of nodes in the network (see Figure~\ref{compare_upper_correct}), to our numerical results for the geometry and dimensionality.

The above argument is independent of the particular geometry that underlies a network, as long as the nongeometric edges are placed uniformly at random. In particular, the inequalities \eqref{inequal} should also hold for the ring lattice in the computations of Taylor et al. \cite{Taylor2015}. Indeed, looking at their results (see Figure~6 in \cite{Taylor2015}), their $T_0^{\rm{ANC}}$ curve (the dotted curve) does seem to be a bit higher than what they observed in their numerical computations, suggesting that $T_0^{\rm{ANC}}$ is indeed bounded above by the idealized bifurcation curve. 
 
If three fifths of the total number of nodes is indeed the correct choice for the maximum number of nodes that activate before a certain time $t$ for the occurrence of ANC to be `significant' at time $t$, then one should expect the actual bifurcation curve $T^{\rm{ANC}}$ to lie somewhere between the red curves in Figure~\ref{compare_upper_correct}.  

To find the actual bifurcation curve $T^{\rm{ANC}}$, we need to find (for a given value of $d^{\rm{NG}}$) the largest threshold $T$ that realistically allows ANC to arise before the active region of the given network is so large that the occurrence of ANC no longer has a significant impact on the spreading dynamics. This amounts to finding a threshold $T$ that is as large as possible, but is still small enough that we obtain a value of at least $1$ for the expected number of inactive nodes with sufficiently many nongeometric edges that reach into the active region. 

To avoid taking into account the activation of nodes via ANC in the boundary of the active region or very close to it, one can count only inactive nodes that lie `sufficiently far away' from active nodes. We use the term \emph{neighborhood}\footnote{Our use of the term `neighborhood' is different from its usual use in graph theory.} and the notation $\mathcal{N}$ for the set of nodes outside which we count nodes being activated via ANC. The neighborhood can consist either of the active nodes only, in which case $|\mathcal{N}|=q(t)$; or it can include some additional nodes around the active region, such that $\mathcal{N} > q(t)$. 

At a given time $t$, suppose that the $q(t)$ nodes that are active at time $t$ were activated at previous time steps by WFP from the cluster seed and that they form an active cluster of roughly square shape. 
We denote this set of nodes by 
\begin{equation*}
	I=\{i \in V \mid \eta_i(t)=1\}  \, ,
\end{equation*}
so $|I|=q(t)$. We define the neighborhood $\mathcal{N}(I)$ of this active cluster to be the cluster itself together with the nodes in its periphery of a certain `width'. That is, $\mathcal{N}(I)$ is the active cluster $I$ itself, its boundary, and (depending on the width) some more nodes around it. Given a width $w$, we approximate the number of nodes in $\mathcal{N}(I)$ as 
\begin{equation}\label{neighborhood_number}
	| \mathcal{N}(I) |  \approx \left(\sqrt{q(t)} +2w \right)^2 \, .
\end{equation}
We can choose any natural number for the width $w$, and one plausible choice is $d^{\rm{G}}/2$. If $w=0$, the neighborhood $\mathcal{N}(I)$ is just $I$ itself (and then $|\mathcal{N}|=q(t)$). To make a sensible choice for the size of the active region of a network at the latest point in the spreading process at which we consider ANC to be significant, we estimate the largest number $q_{\rm{max}}$ of active nodes such that the active region together with a periphery of inactive nodes of width $d^{\rm{G}}$ constitutes at most $90 \%$ of the nodes in the network. We make this estimate by choosing $q_{\rm{max}}$ to be the largest integer such that
\begin{equation}
	|\mathcal{N}_{\rm{max}}|=\left(\sqrt{q_{\rm{max}}} + 2d^{\rm{G}} \right)^2 \leq 0.9 \, N  \, .
\end{equation}
Consequently,
\begin{equation}
	q_{\rm{max}} \leq \left(\sqrt{0.9 \, N} - 2 d^{\rm{G}}\right)^2 \, .
\end{equation}
For $N=2500$ and $d^{\rm{G}}=8$, this gives $q_{\rm{max}}=988$.

If we consider ANC to be significant up to time $t$, the corrected bifurcation curve for ANC is
\begin{equation}\label{TANC}
	T^{\rm{ANC}} = \frac{k}{d^{\rm{G}}+d^{\rm{NG}}} \, ,
\end{equation}
where $k$ is the largest integer such that, at time $t$, the expected number of nodes outside the neighborhood $\mathcal{N}(I)$ of $I$ with more than $k$ active neighbors is at least $1$. 

Let $X_k$ be the number of nodes outside $\mathcal{N}(I)$ with more than $k$ active neighbors. 
It satisfies the binomial distribution
\begin{equation*}
	X_k \sim \mathrm{Bin}\left(N-| \mathcal{N}(I) |,P(d_{\mathrm{in}} > k) \right) \,,
\end{equation*}	
so
\begin{equation*}
	P(X_k=x)= \binom{N-| \mathcal{N}(I) |}{x}P(d_{\mathrm{in}} > k)^x P(d_{\mathrm{in}} \leq k)^{N-| \mathcal{N}(I) |-x} \, ,
\end{equation*}
where $d_{\mathrm{in}}$ is the number of edges of a node outside $\mathcal{N}(I)$ that are incident to an active node. 
Therefore, the expected number of nodes outside $\mathcal{N}(I)$ with more than $k$ active neighbors is 
\begin{equation}\label{expectation}
	\begin{split}
E[X_k]=&\sum_{x=0}^{N-| \mathcal{N}(I) |}P(X_k=x)x \\
&= \left( N-|\mathcal{N}(I)|\right) P(d_{\mathrm{in}}>k) \\
&= \left( N-|\mathcal{N}(I)|\right) \left( 1 -  P(d_{\mathrm{in}} \leq k) \right) \, . 
	\end{split}
\end{equation}
As we argue in a remark at the end of this subsection, it is approximately the case that
\begin{equation}
	d_{\mathrm{in}} \sim  \mathrm{Bin}\left(d^{\rm{NG}},\frac{q(t)}{N}\right)  \,  . 
\end{equation}	
That is, $d_{\mathrm{in}}$ approximately follows a binomial distribution. Its associated (approximate) cumulative distribution function is 
\begin{equation}
	\begin{split}
P(d_{\mathrm{in}} \leq k) & \approx \sum_{d=0}^k \binom{d^{\rm{NG}}}{d}\left(\frac{q(t)}{N}\right)^d\left(1-\frac{q(t)}{N}\right)^{d^{\rm{NG}}-d} \\ \label{dinprob}
& = \left(d^{\rm{NG}}-k\right)\binom{d^{\rm{NG}}}{k} \int_{0}^{1-\frac{q(t)}{N}}s^{d^{NG}-k-1}(1-s)^k ds \,.  
	\end{split}
\end{equation}

\begin{figure}[H]
 \centering
   \leftline{\hskip 1mm (a) \hskip 2.7cm (b) \hskip 2.7cm (c) \hskip 2.7cm (d)} 
    \begin{minipage}{.23\textwidth}
        \centering
        \includegraphics[width=1.00\textwidth]{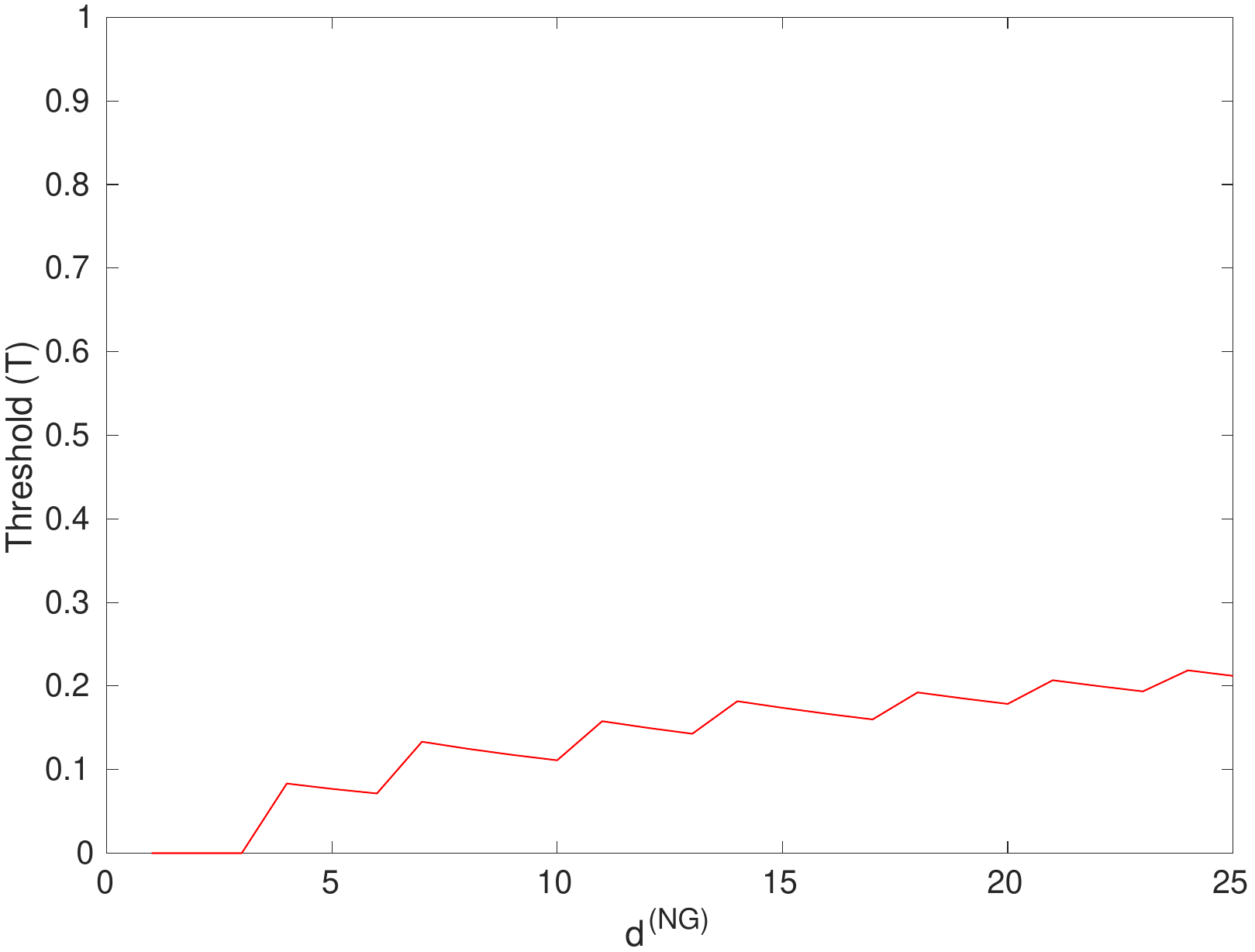}
    \end{minipage}\hfill
    \begin{minipage}{0.23\textwidth}
        \centering
        \includegraphics[width=1.00\textwidth]{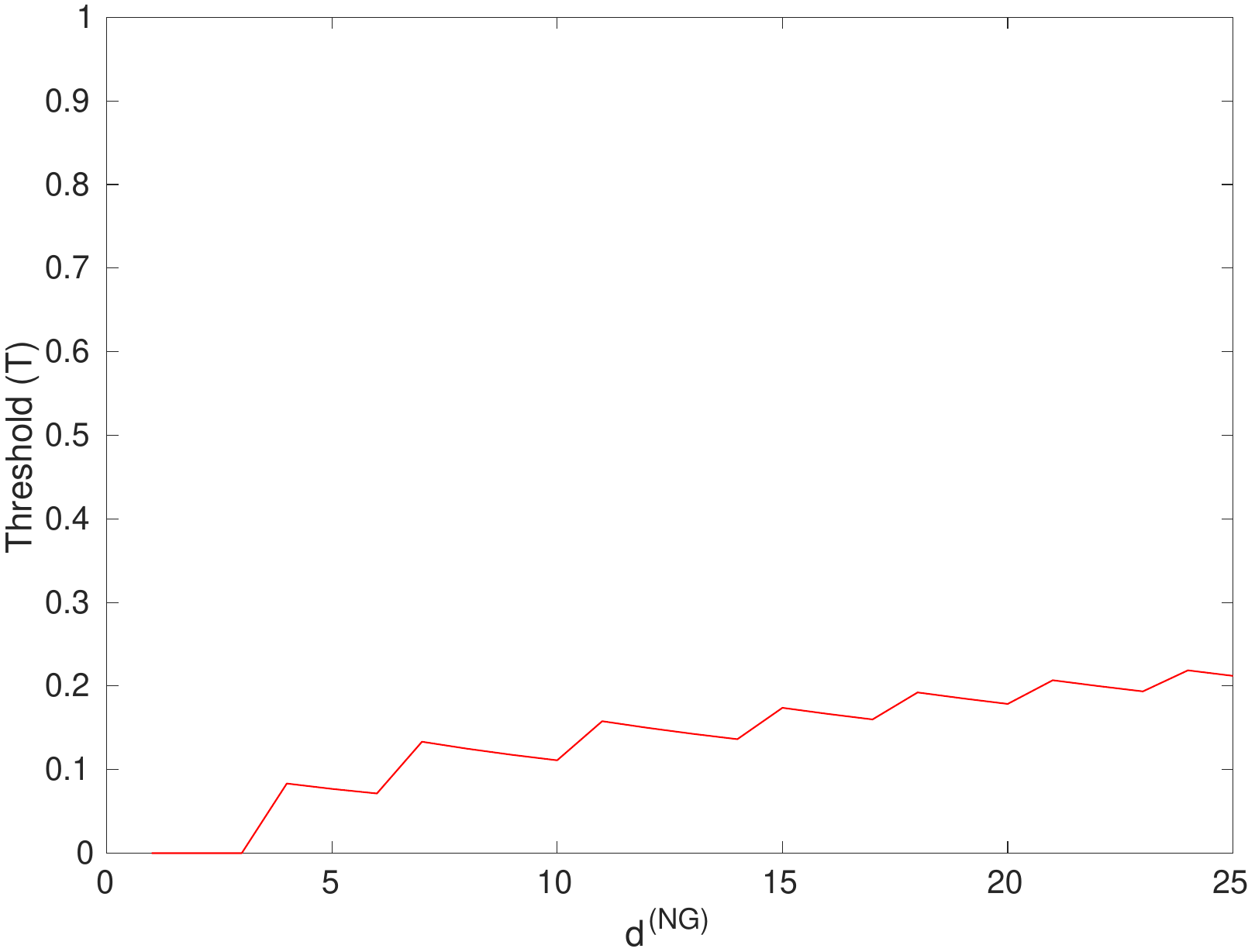}
    \end{minipage}\hfill
    \begin{minipage}{.23\textwidth}
    \centering
        \includegraphics[width=1.00\textwidth]{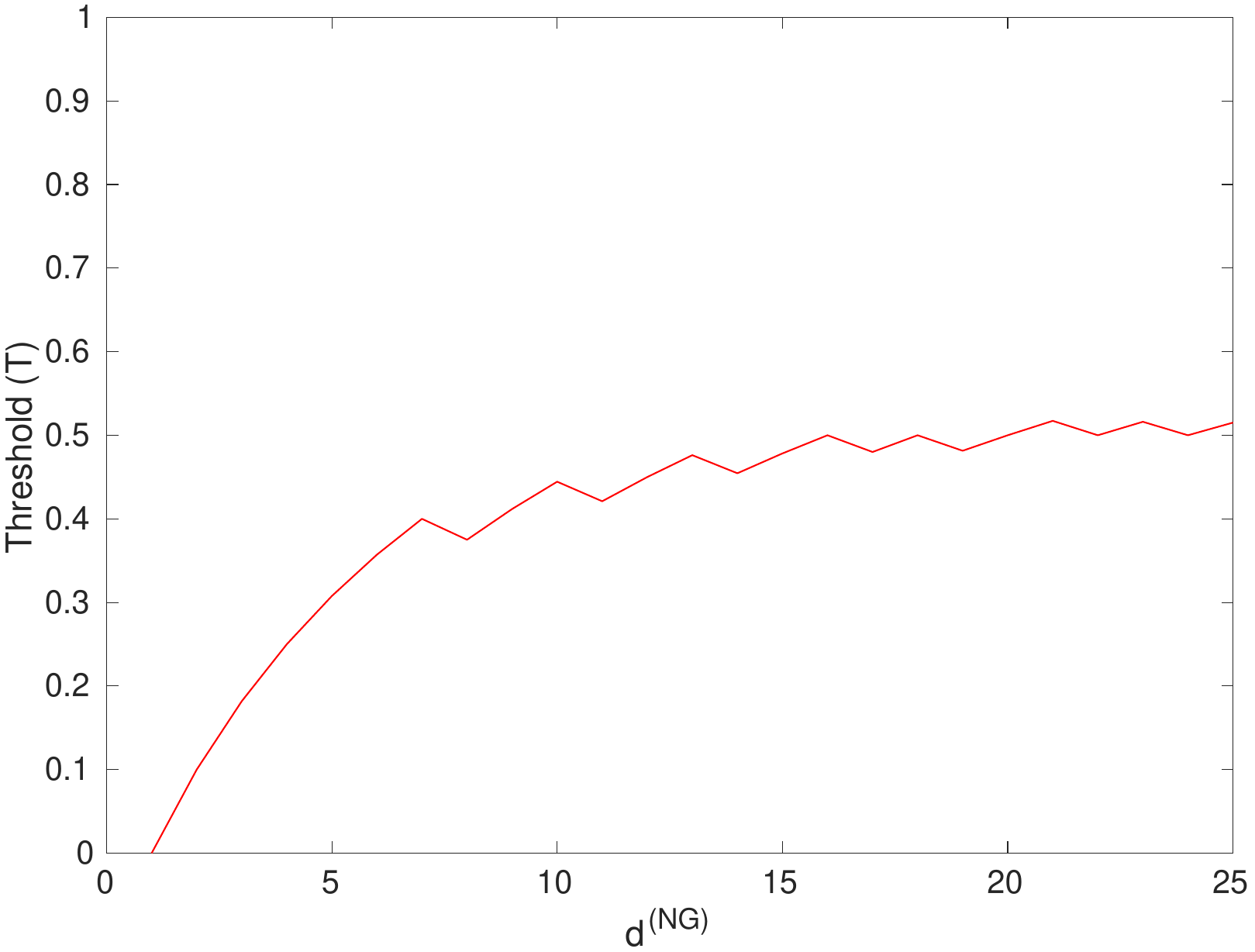}
    \end{minipage}\hfill
    \begin{minipage}{.23\textwidth}
    \centering
        \includegraphics[width=1.00\textwidth]{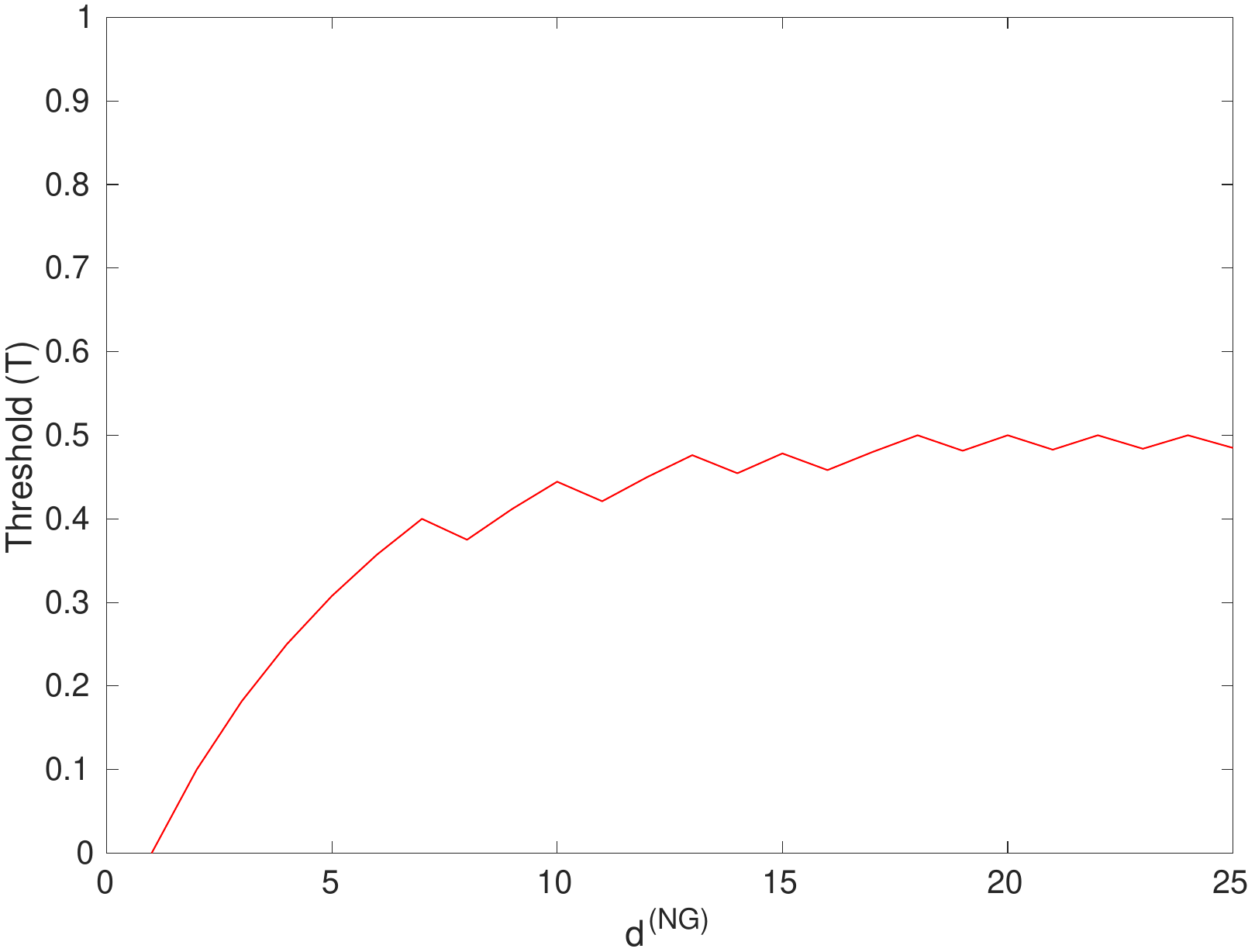}
    \end{minipage}\hfill    
    \caption{Behavior of $T^{\rm{ANC}}$ for a WTM contagion on a Kleinberg-like small-world network with geometric neighbors up to distance $r=\sqrt{2}$ around a node (i.e., with geometric degree $d^{\rm{G}}=8$) for various choices of what constitutes the neighborhood of an active cluster of nodes: (a) $|\mathcal{N}(I)|=q(0)$ (a pathological case), (b) $|\mathcal{N}(I)|=\left(\sqrt{q(0)} +d^{\rm{G}} \right) ^2  $ (a pathological case), (c) $|\mathcal{N}(I)|=988$, and (d) $|\mathcal{N}(I)| = \left(\sqrt{988} +d^{\rm{G}} \right)^2 $.  
}\label{numerical}
\end{figure}

To determine the numerator of $T^{\rm{ANC}}$ (see formula~\eqref{TANC}), we seek the largest integer $k$ such that $E[X_k] \geq 1$. 
That is, using the approximation \eqref{dinprob}, we seek the largest integer $k$ such that 
\begin{equation}\label{largest_k}
	\frac{N-|\mathcal{N}(I)|-1}{N-|\mathcal{N}(I)|} \geq \sum_{d=0}^{k}\binom{d^{\rm{NG}}}{d}\left(\frac{q(t)}{N} \right)^d \left( 1-\frac{q(t)}{N} \right)^{d^{\rm{NG}}-d} \, .
\end{equation}
We can find $k$ for each value of $d^{\rm{NG}}$ and deduce $T^{\rm{ANC}} (d^{\rm{NG}})$ from that value for a given $d^{\rm{G}}$. For $d^{\rm{G}}=8$, this yields the plots in Figure~\ref{numerical} for $T^{\rm{ANC}}$ for various choices of the width of the neighborhood $\mathcal{N}(I)$ and the maximum value of $q(t)$ at which we consider ANC to be significant. 

Observe that the curves are essentially increasing, but in an oscillatory manner, resulting in a staircase-like shape. This shape arises from the fact that, as $d^{\rm{NG}}$ increases from $0$ to $25$ in integer increments, it affects both the largest integer $k$ such that inequality~\eqref{largest_k} is satisfied and the denominator of formula~\eqref{TANC}. The larger the value of $d^{\rm{NG}}$, the larger the value of $\binom{d^{\rm{NG}}}{d}$ and the smaller the value of $\left( 1-\frac{q(t)}{N} \right)^{d^{\rm{NG}}-d}$ in the sum on the right-hand side of inequality~\eqref{largest_k}, with the latter factor being overall dominant. This explains the overall increasing tendency of the curve. For a given value of $k$, an increase of $d^{\rm{NG}}$ leads to a steady decrease of $T^{\rm{ANC}}$, explaining the small intervals of decrease of the curve.

Note (again) that our central reasoning in the above argument is independent of the geometry that underlies our noisy geometric network, provided we place nongeometric edges uniformly at random. The only point at which the particular geometry of the 2D torus comes into play is in our estimation of the approximation~\eqref{neighborhood_number}, which we calculate by assuming that the contagion cluster $I$ is roughly square-shaped and that its neighborhood $\mathcal{N}(I)$ forms a larger square-shaped area of some width around $I$. If we take this width to be $0$ (i.e., if $\mathcal{N}(I)=q(t)$), the formula for the expectation \eqref{expectation} is the same for any noisy geometric network. 

\begin{figure}[H]
\centering
\includegraphics[width=.75\textwidth]{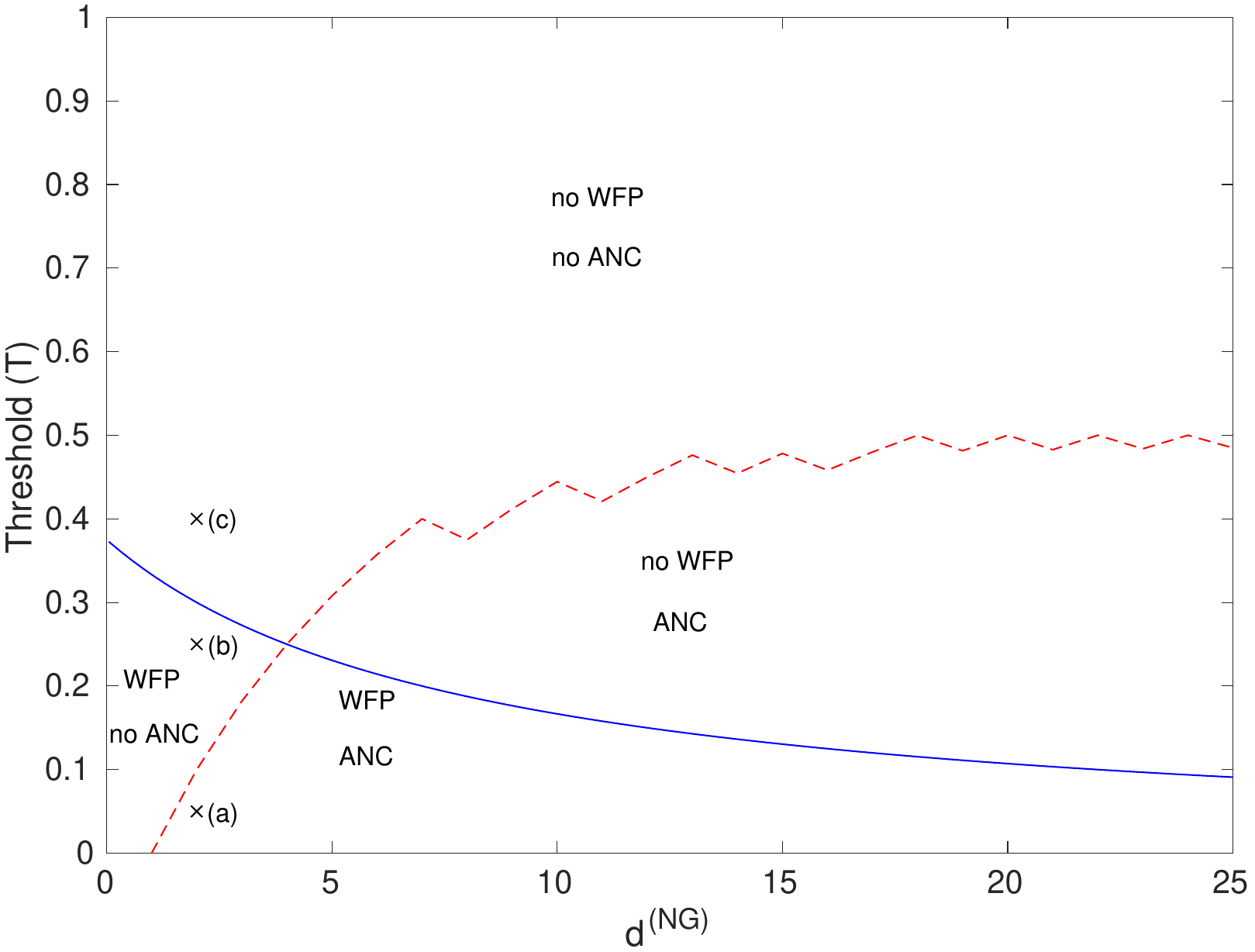}
\caption{Bifurcation diagram for WFP and ANC of a WTM contagion on a Kleinberg-like small-world network with geometric neighbors up to distance $r=\sqrt{2}$ around a node (i.e., geometric degree $d^{\rm{G}}=8$). The blue curve shows $T^{\rm{WFP}}$, and the red curve shows $T^{\rm{ANC}}$. The points $(a)$, $(b)$, and $(c)$ mark parameter combinations that represent three different spreading regimes. 
Point $(a)$ indicates the parameter combination $(d^{\rm{NG}},T)=(2,0.05)$, for which (due to the small value of the threshold $T$) we expect fast spreading via WFP and ANC; point $(b)$ indicates the parameter combination $(d^{\rm{NG}},T)=(2,0.5)$, representing spreading predominantly via WFP; and point $(c)$ indicates the parameter combination $(d^{\rm{NG}},T)=(2,0.4)$, for which we do not expect spreading. If $d^{\rm{G}}=8$ and $d^{\rm{NG}}=2$, the total degree is $10$. This implies that a contagion requires $1$, $3$, and $5$ adjacent activated nodes, respectively, to activate an inactive node for (a) $T=0.05$, (b) $T=0.25$, and (c) $T=0.4$.
}\label{full_bifurcation} 
\end{figure}

We summarize our results in the bifurcation diagram in Figure~\ref{full_bifurcation}. We overlay our results for geometry, topology, and dimensionality on this diagram in Figure~\ref{summary_figure}. 

{\bf Remark:} The random variable $d_{\mathrm{in}}$ follows a binomial distribution only approximately, because --- with our network model avoiding double edges --- the event that a nongeometric edge of a node is incident to an active node is not entirely independent of another one of this node's nongeometric edges being incident to an active node. Consequently, if we want to determine the probability that a node outside $\mathcal{N}(I)$ has $k$ nongeometric edges that reach into the contagion cluster (i.e., $P(d_{\mathrm{in}})=k$), we have to pick $k$ of the node's $d_{\rm{NG}}$ nongeometric edges $\{e_1,e_2,\dots e_k,e_{k+1}, \dots, e_{d^{\rm{NG}}-1},e_{d^{\rm{NG}}}\} $ and calculate consecutively, for each $\ell \in \{1,2,\dots ,d^{\rm{NG}} \}$, the probability that $e_{\ell}$ is incident to an active (if $\ell \in \{1, \dots, k\}$) or inactive (if $\ell \in \{k+1, \dots, d^{\rm{NG}}\}$) node, given the states of the incident nodes of the $e_j$ with $j < \ell$. 

The precise probability of a node having $k$ nongeometric edges that are incident to an active node is thus
\begin{equation*}
	P\left(d_{{\rm in}}=k\right) = \binom{d^{\rm{NG}}}{k} \frac{\left(\prod_{k'=0}^{k-1}q(t)-k' \right)\left(\prod_{k'=0}^{d^{\rm{NG}}-1-k}N-1-q(t)-k'\right)}{\prod_{k'=0}^{d^{\rm{NG}}-1}(N-1-k')} \, .
\end{equation*}
However, for $N \gg k$, it is reasonable to approximate the probability that a given nongeometric edge of an inactive node is incident to an active node as $\frac{q(t)}{N}$ (i.e., the number of active nodes divided by the total number of nodes). Consequently, $d_{\mathrm{in}}$ asymptotically follows the binomial distribution and the approximation \eqref{dinprob} is correct asymptotically. 


\section{Conclusions and Discussion}\label{discussion}

Networks that have some underlying geometry and include both geometric edges (which are short according to that geometry) and nongeometric edges (which can occur between nodes regardless of their distance from each other) arise in many applications \cite{Barthelemy2018}, including modeling of human communication and transportation. The spread of a contagion on such a network can be influenced heavily by the underlying geometry, and it is useful to investigate the strength of such influence. 

To study this problem, we considered a family of networks whose nodes can be considered to be lying on a 2D torus and whose edges include both geometric edges (which are deterministic and close to each other in the underlying geometry) and nongeometric edges (which are formed randomly and can occur between nodes that are far from each other in the underlying geometry). 
Using the Watts threshold model, we investigated the spreading behavior of contagions on this family of networks. We did so by mapping network nodes to a high-dimensional point cloud via a contagion map (following Taylor et al. \cite{Taylor2015}) and analyzing the structure of this point cloud from three perspectives: geometrically, topologically, and in terms of dimensionality. To examine the point cloud's geometry and dimensionality, we calculated a Pearson correlation coefficient and the embedding dimension, which are well-established measures and easy to compute. To study the topology of the point cloud, we computed persistent homology of the Vietoris--Rips filtration and then calculated the Wasserstein distance between the corresponding barcode and a reference barcode. This was the most challenging and time-consuming part of our work, as algorithms for the computation of PH are computationally expensive and software development in the field is still relatively young and evolving. 
We therefore restricted ourselves to computing PH in dimension $1$, although PH in dimension $2$ may also be insightful for our problem. In our analysis of the topological structure of the point clouds, we also illustrated the sensitivity of the Wasserstein distance to the overall scale of barcodes and the ensuing need to correct for geometric factors when using the Wasserstein distance as a measure of purely topological similarity. We proposed a way of calibrating barcodes to eliminate geometric factors and found that our method was effective at quantifying how much a point cloud resembles a torus.

\begin{figure}[H]
\leftline{\hskip .00cm (a) \hskip 3.6cm (b) \hskip 3.6cm (c) } 
\begin{minipage}{0.32\textwidth}
\includegraphics[width=1\textwidth]{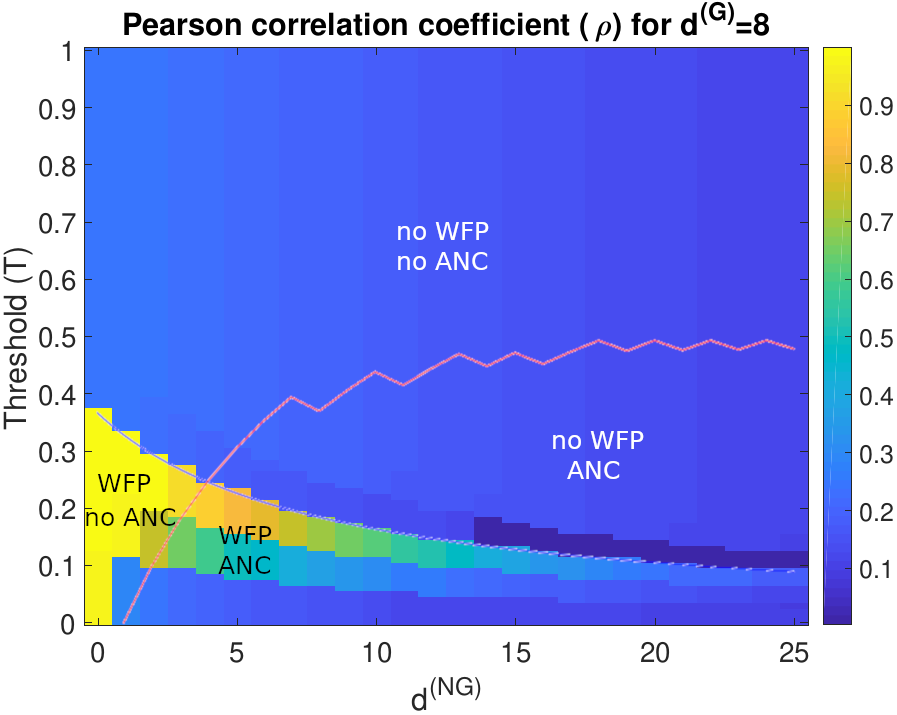}
\end{minipage}
\begin{minipage}{0.32\textwidth}
\includegraphics[width=1\textwidth]{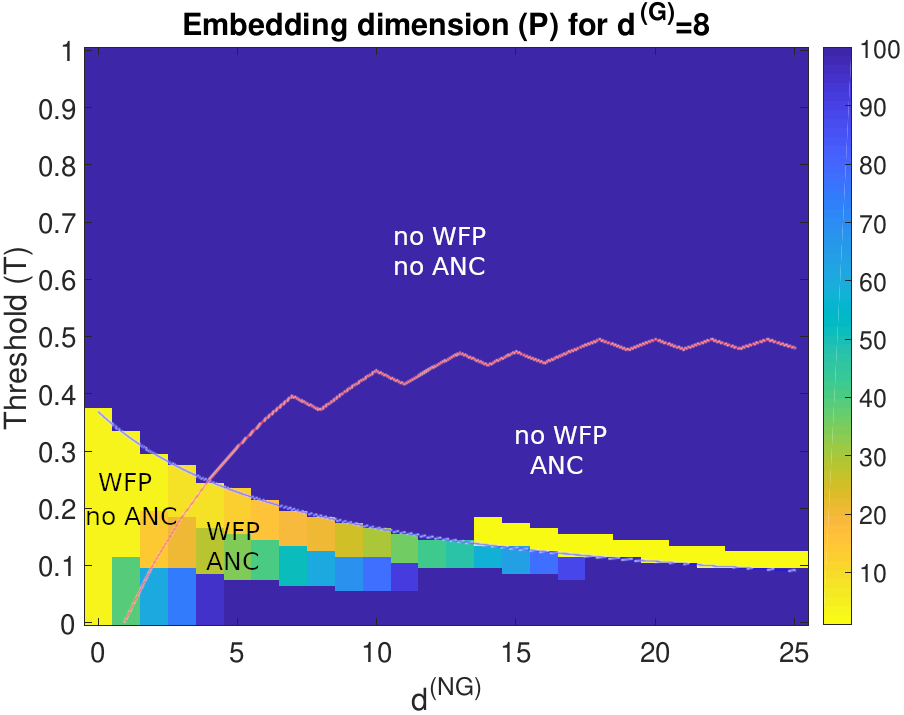}
\end{minipage}
\begin{minipage}{0.32\textwidth}
\includegraphics[width=1\textwidth]{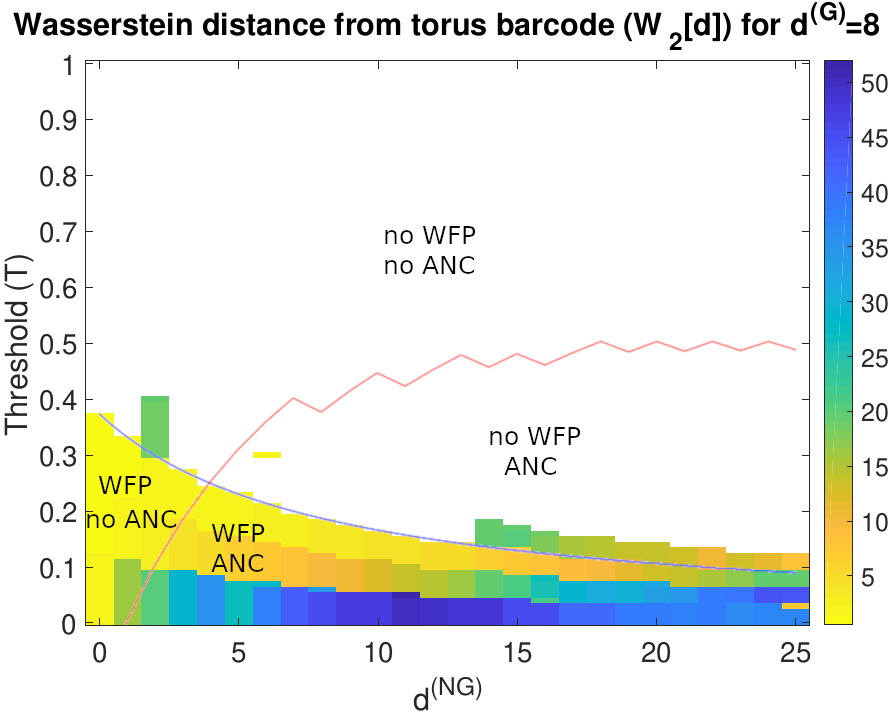}
\end{minipage}
\caption{Our numerical results for (a) geometry, (b) topology, and (c) dimensionality superimposed on the bifurcation diagram for WFP and ANC of a WTM contagion on a Kleinberg-like small-world network with geometric neighbors up to distance $r=\sqrt{2}$ around a node (i.e., geometric degree $d^{\rm{G}}=8$). The transitions between qualitatively different structures in our numerical computations align well with the curves for $T^{\rm{WFP}}$ and $T^{\rm{ANC}}$. In the absence of WFP, our numerical computations cannot distinguish between the presence and absence of ANC. Additionally, in the presence of WFP, the values of our quantifiers that suggest the presence of a toroidal structure of a point cloud are somewhat large even in the region in which we do not expect any ANC, and they become weaker for progressively larger values of $d^{{\rm NG}}$. This is a sensible observation, as one can expect toroidal structure only when WFP is present, and such toroidal structure is disturbed by the presence of ANC to an extent that depends on the rate of ANC.
 }\label{summary_figure}
\end{figure}

Our main finding is that the computational analysis of the point clouds that are the images of contagion maps aligns with the bifurcation analysis of spreading behaviors in the combined parameter space of network and contagion (see Figure~\ref{summary_figure}). We also found that the nature of the nongeometric edges affects contagion maps in the expected way: the shorter these edges are likely to be, the more they appear to contribute to the wave-like spreading of a contagion along the underlying torus and the more torus-like the structure of the point cloud is. This provides empirical evidence to support the effectiveness of contagion maps as a tool for estimating spreading behavior. It also suggests the potential of contagion maps as a manifold-learning technique that has some robustness to noise, a direction that is explored in \cite{Mahler2020}. 


The topological analysis of contagion maps was the most challenging part of our work, and there is room for further investigations of this aspect of our work with respect to both computational experiments and mathematical theory. First, for our focal network family (namely, Kleinberg-like small-world networks), it may be insightful to examine PH in dimension $2$ of the VR filtration, as there are nontrivial topological features to consider in two dimensions. Second, one might challenge our methodology for quantitatively measuring the topology of the contagion map itself: Both the VR complex and the Wasserstein distance --- despite being used as topological tools --- are intrinsically geometric in nature. For the Wasserstein distance, we have addressed this issue by preprocessing barcodes by calibrating them. Regarding the VR filtration itself, we note that its construction is based purely on pairwise distances, just like the Pearson correlation coefficient that we used as a measure of geometry. While the PH of the VR filtration provides a richer, more nuanced summary of the set of pairwise distances than the Pearson correlation coefficient, this information is integrated when we calculate the Wasserstein distance from the associated barcode to a reference barcode. One may therefore question how much additional information one gains from our topological measure, especially in light of its computational complexity in comparison to that of the easily computed Pearson correlation coefficient. To understand this issue, it likely will be useful to explore how our geometric and topological measures relate to each other and if such a relation can be exploited to speed up computations of our topological measure.  

In future work, it will be insightful to use the approach of the present paper to study various other monotonic spreading processes, including those with stochastic update rules (such as compartmental models for diseases), on networks that have some underlying geometric structure (including ones that are more complicated than a torus).

\section*{Acknowledgements}
The author would like to thank Ulrike Tillmann for her advice on this project and for many useful comments on various versions of this paper. In addition, the author would like to thank Florian Klimm and Dane Taylor for helpful discussions at an early stage of this project. 



\newpage


\section*{Supplementary Materials}
We present the mathematical background for the methodology that we used in the main text to construct a topological measure for how closely the point clouds that we obtain from contagion maps (see section~\ref{topology} of the main text). For proofs and further discussion of this theory, see \cite{Edelsbrunner2010}. For a condensed and accessible introduction, see \cite{Otter2017}.

\section{Simplicial Homology}

\begin{definition}
An \emph{abstract finite simplicial complex} is a finite collection $\Sigma$ of finite ordered sets that is closed under inclusion: whenever $\alpha \in \Sigma$ and $\beta \subseteq \alpha$, it follows that $\beta \in \Sigma$. 
\end{definition}

The elements of $\Sigma$ are called \emph{simplices}. A \emph{face} of a simplex $\alpha$ is a {nonempty} proper subset $\beta \subset \alpha$. The dimension of a simplex $\alpha \in \Sigma$ is $|\alpha|-1$. The $0$-dimensional simplices are called \emph{vertices}, and we denote the set of vertices by $V(\Sigma)$. The dimension of a simplicial complex is the maximum of the dimensions of the simplices that it contains. A \emph{simplicial subcomplex} $\Omega \subseteq \Sigma$ of a simplicial complex $\Sigma$ is a subcollection of simplices that is itself a simplicial complex. For $n \in \mathbb{N}$, the \emph{$n$-skeleton} of a simplicial complex is the union of its simplices of dimensions $m \leq n$.

\begin{figure}[H]
 \centering
   \leftline{\hskip 2.9cm (a) \hskip .7cm (b) \hskip 1.1cm (c) \hskip 1.7cm (d)} 
    \begin{minipage}{.5\textwidth}
        \centering
        \includegraphics[width=1.00\textwidth]{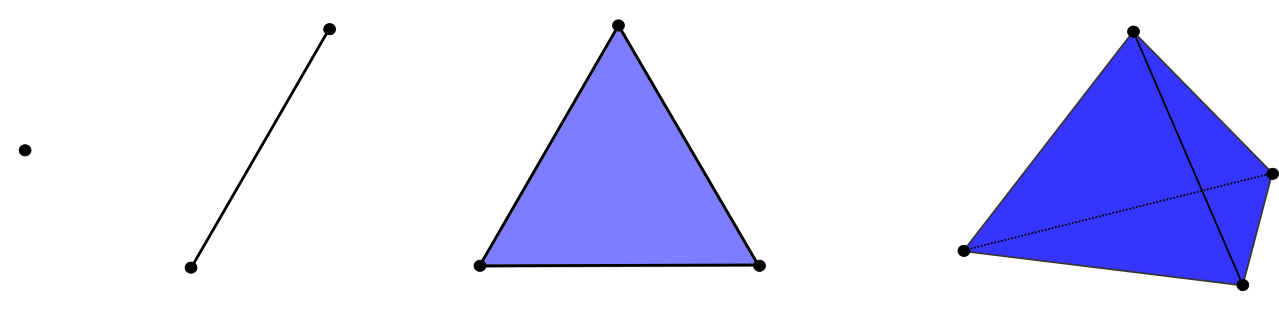}
    \end{minipage}
    \vspace{2mm}
    \leftline{\hskip 3cm (e)} 
    \begin{minipage}{.5\textwidth}
    \centering
        \includegraphics[width=1.00\textwidth]{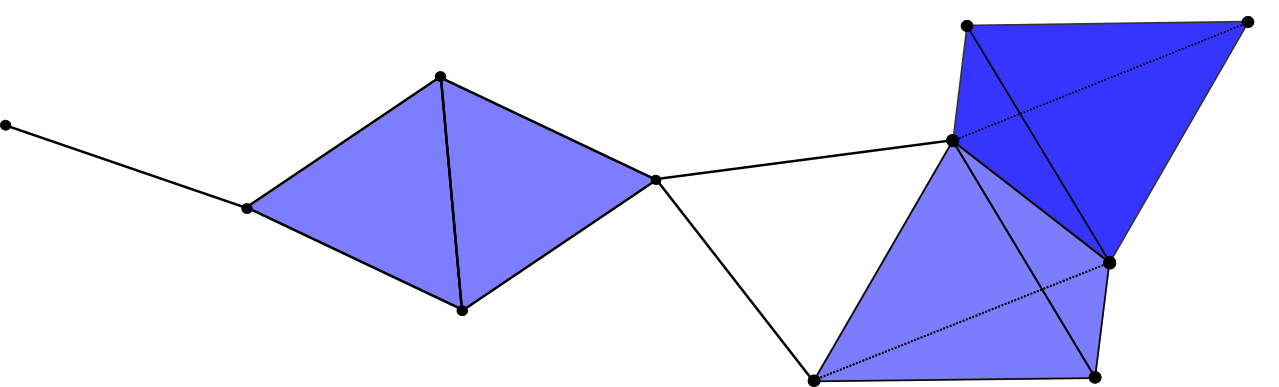}
    \end{minipage}
    \caption{Geometric realizations of (a) a $0$-simplex, (b) a $1$-simplex, (c) a $2$-simplex, (d) a $3$-simplex, and (e) a 3D simplicial complex.}\label{fig:geometric_realizations}
\end{figure}

To each simplex, we can assign a polytope, which is called its \emph{geometric realization}. A $0$-simplex corresponds to a vertex, a $1$-simplex corresponds to an edge, a $2$-simplex corresponds to a triangle, a $3$-simplex corresponds to a tetrahedron, and so on. A simplicial complex $\Sigma$ can thereby be represented as a subset of the simplex that is spanned by its vertices. See Figure~\ref{fig:geometric_realizations} for examples of geometric realizations.

Let $\Sigma$ be a simplicial complex, let $k$ be an integer, and let $\mathbb{F}$ some field. A \emph{$k$-chain} is a linear combination of $k$-simplices in $\Sigma$ over $\mathbb{F}$. We can turn the set $C_k$ of $k$-chains into a vector space by defining addition and scalar multiplication to be component-wise. In topological data analysis, the most common field is $\mathbb{Z}/2\mathbb{Z}$; in this case, a $k$-chain can be interpreted as a collection of $k$-simplices in $\Sigma$. When working over $\mathbb{Z}/2\mathbb{Z}$, addition is equivalent to taking the symmetric difference. It is straightforward to check that $C_k$ satisfies the axioms of a vector space with this definition of addition and scalar multiplication and that the zero vector is the empty set. 

The \emph{boundary} $\partial_k(\alpha)$ of a $k$-simplex $\alpha$ is the alternating sum of its $(k-1)$-dimensional faces. The \emph{boundary} of a $k$-chain is the sum of the boundaries of its simplices. The boundary of a $k$-chain is a $(k-1)$-chain, so the boundary operator $\partial_k$ defines a function $\partial_k : C_k \longrightarrow C_{k-1}$. This function commutes with vector addition and scalar multiplication on $C_k$. That is, $\partial_k$ is a linear map; it is called the \emph{boundary operator}. We thus have a sequence of vector spaces that are connected by boundary operators:
\begin{equation}\label{chaincomplex}
\cdots \xrightarrow{\partial_{k+2}} C_{k+1} \xrightarrow{\partial_{k+1}} C_k \xrightarrow{\partial_{k}} C_{k-1} \xrightarrow{\partial_{k-1}} \cdots .
\end{equation}

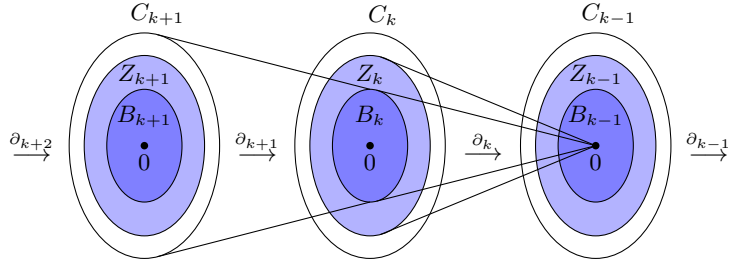
\begin{figure}[!h]
  \centering
  \begin{tikzpicture}[font=\small]
    \draw (0,0) circle (1 and 1.5); 
    \draw (-3,0) circle (1 and 1.5);
    \draw (3,0) circle (1 and 1.5); 
    \filldraw[color=blue!30,draw=black] (0,0) circle (.8 and 1.2); 
    \filldraw[color=blue!50,draw=black] (0,0) circle (.5 and .75); 

    \filldraw[color=blue!30,draw=black] (-3,0) circle (.8 and 1.2); 
    \filldraw[color=blue!50,draw=black] (-3,0) circle (.5 and .75);

    \filldraw[color=blue!30,draw=black] (3,0) circle (.8 and 1.2); 
    \filldraw[color=blue!50,draw=black] (3,0) circle (.5 and .75);

    \fill[color=black] (0,0) node[below] {0} circle (0.05); 
    \fill[color=black] (3,0) node[below] {0} circle (0.05); 
    \fill[color=black] (-3,0) node[below] {0} circle (0.05); 

    \node[above] at (80:1 and 1.5) {$C_k$}; 
    \node[below] at (90:.8 and 1.2) {$Z_k$}; 
    \node[below] at (90:.5 and .65) {$B_k$}; 

    \node[above] at ($(-3,0)+(80:1 and 1.5)$) {$C_{k+1}$}; 
    \node[below] at ($(-3,0)+(90:.8 and 1.2)$) {$Z_{k+1}$}; 
    \node[below] at ($(-3,0)+(90:.5 and .65)$) {$B_{k+1}$}; 

    \node[above] at ($(3,0)+(80:1 and 1.5)$) {$C_{k-1}$};
    \node[below] at ($(3,0)+(90:.8 and 1.2)$) {$Z_{k-1}$}; 
    \node[below] at ($(3,0)+(90:.5 and .65)$) {$B_{k-1}$}; 

    \draw ($(-3,0)+(80:1 and 1.5)$) -- ($(90:.5 and .75)$); 
    \draw ($(-3,0)+(-80:1 and 1.5)$) -- ($(-90:.5 and .75)$); 
    \draw ($(80:.8 and 1.2)$) -- ($(3,0)$); 
    \draw ($(-80:.8 and 1.2)$) -- ($(3,0)$);
    \draw ($(80:.5 and .75)$) -- ($(3,0)$); 
    \draw ($(-80:.5 and .75)$) -- ($(3,0)$); 
    \node at (-4.5,0)  {$\stackrel{\partial_{k+2}}{\longrightarrow}$};
    \node at (-1.5,0) {$\stackrel{\partial_{k+1}}{\longrightarrow}$}; 
    \node at (1.5,0)  {$\stackrel{\partial_{k}}{\longrightarrow}$};
    \node at (4.5,0)  {$\stackrel{\partial_{k-1}}{\longrightarrow}$};
  \end{tikzpicture}
    \caption{A chain complex consists of a sequence of chain, cycle, and boundary spaces that are connected by boundary operators.}
    \label{fig: boundary_map}
\end{figure}

A $k$-chain in the image $B_k$ of $\partial_{k+1}$ is called a \emph{$k$-boundary}. A $k$-chain in the kernel $Z_k$ of $\partial_{k}$ is called a \emph{$k$-cycle} (see Figure~\ref{fig: boundary_map}). A fundamental property of the boundary operator is that the boundary of a boundary is the zero vector. Consequently, the  sequence~(\ref{chaincomplex}) is a \emph{chain complex}.

\begin{lemma}{\rm (Fundamental Lemma of Homology)}
For any an integer $k$ and $(k+1)$-chain $d \in C_{k+1}$, we have that $\partial_k\partial_{k+1}(d)=0$. 
\end{lemma}

That is, the $k$th boundary space $B_k$ is a subspace of the $k$th cycle space $Z_k$, so the following definition makes sense. 

\begin{definition}\label{Betti}
Given a simplicial complex $\Sigma$ and an integer $k$, the \emph{$k$th homology} $H_k(\Sigma)$ is the quotient vector space of the $k$th cycle space $Z_k(\Sigma)$ by the $k$th boundary space $B_k(\Sigma)$: 
\begin{equation*}
	H_k(\Sigma)=Z_k(\Sigma)/B_k(\Sigma) \,.
\end{equation*}	 
The $k$th \emph{Betti number} $\beta_k(\Sigma)$ is the dimension of the $k$th homology of $\Sigma$: 
\begin{equation*}
	\beta_k(\Sigma)=\dim H_k(\Sigma)=\dim Z_k(\Sigma) - \dim B_k(\Sigma) \ .
\end{equation*}	
\end{definition}
Two $k$-cycles represent the same element of the $k$th homology $H_k$ if they differ only by $k$-boundaries. Roughly speaking, $\beta_n(\Sigma)$ is the number of $n$-dimensional ``holes'' of the space $\Sigma$. For example, $\beta_0(\Sigma)$ is the number of connected components, $\beta_1(\Sigma)$ is the number of ``tunnels,'' and $\beta_2(\Sigma)$ is the number of ``voids'' of the geometric realization of $\Sigma$.

\section*{Persistent Homology}

 \begin{definition}
A finite \emph{filtration} of a finite simplicial complex $\Sigma$ is a nested sequence of simplicial subcomplexes of $\Sigma$, such that the $0$th member of the sequence is the empty complex and the last member of it is all of $\Sigma$. That is,
\begin{equation*}
	\emptyset = F_0 \subseteq F_1 \subseteq \cdots \subseteq F_n = \Sigma \, .
\end{equation*}
\end{definition}
A filtration of a simplicial complex $\Sigma$ can be viewed as the construction of $\Sigma$ from the empty set by sequentially adding collections of simplices. 

Given a filtration $F_0 \subseteq F_1 \subseteq \cdots \subseteq F_n$ of a simplicial complex $\Sigma$, for every $i \leq j$ and dimension $k$, the inclusion map from $F_i$ to $F_j$ induces a linear map 
\begin{equation*}
	f_k^{i,j} : H_k(F_i) \longrightarrow H_k(F_j) \, . 
\end{equation*}	
Therefore, there is a sequence of homologies that are related via these linear maps:
\begin{equation*}
	0 = H_k(F_0) \longrightarrow H_k(F_1) \longrightarrow \cdots \longrightarrow H_k(F_n)=H_k(\Sigma) \,.
\end{equation*}
One can track the evolution of $\Sigma$ along the filtration through the algebraic structures of the homologies in this sequence. 

We can generalize the notion of homology in the setting of a filtration of a simplicial complex.

\begin{definition}\label{this}
For an integer $k$, the \emph{$k$th persistent homologies} (PH) are the images of the linear maps induced by inclusion: 
\begin{equation*}
	H_k^{i,j}= \im f_k^{i,j}\,, \,\, \text{   }0 \leq i \leq j \leq n \,, \quad i,j \in \mathbb{N}\,. 
\end{equation*}	
The \emph{$k$th persistent Betti numbers} are the dimensions of these spaces: $\beta_k^{i,j}=\mathrm{dim}(H_k^{i,j})$.
\end{definition} 

Using Definition \ref{this}, we can formalize the notion of birth and death of a homology class. 

\begin{definition}
A nonzero homology class $\xi \in H_k(F_i)$ is \emph{born at} $F_i$ if $\xi \notin H_k^{i-1,i}$, and it \emph{dies at} $F_j$ if $f_k^{i,j-1}(\xi) \neq 0$ but $f_k^{i,j}(\xi) = 0$.
\end{definition}

For $0 \leq i \leq j \leq n$, the $k$th PH $H_k^{i,j}$ can be interpreted as the space that consists of all homology classes that are born at or before $F_i$ and are still alive at $F_j$.

We can now finally give a mathematically rigorous definition of persistence. 
\begin{definition}
Let $\emptyset = F_0 \subseteq F_1 \subseteq \cdots \subseteq F_n = \Sigma$ be a filtration of a simplicial complex $\Sigma$, and let $k$ be an integer.
If $\xi \in H_k(F_i)$ is a homology class that is born at $F_i$ and dies at $F_j$, then its \emph{persistence} (also known as its ``lifespan'') is the difference $\mathrm{pers}(\xi) = j-i$. If $\xi$ never dies, we note its death time as infinite (i.e., $j=\infty$), making its persistence infinite as well (i.e., $\mathrm{pers}(\xi)= \infty$). The interval $[i,j)$ is called a \emph{persistence interval}. 
\end{definition}


\section*{Barcodes, Persistence Diagrams, and Wasserstein Distance}
It is a fundamental theorem \cite{Zomorodian2005} that one can find elements $\{ x_{\ell} \}_{\ell \in \Lambda_k}$ (with $x_{\ell} \in H_k(F_{b_{\ell}})$, where $b_{\ell}\in \mathbb{N}$ is the birth time of $x_{\ell}$ and $d_{\ell}\in \mathbb{N} \cup \{\infty \}$ is the death time of $x_{\ell}$) such that, for each $j \in \{0, \dots , n\}$, we have that $$\left\{f_{k}^{b_{\ell},j}\left( x_{\ell}\right) \ | \ \ell \in \Lambda_k \ , \  b_{\ell} \leq j < d_{\ell}\right\}$$ forms a basis for $H_k(F_j)$. If the number of births (respectively, deaths) of homology classes in $\{ x_{\ell} \}_{\ell \in \Lambda_k}$ exceeds the number of deaths (respectively, births) at $F_j$, then $\beta_k$ increases (respectively, decreases). Tracking the change of Betti numbers during a filtration is thus useful for monitoring the topological evolution of the growing complex. A topological feature that emerges with the birth of a homology class $\xi$ and disappears with the death of $\xi$ is said to ``correspond to'' $\xi$ and has persistence pers$(\xi$). The features that persist for a long time interval are usually considered to be the defining features of the complex, although this is not always the case. 

\begin{figure}[H]
    \centering
     \leftline{\hskip 0.00cm (a) \hskip 6.5cm (b)} 
    \begin{minipage}{.49\textwidth}
        \centering
        \includegraphics[width=1.00\textwidth]{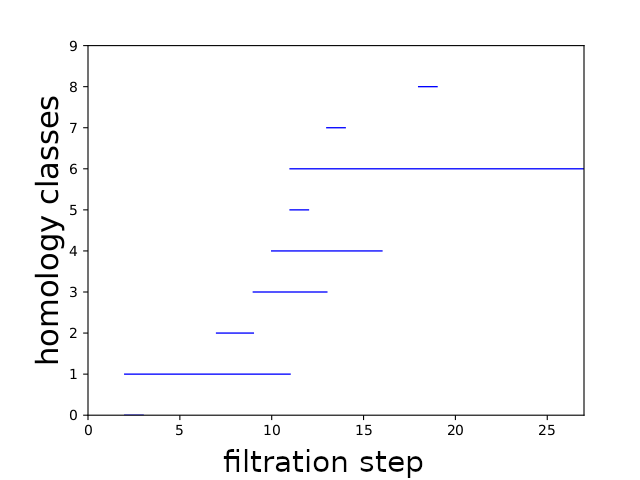} 
    \end{minipage}\hfill
    \begin{minipage}{0.49\textwidth}
        \centering
        \includegraphics[width=1\textwidth]{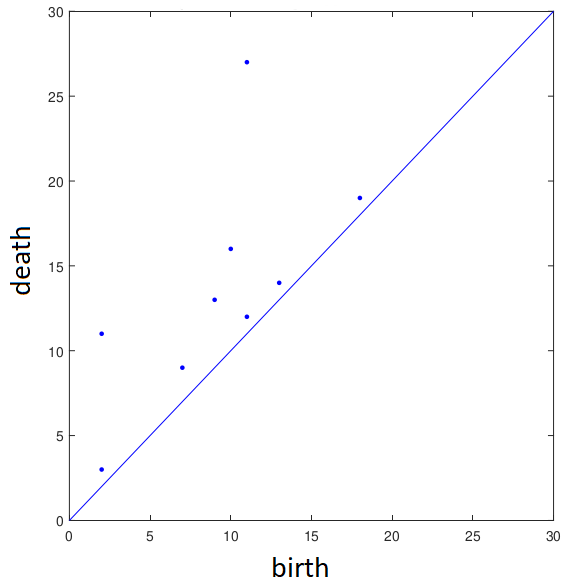} 
    \end{minipage}\hfill
        \caption{(a) An example of a barcode. Each horizontal bar indicates the lifespan of a homology class. (b) Equivalent persistence diagram. Points in the extended plane $\overline{\mathbb{R}}^2$, where $\overline{\mathbb{R}}=\mathbb{R} \cup \{\infty\}$, mark the persistence intervals.}\label{barcode_persistence_diagram}
\end{figure}

The collection of persistence intervals $[b_{\ell},d_{\ell})$ corresponding to elements in $\{ x_{\ell} \}_{\ell \in \Lambda_k}$ can be viewed as the filtered analogue of the Betti number $\beta_k$. Such persistence intervals can be represented as barcodes, or as diagrams (see Figure~\ref{barcode_persistence_diagram}). 

\begin{definition}
The \emph{persistence diagram} of the $k$th PH of a filtered simplicial complex is the multiset 
\begin{equation*}
	\left\{(b_{\ell},d_{\ell}) \in \overline{\mathbb{R}}^2 \ | \ \ell \in \Lambda_k \right\} \cup \left\{(i,j) \in \mathbb{R}^2 \ | \ i=j \right\} \subseteq \overline{\mathbb{R}}^2  \,,
\end{equation*}
where $\overline{\mathbb{R}}=\mathbb{R} \cup \{\infty\}$. 
\end{definition}
Note that all persistence diagrams have equal cardinality. For dimension $k \in \mathbb{Z}$, the collection of persistence intervals $[b_{\ell},d_{\ell})$ corresponding to elements in $\{ x_{\ell} \}_{\ell \in \Lambda_k}$ is called a \emph{barcode} (see Figure~\ref{barcode_persistence_diagram}). 

One can turn the space of persistence diagrams (equivalently, the space of barcodes) into an extended metric space by defining the following notion of distance between two persistence diagrams. 
\begin{definition}\label{Wasserstein}
Given two persistence diagrams $D_1$ and $D_2$, an extended metric $d$ on $\overline{\mathbb{R}}^2$, and a number $p \in [1,\infty]$, the \emph{$p$th Wasserstein distance} between $D_1$ and $D_2$ is 
\begin{equation*}
	W_p[d](D_1,D_2):= \inf_{\phi: D_1 \rightarrow D_2} \left[ \sum_{x \in D_1}d\left[ x,\phi(x)\right]^p\right]^{1/p} \, ,
\end{equation*}
where $\phi$ ranges over all bijections from $D_1$ to $D_2$.  

If $p=\infty$ and $d=L_{\infty}$, where $L_{\infty}\left((x_1,y_1),(x_2,y_2)\right)= \sup\{|x_1-x_2|,|y_1-y_2| \}$, the Wasserstein distance $W_{\infty}[L_{\infty}]$ is called the \emph{bottleneck distance}.

\end{definition}
One property of PH that is central to its utility in applications is that it is stable and therefore robust to noise: A small perturbation of input data induces only a small perturbation of the {corresponding} persistence diagram with respect to the bottleneck distance. 


\section*{\v{C}ech Complex and Vietoris--Rips Complex}

Persistent homology is a useful tool for analyzing point-cloud data. A \emph{point cloud} is a finite set of points, 
\begin{equation*}
	P = \{ x_1, x_2, \dots , x_l \} \subseteq M\,,
\end{equation*} 
in {a metric space} $(M,d)$. One can view $P$ as a sample from some subspace of $M$. There are various ways to construct a simplicial complex from a point cloud. Two of the most common constructions are the \v{C}ech complex and the Vietoris--Rips complex, which we now describe.

Given a nonnegative number $\epsilon$, recall that the closed ball $B_{\epsilon}(x)$ of radius $\epsilon$ around a point $x \in M$ is the set of points within distance $\epsilon$ from $x$; that is, $B_{\epsilon}(x)= \left\{ y \in M : d(x,y) \leq \epsilon  \right\}$.

\begin{definition}
Let $(M,d)$ be {a metric space}, and let $P = \{ x_1, x_2, \dots , x_l \} \subseteq M$ be a point cloud in $M$. For $\epsilon \geq 0$, the \emph{\v{C}ech complex} $\mathcal{C}_{\epsilon}$ at $\epsilon$ associated with $P$ is the simplicial complex whose simplices are sets of points in $P$ whose closed $(\epsilon / 2)$-balls have nonempty intersection. The set $\{y_1, \dots, y_k\} \subseteq P$ is a $(k-1)$-simplex of $\mathcal{C}_{\epsilon}$ if and only if $\bigcap\limits_{i=1}^{k}B_{\epsilon / 2}(y_i) \neq \emptyset$.
\end{definition}
This set of simplices is closed under taking subsets. That is, the conditions for a simplicial complex are indeed satisfied by this definition. 
The $0$-simplices of $\mathcal{C}_{\epsilon}$ correspond precisely to the points $x_1, x_2, \dots , x_l$; the $1$-simplices are the pairs of points that are within distance $\epsilon$ of each other; the $2$-simplices are the triples of points whose $(\epsilon / 2 $)-balls have nonempty intersection; and so on.   
The \v{C}ech complex $\mathcal{C}_0$ is the set of points in $P$. For sufficiently large $\epsilon$ (to be precise, for $\epsilon$ at least as large as the diameter of $P$), the \v{C}ech complex $\mathcal{C}_{\epsilon}$ is the $(l-1)$-dimensional simplex $\{ x_1, x_2, \dots , x_l \}$ together with all of its faces. If $\epsilon_1 \leq \epsilon_2$, then $\mathcal{C}_{\epsilon_1} \subseteq \mathcal{C}_{\epsilon_2}$, and increasing $\epsilon$ incrementally from $0$ to a large enough value gives a filtration of \v{C}ech complexes associated with the point cloud $P$:
\begin{equation*}
	\emptyset \subseteq \mathcal{C}_0 \subseteq \dots \subseteq \mathcal{C}_{\epsilon_{\rm large}}\,.
\end{equation*}

For $M=\mathbb{R}^n$, the following result, known as the \emph{Nerve Theorem}, states that the \v{C}ech complex $\mathcal{C}_{\epsilon}$ associated with a point cloud $P$ is topologically faithful to the union of the closed $(\epsilon / 2$)-balls around the points in $P$ in the sense that it has the same homotopy type. Intuitively, two spaces have the same homotopy type if they can be transformed into each other by bending, compressing, and expanding them (without having to do any cutting or gluing). 

\begin{theorem} {\rm (Nerve Theorem)}.
For a point cloud $P = \{ x_1, x_2, \dots , x_l \} \subseteq \mathbb{R}^n$ and $\epsilon \geq 0,$ the \v{C}ech complex $\mathcal{C}_{\epsilon}$ is homotopy equivalent to the union of the closed $(\epsilon / 2)$-balls around the points in $P$.
\end{theorem}

The Nerve Theorem justifies why, when a point cloud $P$ is a sample of some subspace of $M$, the \v{C}ech filtration can reveal features of this subspace. We expect that features that have a long lifespan in the \v{C}ech filtration are likely to correspond to features of the underlying space.

Whether the balls of a certain radius around a set of points in $P \subseteq M$ have a point of common intersection depends on the entire metric space $M$ and the position of $P$ in it. Checking for a point of common intersection is computationally intensive, so it can be impractical (or even infeasible) to construct the \v{C}ech filtration associated with a point cloud. 

The following construction of a simplicial complex from a point cloud depends only on the pairwise distances of the points. Therefore, it is computationally more efficient than constructing a \v{C}ech filtration and hence more useful in practice. 

\begin{definition}\label{Rips_definition}
Let $(M,d)$ be {a metric space}, and let $P = \{ x_1, x_2, \dots , x_l \} \subseteq M$ be a point cloud in $M$. For $\epsilon \geq 0$, the \emph{Vietoris--Rips (VR) complex} $\mathcal{R}_{\epsilon}$ at $\epsilon$ associated with $P$ is the simplicial complex whose simplices are sets of points in $P$ that are pairwise within distance $\epsilon$. The set $\{y_1, \dots, y_k\} \subseteq P$ is a $(k-1)$-simplex of $\mathcal{R}_{\epsilon}$ if and only if $d(y_i,y_j) \leq \epsilon$ for all $i,j \in \{1, \dots , k\}$.
\end{definition}

As with $\mathcal{C}_{\epsilon}$, the $0$-simplices of $\mathcal{R}_{\epsilon}$ correspond precisely to the points $x_1, x_2, \dots , x_l$, and the $1$-simplices of $\mathcal{R}_{\epsilon}$ are the pairs of points that are within distance $\epsilon$ of each other. Therefore, the $1$-skeleton of $\mathcal{R}_{\epsilon}$ is the same as that of $\mathcal{C}_{\epsilon}$. In the definition of higher-dimensional simplices, only the pairwise distances of points play a role. The simplicial complex $\mathcal{R}_{\epsilon}$ is the maximal simplicial complex can be built on its $1$-skeleton; its $k$-simplices are the $(k+1)$-cliques of its $1$-skeleton. Consequently, the $1$-skeleton of $\mathcal{R}_{\epsilon}$ completely determines the entire simplicial complex. This is an attractive quality from a computational point of view, because it implies that it is possible to store a VR complex as a graph (its $1$-skeleton).

If $\epsilon_1 \leq \epsilon_2$, then $\mathcal{R}_{\epsilon_1} \subseteq \mathcal{R}_{\epsilon_2}$; increasing $\epsilon$ incremently from $0$ to a value larger than the maximum distance beween any pair of points in a point cloud gives a filtration of VR complexes whose $0$th member is the collection of $0$-simplices and whose final member is the $(l-1)$-dimensional simplex $\{ x_1, x_2, \dots , x_l \}$ together with its faces. That is,
\begin{equation*}
	\emptyset \subseteq \mathcal{R}_0 \subseteq \dots \subseteq \mathcal{R}_{\epsilon_{\rm large}}\,.
\end{equation*}

Let's now return to the special case $M=\mathbb{R}^n$. Although a VR complex $\mathcal{R}_{\epsilon}$ associated with a point cloud $P \subseteq \mathbb{R}^n$ is not a faithful representation of the union of balls around the points in $P$ and may not even be topologically equivalent to a subspace of $\mathbb{R}^n$, VR complexes provide a good approximation in the light of persistence, as the following lemma, due to de Silva and Ghrist \cite{Silva2007}, shows.

\begin{lemma}\label{inclusion} 
For $M=\mathbb{R}^n$ and any $\epsilon \geq 0$, we have that
\begin{equation*}
	\mathcal{R}_{\epsilon} \subseteq \mathcal{C}_{\epsilon \sqrt{2}} \subseteq \mathcal{R}_{\epsilon \sqrt{2}}\,.
\end{equation*}
\end{lemma}
Consequently, any topological feature that persists between $\mathcal{R}_{\epsilon}$ and $\mathcal{R}_{\epsilon \sqrt{2}}$ in the VR filtration is also a feature of the \v{C}ech complex $\mathcal{C}_{\epsilon \sqrt{2}}$.


\end{document}